\documentclass[12pt, twoside]{book}

\usepackage{changepage}
\usepackage{amsmath}
\usepackage{amssymb}
\usepackage{graphicx}
\usepackage[textfont={small,it}]{caption}
\usepackage{float}
\usepackage{hyperref}
\usepackage{xcolor}
\usepackage[style=authoryear]{biblatex}

\usepackage[normalem]{ulem}
\usepackage{subcaption}
\usepackage[toc,page]{appendix}
\usepackage{pdfpages}
\usepackage{mhchem}
\usepackage{braket}

\begin{document}

\frontmatter

\includepdf{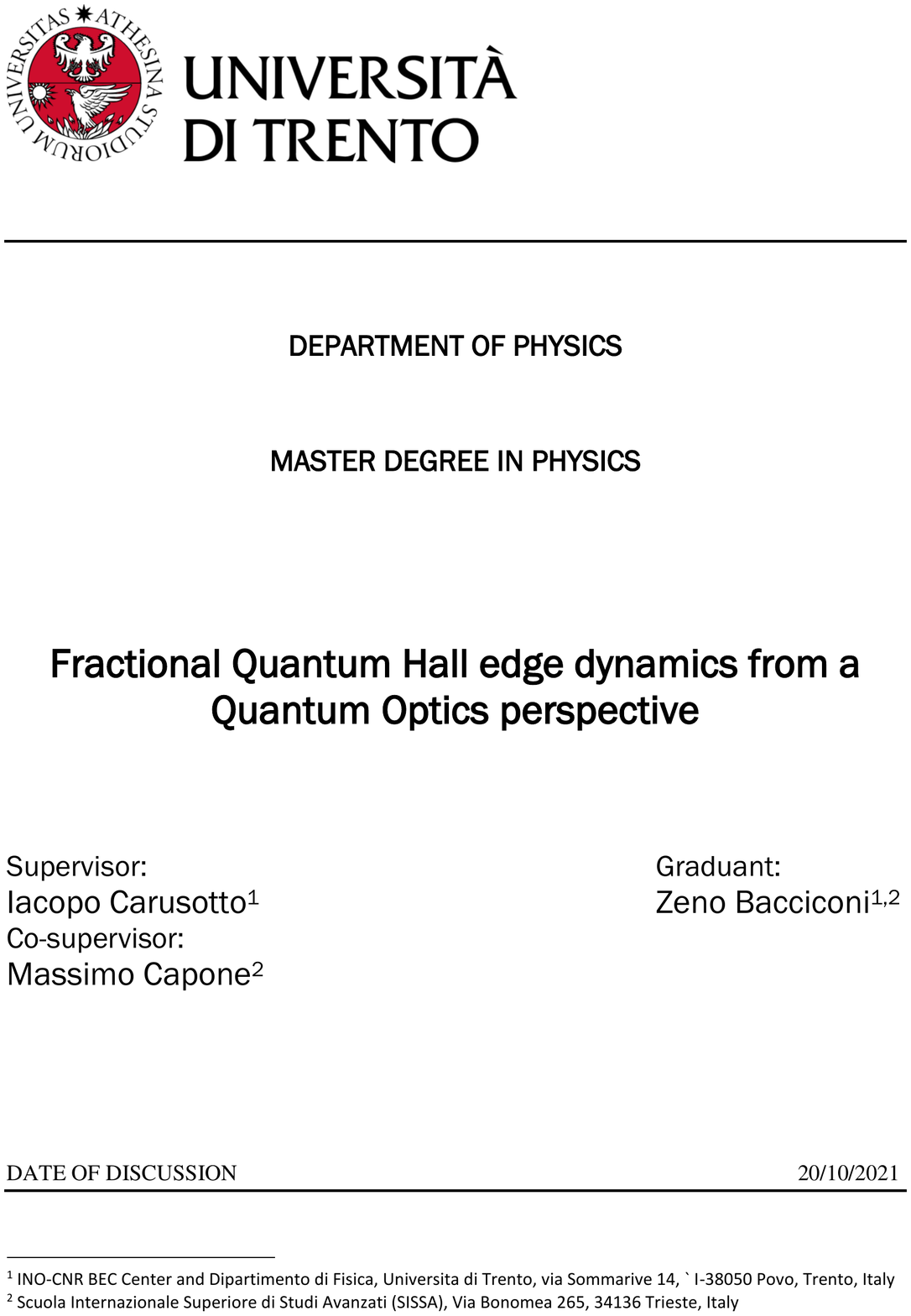}

\newpage
\pagestyle{empty}

\tableofcontents \thispagestyle{empty}

\mainmatter

\pagestyle{myheadings}

\section*{Abstract}
The goal of this master thesis is to give a different perspective on the problem of quasi-particle tunneling in Fractional Quantum Hall liquids by using a typical Quantum Optics tool such as the Truncated Wigner Approximation. Our novel approach provides a mapping of the quasi-particle tunneling problem into a stochastic classical field problem. This offers an intuitive physical picture which allows us to recover non-trivial features of FQH physics and predict novel phenomena.

\chapter*{Introduction}
\begin{figure}[!h]
    \centering
    \includegraphics[width=10cm]{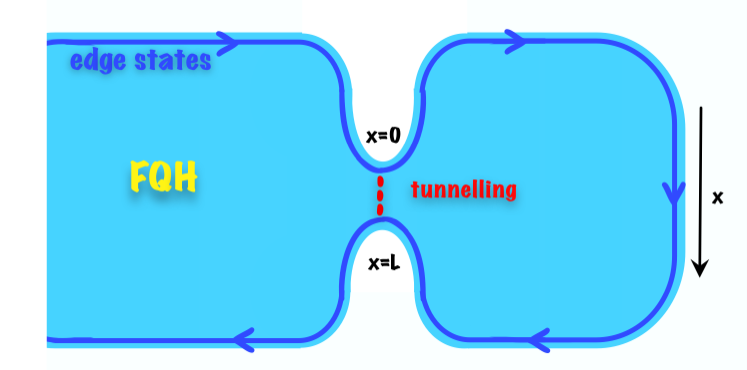}
    \caption{Pictorial representation of tunneling experiments}
    \label{fig:intro_geometry}
\end{figure}

\addcontentsline{toc}{chapter}{Introduction}
The Quantum Hall Effect (QHE) is probably one of the richest physical systems in terms of exotic phenomena and has been the drive for countless fields of research. It has been discovered in 1980  by looking at the transport properties of two dimensional electron gases (2DEG) at high magnetic fields and low temperatures \cite{Klitzing1980}. In these regimes a new phase of matter arises whose most evident feature is the quantized Hall conductance:
\begin{equation*}
    \sigma_H = \nu \frac{e^2}{2\pi\hbar}
\end{equation*}
with $\sigma_0= e^2/2\pi\hbar$ the conductance quantum and $\nu$ the so called filling factor. The first observations measured $\nu$ as an integer but soon after in 1982 some particular rational fractions were also observed. We then distinguish two cases based on $\nu$, the integer quantum Hall effect (IQHE) when $\nu$ is an integer and the fractional quantum Hall effect (FQHE) when $\nu$ is a fraction. In both cases the crucial ingredient behind such quantization is the topological nature of these phases of matter (\cite{tonglectures}). Although similar from a phenomenological point of view, the integer and fractional cases require different descriptions: while the former can be perfectly explained with non-interacting particles, the latter is a strongly correlated phase where the interplay between interactions and topology gives rise to even more fascinating and exotic phenomena ruled by a new kind of order, the \textit{topological order}. One of the most intriguing features is the emergence of excitations with fractional charge which are neither boson nor fermions but anyons, pick up a non-trivial phase upon an exchange (\cite{Leinaas_Myrheim}). Another important property of FQH states is their incompressibility which, together with confinement, makes bulk excitations gapped and leaves gapless excitations at the edges. The edge states of a FQH propagate \textit{chirally}, i.e. they move in one direction only, but, differently from the simple IQH case, these states are not single-particle ones on the basis of free particles and are intrinsically strongly interacting as the bulk. \\
The theoretical framework which best describe such FQH edge states is that of a chiral Luttinger Liquid ($\chi$LL) as pointed out by Xiao-Gang Wen in the beginning of the 90s \cite{Wen1990,Wen1990_2,Wen_pert_theory}. The Luttinger Liquid (LL) theory had been set up some years before Wen's works in order to describe locally interacting electrons in 1D which even for small interactions falls outside the Fermi-Liquid picture \cite{Haldane_LL}. In the LL formalism there is one parameter $g$ that governs the strength of interaction; in particular it is greater than one for attractive interactions and smaller than one for repulsive interactions. Wen's hypothesis was that the edges of a FQH liquid can be described by "half" of a LL, only right or left moving part, with a specific $g=\nu$ dictated only by the bulk filling $\nu$\footnote{The picture is a bit more complicated when $\nu$ is not in the Laughlin sequence $1/m$ with $m$ integer and multiple edge $\chi$LL are needed with non-universal coupling between them}. In the years the $\chi$LL description have revealed itself as an almost exact picture arriving to our days as the commonly accepted framework \cite{Review_FeldmanHalperin2021,IsLL_universal,Exact_reviewSaleur2002,tonglectures,Kane_review1996,Giamarchi2003}. An important outcome of such theory is that the edge is well described by free bosonic modes which describe the plasmonic excitations of the charge density on the edge and move chirally at a constant velocity. Moreover, within the $\chi$LL, it is possible to describe charged excitations which carry exactly the fractional charge of the bulk excitations and also have the same quantum statistics. \\
The main probe for the $\chi$LL predictions and the non-trivial properties of FQH quasi-particles have been the tunneling between two counterpropagating chiral edge states as depicted in figure \ref{fig:intro_geometry} .
The $\chi$LL predicts particular scaling for DC transport properties with temperature and incoming current that have been widely tested at the end of the 90s and beginning of years 2000s \cite{Milliken1996_resonances,chiLL_extelectron_tunn}. Soon after, using DC transport properties once again, it has been possible to confirm the fractional charge using shot noise (\cite{DePicciotto1997}) and to provide evidence of the quantum statistics of the quasi-particles via interferometer experiments (\cite{Nakamura2020}). If on one side the DC scattering in $\chi$LL has been explored and exploited in all of its glory, on the other side time dependent scattering experiments have not been a main focus. We identify two main motivations for this:
\begin{itemize}
    \item \textbf{experimental accessibility}: Up to now FQH have been realized only in solid state systems where it requires ultralow temperatures and time-dependent measurements are challenging. Indeed one of the first successful AC experiment in FQH has been done only in 2019 \cite{PASN_fqh}
    \item \textbf{theoretical convenience}: Even though predictions about AC properties were investigated from the beginning with Wen's works, the lack of experimental accessibility probably pushed also the theoretical work towards DC properties.  
\end{itemize}
However, in the last 5 years the level of attention paid to time dependent situations in FQH has been growing (\cite{FQH_quantumdot,QuasiparticleEmitter_Ferraro2015}) and a first investigation of scattering of wavepacket-like excitations has been carried out showing a non-trivial effect named "crystallization" \cite{Cristallization} . A strong drive for this recent interest is given by the achievements of the Electron Quantum Optics (EQO) \cite{EQO_review2014,EQO_levitonsReview2017,EQO_wigner_interf_2017}. EQO deals with the coherent transport of electron at the edges of IQH system where time resolved sources of electrons have been realised (\cite{EQO_coherence_SingleElectronSource}) and Hong-Ou-Mandel type of experiments have been successfully carried out with geometries similar to that of figure \ref{fig:intro_geometry} (\cite{EQO_review2014}). Translating EQO to FQH edge states would mean to have coherent and time resolved anyon sources and the possibility to control them as they travel on the chiral edge states. This is a very fascinating idea \textit{per se} but it could also find strong applications in the realization of topologically protected quantum computers \cite{TopoQC_review}. Differently from the IQH case where the edges are simple free electronic states in the FQH case the edges are described by a $\chi$LL and the solution of the time dependent scattering problem beyond perturbation theory is highly non-trivial and the physical interpretation of the result is not yet clear. \\
Another justification for the exploration of such physics comes from the fact that a whole new generation of experimental platforms realizing FQH states are under intense study. These are generally referred to as syntetic materials and realistic proposals for the realization of FQH exists in cold atoms \cite{Cooper_topoatoms}, circuit QED \cite{Carusotto_PhotonicMaterials2020,Roushan2017_cQED_FQH} and Rydberg polaritons in twisted cavities \cite{Simon2020_Rydberg_fqh}. All these platforms are characterized by much more accessible time domain measurements and in general a wider range of possible observables to look at. Actually edge dynamics experiments are probably better suited to these type of platforms than the DC transport properties which have been so successfully measured in electronic samples. The problem with synthetic materials may be the scalability of the systems that may not reach the thermodynamic limit where the $\chi$LL is valid. \\
In this context the work of this thesis tries to give a new perspective to the problem and a clear physical interpretation of time dependent scattering process within FQH edges. The central point of view of our approach is to treat the linear dispersion of $\chi$LL excitations as free bosons and the quasi-particle tunneling hamiltonian as a non-linearity over the free $\chi$LL theory. As a starting point we solved the classical limit of the $\chi$LL theory with quasi-particle tunneling and already at this level interesting and non-trivial time dependent scattering phenomena are present. For example it is possible to recover in very simple terms the crystallization phenomenology obtained with a full quantum perturbation theory in \cite{Cristallization}. Then, the main part of the thesis concerns the treatment of the quasi-particle tunneling non-linearity with a Truncated Wigner Approximation (TWA), a tool widely used in the quantum optics community. The TWA provides a mapping of a complicated quantum field problem onto a more tractable classical stochastic problem that can be solved by integrating a Langevin-like equation. The TWA for a quadratic theory as the free $\chi$LL is exact and the non-linear term is usually treated within an approximation. In our case the non-linear element is the quasi-particle tunneling Hamiltonian. Within the TWA we found that the Langevin equation that comes out only include drift terms corresponding to the classical evolution. The stochasticity in the Langevin equation then comes only from the initial condition that must reproduce the non-trivial correlations present in the free $\chi$LL. In contrast to other typical situations where the TWA has been applied, here the TWA is non-perturbative in the tunneling Hamiltonian. The small parameter of the truncation is the filling $\nu$ so that the theory becomes exact in the limit $\nu\rightarrow 0$ or equivalently $m=\nu^{-1} \rightarrow \infty$. The approximation is then exact at zero tunneling and in the limit of $m\rightarrow\infty$ but it turns out to be able to recover known results in the whole range of tunneling strengths and even for $m\simeq 1$. We are confident that the approximation can be trusted also for other non-trivial effects we found in the time dependent scattering problem. In particular, the TWA allows us to get a physical understanding of what is happening in the scattering processes. The quantum evolution can be understood as the classical evolution with the addition of fluctuations of strength proportional to $\nu$. Then as $\nu\rightarrow0$ the classical trajectories become the actual quantum dynamics and indeed this is what we find for $\nu\lesssim 1/2$. On the contrary at $\nu =1$ the fluctuations are large enough to wash out the classical trajectories which indeed do not reproduce the free particle picture of the IQH. The interesting fact is that even at $\nu=1$ the TWA works and predicts a very similar dynamic to the one that is obtained with a free particle picture. A deeper analysis of why the TWA works so well even when the small parameter $\nu$ is not so small will be object of future studies. \\
\newline
The structure of thesis is the following: first two chapters review the literature while the last three ones report our original results:
\begin{itemize}
    \item \textbf{Chapter 1} We give a general introduction on QH physics, from Landau Levels to Laughlin states. In the end we will focus on the edge states of the FQH effect and describe the chiral Luttinger Liquid ($\chi$LL) desccription. 
    \item \textbf{Chapter 2} This chapter introduces the tunneling or scattering in chiral Luttinger Liquid. It is meant to give an introduction to the problem and to the various results that have been obtained in the literature using the $\chi$LL theory. 
    \item \textbf{Chapter 3} Here we solve the tunneling problem at the classical level. We write down the equation of motion for the charge density and study the scattering of charge wavepackets off a constriction evidencing the highly nonlinear behavior of the quasi-particle tunneling hamiltonian
    \item \textbf{Chapter 4} The truncated Wigner approximation (TWA) is explained and applied to the $\chi$LL with tunneling. The second part of the chapter is devoted to a detailed test of the TWA by recovering from a novel point of view most of the very non-trivial DC transport properties discussed in chapter 2
    \item \textbf{Chapter 5} Here we use the TWA to solve the dynamics of wavepacket scattering in both the IQH and FQH cases.
    \\
    \item \textbf{Conclusions} We provide conclusions and prospects for future work.
\end{itemize}

\chapter{Quantum Hall effect}\label{chap:1}
The discovery of the Quantum Hall Effect (QHE) dates back to 1980 when Von Klitzing (\cite{Klitzing1980}) realized that 2D electron gases in high magnetic fields showed peculiar magnetoresistence properties. The first experimental platforms where Quantum Hall physics has been witnessed are semiconductor heterostructures of GaAs. Today Quantum Hall physics does not only belong to solid state systems but plays its important role also on cold atoms \cite{Cooper_topoatoms} and photonic systems \cite{Topophotonics}. In our treatment we will mainly refer to electronic systems but most of the discussed physics is valid also for synthetic systems.   \\
This chapter should give an introduction to the basics physics of the QHE. We will start from a microscopic description of the Integer and Fractional Quantum Hall effects and at the end we will discuss why chiral Luttinger Liquids are a good description of FQH edge states which are the main focus of this thesis. 

\section{Integer Quantum Hall}
\begin{figure}
\begin{subfigure}{.5\textwidth}
\centering
    \includegraphics[width=7cm]{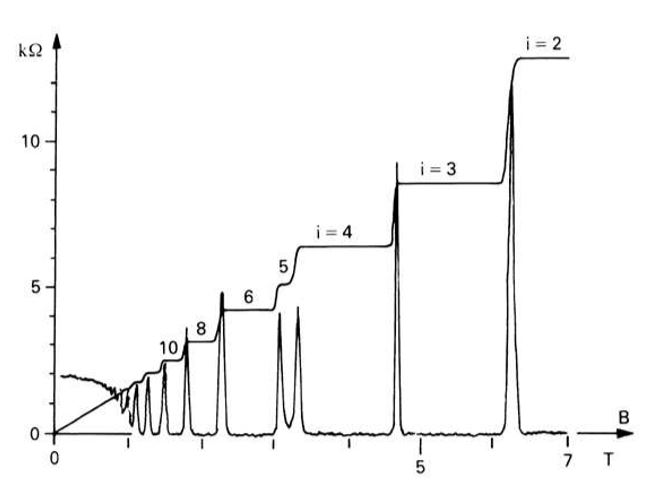}
    \caption{Magnetoresistence properties}
    \label{fig:1_iqh_a}
\end{subfigure}
\begin{subfigure}{.5\textwidth}
  \centering
  \includegraphics[width=6.5cm]{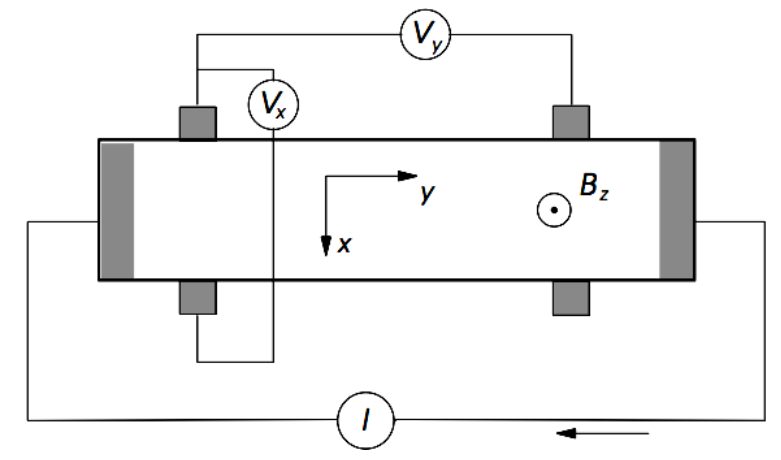}
  \caption{Hall bar geometry}
  \label{fig:1_iqh_b}
\end{subfigure}
\caption{On the left: Hall resistivity $\rho_{xy}$ and longitudinal resitivity $\rho_{xx}$ showing respectively plateaus at quantised values and a constant zero value as the magnetic field is varied. On the right: typical Hall bar geometry.Taken from \cite{tonglectures}}
\label{fig:1_iqh}
\end{figure}

In the Integer Quantum Hall (IQH) effect the Hall conductivity takes precisely quantized values as shown in figure \ref{fig:1_iqh}.
The Hall conductivity is measured by applying a current in one direction and measuring the voltage difference in the other. For example consider the Hall bar geometry in figure \ref{fig:1_iqh_b}, the Hall conductivity is defined as the ratio between the current along $y$ and the voltage difference along $x$, $\sigma_{xy}=I_y/V_x$. These plateaus are at values of:
\begin{equation}
    \sigma_{xy} = \frac{e^2}{2\pi\hbar} n 
\end{equation}
with n integer. The physics behind this is very rich and simple in its basic constituencies since it only need a non-interacting picture. In the following we will explain the main points, for a more detailed discussion see \cite{tonglectures}
\subsection{Landau Levels}\label{sec:1_Landau_levels}
In order to understand Quantum Hall physics is important to know what happens to charged particles in strong magnetic fields confined to 2 dimensions. The single particle Hamiltonian is the following:
\begin{equation}
    H_0 = \frac{\hbar^2}{2 m} (\bold{p}+\frac{e}{c} \bold{A})^2
\end{equation}
where $-e$ is the electron charge, $c$ the speed of light and $\bold{A}$ the vector potential. As we will see the eigenstates are organized in heavily degenerate levels known as Landau Levels which in the limit of strong magnetic field become more and more degenerate. We keep the gauge unspecified for the moment, we just have $\bold{B}= \nabla \times \bold{A}= B\,\hat{z}$ so that the magnetic field is perpendicular to the $x-y$ plane where particles are confined. The Hamiltonian can then be rephrased into the one of an harmonic oscillator by passing through the mechanical momentum operator which is a gauge invariant quantity:
\begin{equation}
   \boldsymbol{\pi} = \bold{p}+\frac{e}{c} \bold{A} 
\end{equation}
and satisfies the following commutation relations:
\begin{equation}
    [\pi_x,\pi_y]= -ie \hbar B
\end{equation}
Now the Hamiltonian becomes quadratic in the two components of the mechanical momentum. The ladder operators can be defined as:
\begin{equation}
    a = \frac{1}{\sqrt{2e\hbar B}}(\pi_x -i\pi_y) \qquad a^{\dagger} =  \frac{1}{\sqrt{2e\hbar B}}(\pi_x +i\pi_y) \qquad \textit{with}\qquad [a,a^\dagger]=1
\end{equation}
and the Hamiltonian is:
\begin{equation}
H = \frac{\hbar^2}{2 m} \boldsymbol{\pi}\cdot \boldsymbol{\pi} = \hbar \omega_B (a^\dagger a + \frac{1}{2})
\end{equation}
The cyclotron frequency $\omega_B= eB/m$ sets the separation between different levels of the harmonic oscillator and in the Quantum Hall regime will usually be the highest energy scale. \\
We have found an easy form of the Hamiltonian which includes only one quantum number for a system in two dimensions, where is the other degree of freedom? It does not enter in the single particle Hamiltonian and thus it must provide a degeneracy of the levels we found! \\

To make this degeneracy explicit we must choose a specific gauge for the magnetic field by fixing the vector potential. This will highlight different geometrical symmetries of the problem, here we briefly illustrate two possible choices.
\paragraph{Symmetric gauge}
The symmetric gauge keeps the rotational symmetry of the plane but breaks the translational invariance problem, the vector potential is:
\begin{equation}
    \bold{A}= -\frac{1}{2} \bold{r} \times \bold{B}= -\frac{yB}{2}\hat{\bold{x}} +\frac{xB}{2}\hat{\bold{y}}
\end{equation}
We can now define new gauge dependent operators which only in the symmetric gauge commute with $\bold{\pi}$ and hence with the Hamiltonian, these are:
\begin{equation}
    \Tilde{\boldsymbol{\pi}} = \bold{p} -\frac{e}{c}\bold{A} \qquad \textit{with} \qquad [\pi_i,\Tilde{\pi}_j]=0
\end{equation}
New ladders operators can then be defined:
\begin{equation}
    b = \frac{1}{\sqrt{2e\hbar B}}(\Tilde{\pi}_x +i\Tilde{\pi}_y) \qquad b^{\dagger} =  \frac{1}{\sqrt{2e\hbar B}}(\Tilde{\pi}_x -i\Tilde{\pi}_y) \qquad \textit{with}\qquad [b,b^\dagger]=1
\end{equation}
We then found a set of eigenstates for the single particle Hamiltonian which span the whole single-particle Hilbert space:
\begin{equation}
    \ket{n,m} = \frac{{a^{\dagger}}^n {b^\dagger} ^m}{\sqrt{n! m!}}\ket{0,0} \qquad E_{n,m}=E_n = \hbar \omega_B\; (n + \frac{1}{2})
\end{equation}
The eigenvalue only depends on $n$ thus giving degenerate levels of states with an energy spacing proportional to the magnetic field. The second quantum number $m$ represent the angular momentum, indeed the angular momentum operator is diagonal on this basis:
\begin{equation}
    \hat{J}_z\ket{n,m}=\hbar \,m \ket{n,m}
\end{equation}
Within the Lowest Landau Level (LLL) $n=0$ the wavefunctions of the eigenstates, using complex coordinates $z=x-iy$, are:
\begin{equation} \label{eq:1_LLLwavefunction}
\psi_{LLL,m}(z,\Bar{z}) = \Bigl(\frac{z}{l_B}\Bigl)^m \, e^{-|z|^2/4 l_B^2}
\end{equation}
where $l_B=\sqrt{\hbar /eB}$ is the magnetic length and sets the length scale of the single particle wavefunctions . The eigenstates are exponentially localized on a ring of radius $r_m = l_B\sqrt{2m}$ and width $\Delta_m\simeq l_B$. The number of states $\mathcal{N}$ within a disk of area $A$ and the density of states then are:
\begin{equation}
    \mathcal{N}= \frac{A}{2 \pi l_B^2} = e \frac{AB}{2\pi\hbar} \qquad\qquad n = \frac{1}{2\pi l_B^2}
\end{equation}
\paragraph{Landau gauge}
In the Landau gauge the vector potential is not rotationally invariant but it preserves the translational symmetry along one axis. A possible choice is:
\begin{equation}
    \bold{A} = xB\;\hat{\bold{y}}
\end{equation}
which is invariant under translations on $y$. The Hamiltonian then commute with the momentum along y giving us the second quantum number to span the Hilbert space. More explicitly we have: 
\begin{equation}
    H = \frac{1}{2m}\Bigl(p_x^2+ (p_y+eBx)^2 \Bigl)
\end{equation}
The eigenstates are then found via a separation of variables, on the $y$ axis we have plane waves: 
\begin{equation}
    \Psi_k(x,y) = e^{iky} f_k(x)\qquad \longrightarrow \qquad H\Psi_k(x,y) = H_{p_y=k}\Psi_k(x,y) 
\end{equation}
and the remaining part on the x variable is an harmonic oscillator:
\begin{equation}
    H_k = \frac{1}{2m}p_x^2 + \frac{m \omega_B^2}{2}(x + k l_B^2)^2
\end{equation}
where again $\omega_B$ and $l_B$ are the two magnetic energy and length scales. The solutions are characterized by two quantum numbers $(n,k)$, $k$ denotes momentum along y and $n$ is the level of the harmonic oscillator. The spectrum again reproduces Landau Levels with the degeneracy taken into account by the momentum quantum number. The eigenfunctions looks like strips parallel to the $y$ axis exponentially localized within $l_B$ around the center of the harmonic potential at $x = -k l_B^2$. Within the LLL at $n=0$ the eigenfunctions are:
\begin{equation}
    \Psi_{LLL,k} =  e^{iky} e^{-(x+kl_B^2)^2/2l_B^2}
\end{equation}
We can again count the number of states within an area A, but now a rectangular shape is better suited for the counting. Momentum is then quantised in units of $2\pi/L_y$ and is restricted to a range dictated by the $x$ axis confinement. Take for example $x\in[0,L_x]$ then the momentum must be $k\in[-L_x/l_B^2,0]$. This gives the following number of states and a density of states:
\begin{equation}
    \mathcal{N}= \frac{L_x/l_B^2}{2\pi/L_y} = \frac{L_xL_y}{2\pi l_B^2} = \frac{eAB}{2\pi \hbar} \qquad \qquad n = \frac{1}{2\pi l_B^2}
\end{equation}
Note that the two gauge choices give the same degeneracy for the LL, indeed this is a gauge invariant quantity. \\

The resulting picture is then the one of highly degenerate levels. The filling of such levels gives rise to the so called Quantum Hall effect but before explaining it we take a look at what happens when we put a confining potential.
\subsection{Edge modes} 
\begin{figure}[h!]
    \centering
    \includegraphics[width=10cm]{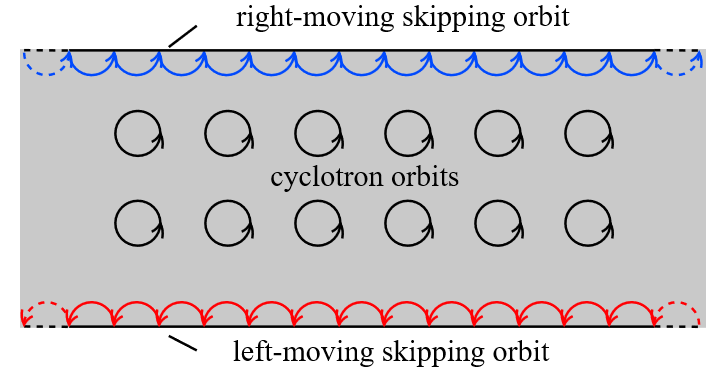}
    \caption{Semiclassical picture of the cyclotron orbits followed by charged particles in a magnetic field. The orbits near the edge effectively drift in a precise direction. Figure taken from \cite{Online_topocond}}
    \label{fig:1.1_skipping_orbits}
\end{figure}

An important role in the QHE is played by the edge modes. In the above description we did not mention any confinement of the electrons, but even at the semiclassical level one can get a good hint on what happens (Fig \ref{fig:1.1_skipping_orbits}).
Charged particles in the presence of a magnetic field want to move along circles in a specific direction, say anti-clockwise, but what happens if we put an hard wall confinement? Away from the edge particles will be undisturbed but near the edge they will bounce back every time they hit the boundary, hence effectively drifting in a \textit{chiral} manner along the the edge with a skipping motion, in opposite direction on opposite sides. \\
This can be put down in a more "quantum" way. Consider a confining potential along the $x$ direction $V(x)$ so that the Hamiltonian becomes in the Landau gauge:
\begin{equation}
     H = \frac{1}{2m}\Bigl(p_x^2+ (p_y+eBx)^2 \Bigl) + V(x)
\end{equation}
As we saw above the solutions to the bare problem are stripes along $y$ with a gaussian localization along $x$. Given a potential which is sufficiently smooth on the scale $l_B$ the potential is just a momentum dependent shift of the energy, since the momentum $k$ along $y$ gives the localization of the state along $x$. The energy spectrum will then be that of LL plus a momentum dispersion which will follow the confinement:
\begin{equation}
E(n,k) \simeq \hbar\omega_B(\frac{1}{2}+n) + V(-k l_B^2)  
\end{equation}
Then a wavepacket created near the edge will have a group velocity along y of:
\begin{equation}
    v_y = \frac{\partial E(n,k)}{\partial k} = - \frac{1}{eB}\frac{\partial V(x)}{\partial x}
\end{equation}
The velocity will then have opposite signs for the two opposite edges, as we anticipated from a semiclassical analysis! 
\subsection{Filling the states} 
\begin{figure}[h!]
    \centering
    \includegraphics[width=7cm]{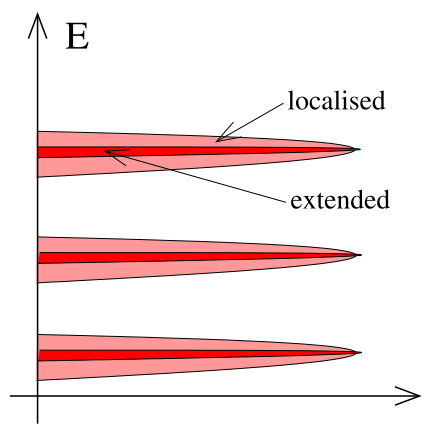}
    \caption{Pictorial representation of the density of states. Taken from \cite{tonglectures}}
    \label{fig:1_dos_disorder}
\end{figure}

When we consider non-interacting electrons the state of the system at low enough temperature will be characterized by a Fermi sea where all the states up to the Fermi energy $E_F$ are occupied. Take for example the confined geometry we analyzed above, the resulting state will look like a metal on the edges where the Fermi energy cross the LL and like an insulator in the bulk. We can immediately compute the the Hall response of the system by considering an Hall potential on the $x$ direction $V_H$ which in the band dispersion acts effectively as a different chemical potential on the two opposite edges $\delta \mu = V_H/e$. The current carried in the y direction can be naively calculated by adding up all the contributions of the filled states as:
\begin{equation}\label{eq:1_iqh_current}
    I_y = -e \int \frac{dk}{2\pi} v_y(k) = \frac{e}{2\pi l_B^2}\int d    x\frac{1}{eB}\frac{\partial V(x)}{\partial x} = \frac{e}{2\pi \hbar}\delta \mu
\end{equation}
Which gives a Hall conductivity of:
\begin{equation}
    \sigma_{xy} = \frac{I_y}{V_H} = \frac{e^2}{2\pi \hbar}
\end{equation}
This reasoning applies also when the filled LL are $n$ so to get a quantum of conductance for every filled level. However the problem with this simple argument is that is that it does not explain the plateaus in the conductance as a function of the magnetic field of figure \ref{fig:1_iqh_a} and it seems to rely on a full occupation of each LL. Imagine for example that the degeneracy in the LLs, which fixes $B$, is such that the number of particles fills the LLL and $3/4$ of the second LL. Clearly the formula in equation \ref{eq:1_iqh_current} cannot be applied since not all the states are filled in the second LL, still the same value of Hall conductivity is observed in experiments.
When discussing real samples our story lack an important ingredient, the presence of disorder. Indeed disorder, if not stronger than the gap between LL $V_{dis}<<\hbar \omega_B$, is able to localize most of the states which are present in the bulk. The same semiclassical analysis on the drifting orbits of particles tells us that they drift along equipotential lines as they do along the equipotential of the edges\footnote{This can be shown at the quantum level, see for example \cite{tonglectures}}. The bulk of such system will then mostly be a collection of localized states which drift around the peak and valleys of the disorder potential while at the edges the states are delocalised since the equipotential lines encircle the whole sample. It is important to remark that these edge states are not backscattered by disorder because of the chiral nature of this drift, there is  only one direction allowed, the only possibility is to tunnel to the other counter propagating edge of the sample. \\
The density of states then look something like figure \ref{fig:1_dos_disorder} and the existence of plateaus can be explained. 
When increasing the magnetic field more and more localized states becomes available, the degeneracy of LL increases, and then the most energetic states in the highest occupied LL gets emptier while delocalised states on the edges are still active. At a certain point the Fermi energy is so low that only states trapped in minima of the disorder potential remain available and the delocalised states of the highest LL becomes empty giving a jump in the transport properties. By further increasing the magnetic field more states becomes available in the lower LL and the remaining electrons trapped in the localized states of the higher LL function as reservoir of particles. The transport properties are dictated only by the delocalised states. \\
Now the argument for the quantisation of the conductivity we gave in equation \ref{eq:1_iqh_current} is not fully satisfying since it relied on having all states in each LL delocalised. The deep answer to this question is that the Hall conductivity its strictly related to a topological invariant of the filled bands, the Chern number. We will not go into the detail of the topological properties but we will give an argument for this quantisation that is strictly related to it. 
\subsection{Gauge invariance and conductivity}\label{sec:1_corbino}
\begin{figure}
    \centering
    \includegraphics[width=10cm]{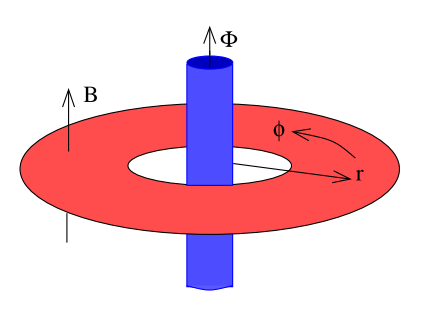}
    \caption{Schematic representation of the Corbino disk geometry threaded by a flux. Taken from \cite{tonglectures} }
    \label{fig:1_corbino}
\end{figure}

The first argument on why the Hall conductivity has such quantised values is due to Laughlin in one of the first pioneering works on the QH physics (\cite{Laughlin_quant_cond}) and invoke gauge invariance. Consider a Corbino disk, as shown in figure \ref{fig:1_corbino}, with a solenoid carrying a flux $\Phi$ threading the hole of the Corbino.
Particles will feel the flux $\Phi$ through the Aharonov-Bohm effect, which will depend only on $\Phi$ modulo the flux quantum $\Phi_0$. Now if we imagine to have a time dependent flux $\Phi(t)$ which goes from 0 to $\Phi_0$ adiabatically in a time $T\gg 1/\omega_B$ we would induce an electromagnetic force around the ring equal to $\mathcal{E}= -\partial \Phi(t) / \partial t = - \Phi_0 /T$. But what is happening to the electrons which are sitting in their states? The eigenstate of the Hamiltoniana are evolving in time and they undergo a \textit{spectral flow}. The vector potential of a solenoid in polar cylindrical coordinates has just an azimuthal component $A_{\phi} = \Phi/2\pi r$ so that the effect in the Hamiltonian is just a shift of the azimuthal component of the momentum or equivalently the angular momentum. Since the degeneracy in the LL, when working in the symmetric gauge, is spanned by the angular momentum we can conclude that by increasing $\Phi$ we are shifting the eigenstates toward higher angular momentum:
\begin{equation}
    \psi_m(\Phi=0) \;\;\longrightarrow\;\; \psi_m(\Phi = \Phi_0) =\psi_{m+1}(\Phi=0)
\end{equation}
Threading a flux then correspond to an increase in the angular momentum. The same idea of having a spectral flow remains when we consider disorder. Even though the angular momentum is no longer a good quantum number having delocalised states which wrap around the ring is sufficient for the argument to work. These state must satisfy a single-valued condition at $\phi=0$ and $\phi=2\pi$ and when we undo the flux insertion trough a gauge transformation which add a phase $e^{-ie\Phi \phi/2\pi\hbar}$ we see that the condition is satisfied only when $\Phi=\Phi_0$. Hence we conclude that these delocalised states must perform a spectral flow because the spectrum is left unchanged. The existence of such states is again guaranteed by the confining potential. Now we know that every time we insert a flux quantum we are bringing one electron for each LL from the inner to the outer ring, as long as we do it adiabatically. This phenomena is showing that by applying the voltage in one direction we are generating a current in the perpendicular one. The Hall conductivity is then:
\begin{equation}\label{eq:1_corbino_cond}
    \sigma_{xy} = \frac{I_r}{\mathcal{E}}= \frac{ne/T}{\Phi_0/T} =n\frac{e^2}{2\pi\hbar}
\end{equation}
where $n$ is the number of filled LL that carries one electron each.
This is indeed the quantised Hall conductivity we expected. The gauge invariance condition for the delocalised state is telling us that each filled LL must bring a quantum of conductivity.\\

\section{Fractional Quantum Hall}
The phenomenology of the fractional quantum Hall effect is not that different from the integer quantum Hall effect, the Hall conductivity is now quantised at specific rational values of the conductance quantum:
\begin{equation}
    \sigma_{xy} = \nu \frac{e^2}{2\pi \hbar}
\end{equation}
\begin{figure}[h!]
    \centering
    \includegraphics[width=10cm]{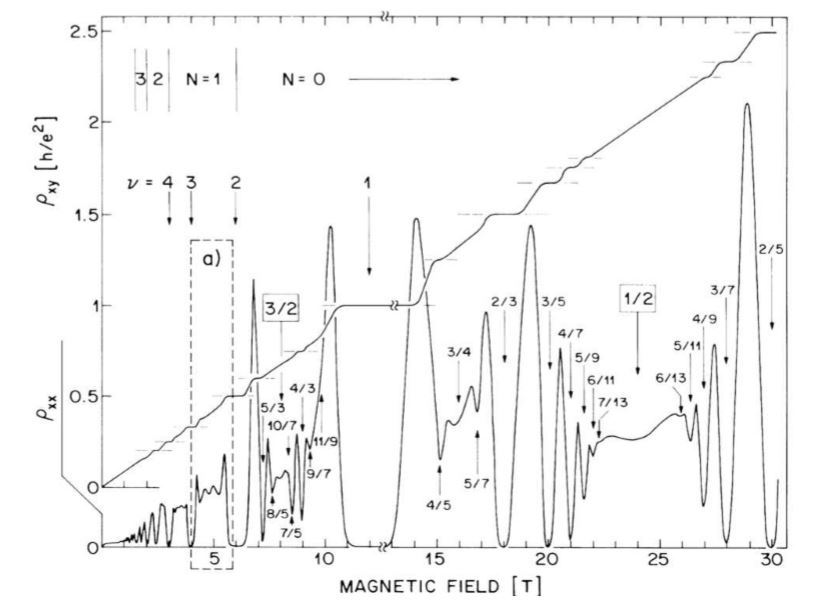}
    \caption{Hall resistivity $\rho_{xy}$ and longitudinal resitivity $\rho_{xx}$ showing respectively plateaus at fractional values and a constant zero value as the magnetic field is varied. Differently from figure \ref{fig:1_iqh} here also fractional plateaus are present because the interaction becomes the dominant energy scale below the cyclotron energy. Taken from \cite{tonglectures}}
    \label{fig:1_fqh}
\end{figure}
The fractions $\nu$ that appear in the experimental samples are not always the same but they can be categorized in specific categories, an example is in figure \ref{fig:1_fqh}. Most of the fractions have an odd denominator but also even denominator fractions exist. The physics of such states can only be captured by the inclusion of interaction in the game. When talking about electrons the interactions are not an adjustable parameter, hence only the disorder energy scale can be tuned. Indeed in order to observe FQH states in 2D electron gases the samples must be much more clean than those showing only IQH plateaus. The idea is that when the energy scales of the disorder is higher than the interaction energy the latter does not spoil the single particle picture within each LL. On the contrary when the interaction is stronger than the disorder the ground state will be a new many-body state as FQH are. The new energy scale hierarchy in FQH systems then is:
\begin{equation}
    \hbar\omega_B \gg E_{int} \gg V_{dis}
\end{equation}
The role of disorder for the presence of FQH plateaus is similar, it acts as a reservoir of particles so that changing the magnetic field we are able to localize particles around the disorder peaks and valleys before a transition to another many-body state happens. \\
In the following we will focus on the description of some of these exotic many body states which live entirely in the LLL.
\subsection{Laughlin wavefunction}\label{sec:1_Laughlin_wavefunction}
The first description of such state came with the Laughlin wavefunction guess \cite{Laughlin_wavefunction}, it describes the LLL with a filling of $\nu=1/m$ with $m$ an odd integer for fermions and even integer for bosons. Laughlin many-body wavefunction for filling $\nu=1/m$ reads:
\begin{equation}\label{eq:1_Laughlin_wavefunction}
    \Psi(\{z_i\}) = \prod_{i<j} (z_i -z_j)^m e^{-\sum_i |z_i|^2 /4l_B^2}
\end{equation}
As a first check we see that the statistic under the exchange of particle positions is fermionic if $m$ is odd and bosonic id $m$ is even. Then we can also argue that it recovers the desired filling fraction. Each particle appear in the polynomial prefactor with a power of at most $m(N-1)$ and by remembering that the single particle state with a power $M$ is localized at a radius $R\simeq \sqrt{2M}l_B$ this gives the radius of the droplet where particles are confined as $R= \sqrt{2m(N-1)} l_B$. The density will then be:
\begin{equation}
    n = \frac{N e}{\pi R^2}=  \frac{e}{m 2\pi l_B^2} 
\end{equation}
which is exactly $1/m$ the density of the fully filled LL. The density of such many-body state is actually constant in the bulk and goes to 0 exponentially at $R= \sqrt{2m(N-1)} l_B$ within $l_B$.\\

The zeros of order $m$ that is present whenever two particles come close to one another are responsible for the minimization of the interaction energy. Note that when $m=1$ we should recover a fully filled LLL and this is actually what happens. The polynomial prefactor of the Laughlin is just the polynomial that one would obtain by computing the Slater determinant of the first $N$ single particle states we previously determined in equation \ref{eq:1_LLLwavefunction} \\  
There is another point of view that helps explaining the variety of possible filling fractions we will only mention here that calls for new effective entities called \textit{composite fermions}, see for example \cite{composite_fermions}. The idea is that the electrons "dress" themselves with an odd number of vortexes reducing the magnetic field these new composite particles experience. After that, they behave as weakly interacting particles filling an integer number of LL of the reduced magnetic field.  \\
\subsection{Bulk charged Excitations}
The Laughlin state has some particular elementary charged excitation which carry a fractional charge charge $q=\pm 1/m$ and have a fractional quantum statistics, they are neither boson nor fermions but $any$ons since upon exchange they pick up a phase $e^{\pm i\pi/m}$. The existence of such exotic particles have been suggested for the first time in Laughlin states in \cite{fractional_statistics}. In 2 dimensions the concept of braiding the position of two particles is quite tricky as it can be done in many topologically different way, braiding once, twice or $n$ times the trajectory of the particles. It can be shown that this peculiarity, absent in 3 or more spatial dimensions, allows the existence of \textit{any}ons, neither bosons nor fermions, which upon an exchange of position acquire a phase of $e^{i\pi \alpha}$. This was first pointed out in \cite{Leinaas_Myrheim} before the discovery of FQH excitations. In the case of Laughlin states quasi-particles $\alpha=1/m$. Note that since we live in 3 dimensions real particles cannot show anyonic statistics and this possibility is left only to emergent particles in two dimensional systems as is the case of FQH states. Although we are not going into the details of the derivation of the charge and statistics of such excitations it is easy and instructive to write down the wavefunction of a quasi-hole state on top of a Laughlin, it reads:
\begin{equation}
    \Psi_{hole}(z_i;\eta) = \Pi_{i} (z_i - \eta) \Pi_{k<l} (z_k - z_l)^m
\end{equation}
The quasi-hole wavefuntion is a \textit{vortex} state, the density is depleted and the phase winds once. The existence of these particular excitations is one of the main property of a particular class of phases of matter characterized by a so called \textit{topological order}. The Laughlin state has the fundamental property of capturing this. It is then possible to explain most of the other observed plateaus by constructing a hierarchical structure of Laughlin states made up by the quasi-particles or quasi-holes of the previous Laughlin, which we will not discuss here. It is important to remark that a FQH state is not characterized only by its filling factor but one need to specify the topological properties of its elementary excitations which define the topological class of the state. In some even more exotic FQH states, such as the $\nu=5/2$, non-abelian anyonic excitations are predicted. Upon exchange of position, the state of non-abelian quasi-particles does not change by a phase factor but by a unitary transformation within the degenerate subspace where they live. The possibility of encoding and manipulating quantum information by means of "topological" actions resilient to errors, as exchanging two particles position, gave birth to the field of \textit{topological quantum computing}. This has been a drive for a lot of investigations regarding the fascinating physics of quasi-particles in FQH states. It must be mentioned that although FQH is not the only system where anyonic excitations may exists, it is historically the first and at present the only one where they have been observed (\cite{anyons_braiding_obs}) after almost 20 years of attempts by means of interferometry experiments using edge states and anyons localized in the bulk.

\subsection{Toy Hamiltonian}\label{sec:1_toyH}
A good insight in the physics described the Laughlin wavefunction can be obtained by finding a simple Hamiltonian for which the Laughlin is the exact ground state, a \textit{toy} Hamiltonian. To do so we consider a central potential $V(|z_i-z_j|)$ which depend only on the relative distance between two particles as is the case for the Coulomb interaction. Let us work in the symmetric gauge and analyse the two particle problem in the LLL. Here we can immediately identify two good quantum numbers which commute also with the interaction Hamiltonian we are considering, the total and relative angular momentum between the particles. Then, if we restrict to the LLL, all central potential $V(r)$ in the two particle problem has the same eigenstates:
\begin{equation}
    \ket{M,m} = (z_1+z_2)^M(z_1-z_2)^me^{-(|z_1|^2+|z_2|^2)/4l_B^2}
\end{equation}
where $(M,m)$ are respectively total and relative angular momentum. This wavefunction is exponentially peaked at a relative distance $|z_1-z_2|\simeq\sqrt{2m}l_B$ so that the interaction energy is directly related to the relative angular momentum $m$. If we restrict to the LLL all we need to know are the so called \textit{Haldane pseudopotentials}:
\begin{equation}
    v_m = \frac{\bra{M,m}V\ket{M,m}}{\braket{M,m}{M,m}} \simeq V(r=\sqrt{2m}l_B)
\end{equation}
Note that they do not depend on the total angular momentum. When passing to the $N$ particle case the two body interaction Hamiltonian can be written starting from these pseudopotentials as:
\begin{equation}
    H = \sum_{m'=1}^\infty \sum_{i<j} v_{m'} \mathcal{P}_{m'}(ij)
\end{equation}
where $\mathcal{P}_{m'}(ij)$ project the wavefunction onto the LLL states where particles $i$ and $j$ have relative angular momentum $m'$. Of course $v_m$ depend on the true interaction but it is reasonable to take them equal to a positive constant up to $m$ and zero for higher momenta. The picture is that of particles interacting only when they get close enough, something like little disks that pay an energy cost to overlap. The ground state of these Hamiltonian can be obtained just by adding $m$ units of angular momentum to each pair of particles in the LLL:
\begin{equation}
    \Psi(\{z_i\}) = s(z_i)\Pi_{i<j} (z_i -z_j)^m e^{-\sum_i |z_i|^2 /4l_B^2}
\end{equation}
For $m$ odd these looks really similar to the Laughlin state. There is still the freedom of having a symmetric polynomial  $s(z_i)$ as a prefactor which has the effect of spreading out the particles. An example of such symmetric polynomial is $\prod_i (z_i -\eta)$ which a quasi-hole in $\eta$. If a confinement potential is present than the most localized wavefunction will be the preferred one, hence the Laughlin is the ground state of such a system. Another important fact is that it costs a finite amount of energy to reduce the area occupied by the Laughlin, this makes the Laughlin state \textit{incompressible}. However if the confinement isn't too steep and allows the spreading of the wavefunction over a slightly bigger area then we can have low energy edge excitations which are deformations of the Laughlin droplet. We will see more of these later.

\subsection{Hall response}
So far we did not mention the role of edge states or the reason why the Hall conductivity takes the precise value proportional to the filling. The edge states in an interacting system such as the Laughlin states are not a straightforward story. While in the IQH the whole discussion was based on basically having a 1D non-interacting chiral edge here interactions modify this picture. We can however use the example of the Corbino disk we gave for the IQH in section \ref{sec:1_corbino} . Instead of the hole in the Corbino disk we can consider a infinitesimally small solenoid. Again by threading one flux quantum $\Phi_0$ adiabatically we are inducing a spectral flow so that the angular momentum of each particle is increased by one. This in the wavefunction correspond to an additional prefactor $\Pi_i z_i$ which can be seen as a quasi-hole state in $\eta=0$. Then when we add a flux quantum we are depleting the center of the disk by a charge $e/m$ which then ends up on the outer edge. The Hall conductivity can then be calculated similarly to equation \ref{eq:1_corbino_cond} with the slight modification that the charge transferred for every flux quantum $\Phi_0$ is smaller:
\begin{equation}
        \sigma_{xy} = \frac{I_r}{\mathcal{E}}= \frac{e/mT}{\Phi_0/T} =\frac{1}{m}\frac{e^2}{2\pi\hbar}
\end{equation}

\section{Edge states}
In the last sections we presented a lot of interesting physics taking place in FQH systems but still a satisfactory theoretical framework for the edge states of this systems is lacking. We now have different possibilities of attacking the problem and we will see that all approaches gives a similar answer, the low energy excitations on the edge are described by a chiral Luttinger Liquid ($\chi LL)$). As Laughlin states will be the main focus of the thesis we will refer to these if not stated differently. \\
\subsection{Hydrodynamical approach} \label{sec:1_hydrodinamical}
\begin{figure}
    \centering
    \includegraphics[width=7cm]{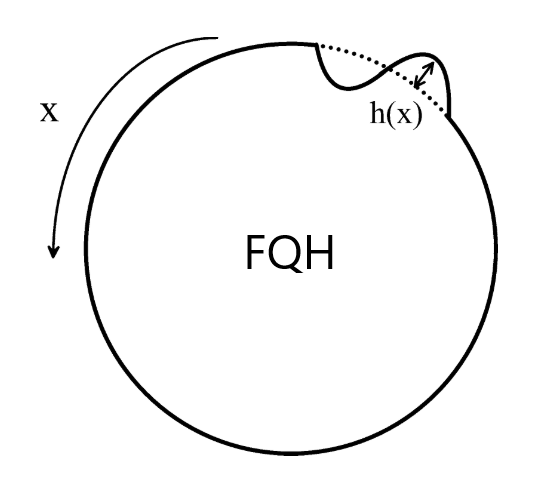}{h}
    \caption{Schematic picture of an incompressible droplet of FQH with an edge excitation}
    \label{fig:1_hydrodinamical}
\end{figure}

Laughlin state is an incompressible liquid with constant density $n= \frac{e\nu}{2\pi l_B^2}$. With a semiclassical picture in mind we can argue that the a finite droplet of such liquid will have some low-energy excitations which correspond to a shape modification of a such droplet as depicted in figure \ref{fig:1_hydrodinamical} . 
If a confinement $V$ is present and it is possible to linearise it around the shape of the droplet at rest as $V(y)\simeq V_0 +y\partial V / \partial $ with $\partial V/\partial y =E$ the electric field perpendicular to the boundary, the energy paid to have an excitation on the boundary is: 
\begin{equation}
    U = \int dx \int_0^{h(x)} dy \; n e V(y) \simeq \frac{neE}{2} \int dx \; h^2(x) 
\end{equation}
where $h(x)$ is the displacement from the the rest position. Given that we have a Hall response $\sigma_{xy} = \nu e^2/2\pi \hbar$ the confinement electric field $E\hat{y}$ generate a persistent chiral current and the drifting velocity of the charge is $v= \frac{E}{B}c$. The equation describing this chiral motion is a chiral wave equation:
\begin{equation}
    \partial_t\rho(x,t) = -v \partial_x \rho(x,t)
\end{equation}
By using the velocity of the chiral edge excitations as our free parameter rather then the confinement potential and excess charge density $\rho(x) = n \,h(x)$ to specify the excitation energy we express the energy as:
\begin{equation}
    U = \frac{\pi v }{\nu}\int dx \rho^2(x) 
\end{equation}
Now we specify a finite size of the droplet $x\in[0,L)$ and go to Fourier space with $k = n_k2\pi/L$:
\begin{equation}
    \rho(x) = \sum_{n_k=1}^\infty \frac{1}{\sqrt{L}} \rho_k e^{ikx} + \rho_{-k}e^{-ikx}  \qquad \rho_k = \frac{1}{\sqrt{L}}\int dx \rho(x) e^{-ikx} 
\end{equation}
And 
\begin{equation}
    \dot{\rho}_k = -i v k \rho_k \qquad U = 2\pi \frac{v}{\nu} \sum_{k=1}^\infty \rho_{-k}\rho_k
\end{equation}
Here we are in the good position to pass to quantum mechanics and quantise these excitations. The relation given by the chiral wave equation in momentum space allows us to identify each momentum component of $\pm|k|$ as a pair of conjugate Hamiltonian coordinates. For notational convenience we identify $k$ with its modulus:
\begin{equation}
    q = \rho_{k}\qquad p = i\frac{2\pi}{\nu k}\rho_{-k}
\end{equation}
 To check that our identification for the Hamiltonian variables is correct we compute the Hamilton equation and verify that they coincide with the chiral wave equation:
\begin{align}
    \begin{cases}\dot{q} = \frac{\partial H}{\partial p} \\
    \dot{p} =- \frac{\partial H}{\partial q} \end{cases} \qquad \rightarrow\qquad \begin{cases} \dot{\rho}_{k} =  \frac{\nu k}{2\pi i} \frac{\partial H}{\partial \rho_{-k}} = -iv k \rho_k \\ \dot{\rho}_{-k} = - \frac{\nu k}{2\pi i}\frac{\partial H}{ \partial\rho_{k}} = -iv(- k) \rho_{-k} \end{cases}
\end{align}
This allows us to canonically quantise the system by imposing the commutation prescription to each $k$ mode independently:
\begin{equation}
    [q,p]= i\qquad \rightarrow\qquad [\rho_k,\rho_{-k}] = \frac{\nu}{2\pi} k
\end{equation}
Now $\rho_k$ and $\rho_{-k}$ can be identified with creation and annihilation operators of an harmonic oscillator due to their commutation relation and the form of the Hamiltonian. Since working with a normalized commutation relation is easier we normalize each $k$ mode:
\begin{align}\label{eq:1_ladder_a}
    a_k= \sqrt{\frac{2\pi}{\nu k}}\rho_{k}  \qquad a^\dagger_k = \sqrt{\frac{2\pi}{\nu k}} \rho_{-k} \qquad \longrightarrow \qquad [a_k,a_k^\dagger]=1
 \end{align}
 So the Hamiltonian becomes a collection of independent harmonic oscillators with a linear dispersion:
 \begin{equation}
     H = \sum_{k>0} v\hbar k\;\;a^\dagger_k a_k
 \end{equation}
 where $v$ is the velocity of these excitations. Note that the chirality constraint has the effect of allowing only positive momenta. The charge density operator is written as:
 \begin{equation}
     \rho(x) = \sum_{k>0} r_k  a_k e^{ikx} + r_k a^\dagger_k e^{-ikx} \qquad \qquad \text{with} \qquad r_k =\sqrt{\frac{\nu k}{2\pi L}}
 \end{equation}
 However we must a little bit more careful with this expression since we cannot ask our effective theory to work at too high energy and so there must be a cutoff in the momentum. This can be achieved by smearing the charge density operator over a window of size $a$ by convolving it with an appropriate localized function. This is equivalent in Fourier space to multiply each mode by a localized function on $1/a$. By choosing an exponential in Fourier space and then a Lorentzian in real space we have:
 \begin{equation}
          \rho_a(x) = \sum_{k>0} r_k  a_k e^{ikx}e^{-ka} + r_k a^\dagger_k e^{-ikx}e^{-ka} \qquad \text{with} \qquad r_k =\sqrt{\frac{\nu k}{2\pi L}}
 \end{equation}
This means that our theory cannot describe physics happening at length scales which are smaller than $a$. From now on we will drop the subscript $a$ and instead put an apostrophe in the summation to remind us of the cutoff. 
 
 \subsection{Microscopic approach}
\begin{figure}[h!]
     \centering
     \includegraphics[width=10cm]{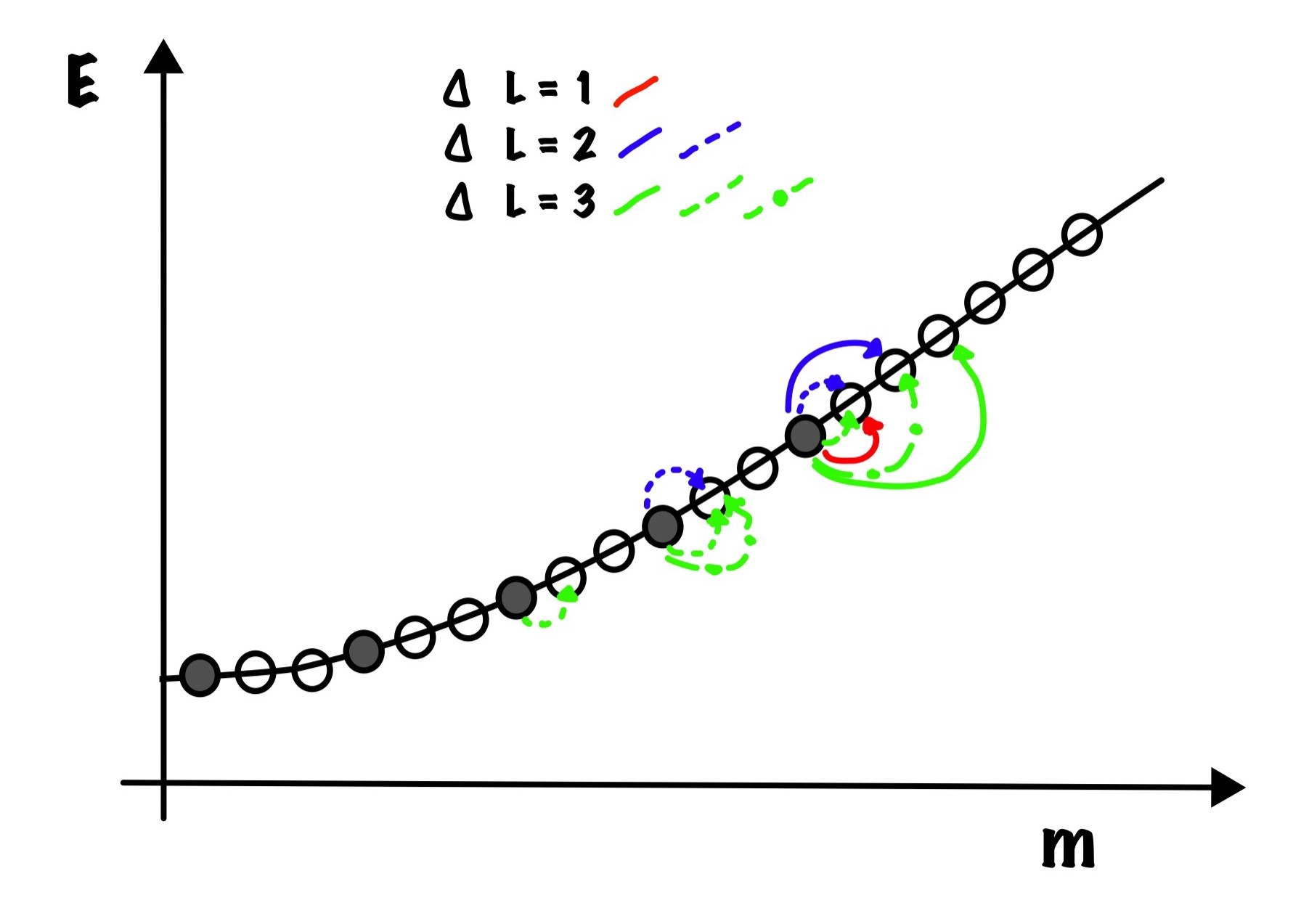}
     \caption{Single particle spectrum with the first edge excitations which bring momentum $\Delta L=1,2,3$ for a Laughlin $m=3$}
     \label{fig:1_spectrum_cartoon}
 \end{figure}

 This latter approach may not seem complete, in the end semiclassical pictures don't always capture the correct description of a quantum system. However we will show that indeed at the microscopic level we do have plasmonic type of excited states which live on the edge of confined FQH liquids. The requirement is that of low energy and/or smooth enough edge. As we saw in section \ref{sec:1_toyH} the Laughlin state is basically the most compact state which respects a lower bound on the relative angular momentum between two particles. The states where particles are too close are above the many-body gap set by the interaction $V_{int}$ but states where particles just take more space will only pay an energy due to the confinement. Consider a rotationally invariant confinement so that we can use total angular momentum as a good quantum number, the single particle spectrum will look something like figure \ref{fig:1_spectrum_cartoon}.
 The Laughlin state then will consist of filling these states with particles respecting the relative angular momentum bound which result in an exclusion statistics, particles must stay at least $m$ orbitals from each other. The first excited state that we can build is by moving the last occupied orbital from $M$ to $M+1$; no interaction energy will be paid, only the difference in the confinement energy $\Delta$. The second excited state can be build either by moving the last occupied orbital from $M$ to $M+2$ or by moving the last two occupied from $M$ and $M-m$ to $M+1$ and $M-m+1$. All this excited states are organized in subspaces of eigenstates of the angular momentum operator with a definite angular momentum $L = L_L+\Delta L$ where $L_L$ is the angolar momentum of the Laughlin. As we anticipated in section \ref{sec:1_toyH} all the excited states correspond to a specific symmetric polynomial which multiplies the Laughlin wavefunction. It is possible to generate the subspace at $\Delta L$ with a linear combination of symmetric polynomials $S_{d_l}$ defined as:
 \begin{equation}
     S_{d_l}(\{z_i\}) = \prod_{l=1}^{\infty}(\sum_i z_i^l)^{d_l}\qquad\qquad\text{with}\qquad \sum_l d_l l = \Delta L
 \end{equation}
 The degeneracy at $\Delta L$ then correspond to the number of its partitions in term of positive integers characterized by $\{d_l\}$. Note that this degeneracy correspond to the degeneracy we found in the Hilbert space obtained in the hydrodynamical approach and we can make the identification $\sum_i z_i^k \leftrightarrow a^\dagger_k$ since they both add the same quantity of (angular) momentum. However the excited states showed in figure \ref{fig:1_spectrum_cartoon} that have a nice representation in terms of filling different single particle states do not in general correspond to the $S_{d_l}$ but to combination of those. Consider the simple case of two particles and the subspace $\Delta L= 2 $, the basis will be:
 \begin{align}
     S_{\{2,0,0...\}} = (z_1+z_2)^2\qquad S_{\{0,1,0...\}} = (z_1^2+z_2^2)
 \end{align}
 while the state that we would build following the blue lines of figure \ref{fig:1_spectrum_cartoon} are:
 \begin{equation}
     s_1 = z_1z_2 \qquad\qquad s_2 = (z_1-z_2)^2
 \end{equation}
 The first is a quasi-hole state and the second is equivalent to forming a $m=5$ Laughlin since the particles are separated by 5 units of angular momentum. There is a simple and systematic description of the latter polynomials as Jack polynomials which we will not discuss here, see for example \cite{Jack_Thomale2011}  .
 Note that in the simple case of IQH with $m=1$ these excited states correspond to particle-hole excitations. In all this discussion we must however be careful to the maximum $\Delta L$ we consider since after a certain point we may pick up non-linearities of the excitation spectrum, deep in the bulk or far away from the edge, so that the subspaces we described would not be degenerate anymore\footnote{Non linearities are very often neglected, an example of their physical consequences can be found in \cite{Nardin2020}}.
 \\
 We then have an effective theory for our edge which consist of bosonic excitations chirally propagating in one direction with a linear dispersion . There is still something missing in the picture: we have not allowed any charged excitation to live on the edge. We would like to have something in our effective theory that describe quasi-particles and quasi-holes and the best way to introduce them is yet in another picture.
 \subsection{Boundary effective theory}\label{sec:1_effective_theory}
 There is then a last approach to the description of FQH edge states based on an effective field theory description of such states. This is actually the route which brought Wen (\cite{Wen1990} \cite{Wen1990_2}) to the description of the edge states as a chiral Luttinger Liquid. \\
 Here we will only state some basics concepts and we refer to the literature for a more comprehensive treatment. The effective field theory describing Laughlin states is the Chern-Simons theory, whose action is:
 \begin{equation}
S_{CS}[a] =  -\frac{m}{4\pi} \int d^3x \epsilon^{\mu\nu\rho} a_\mu \partial_\nu a_\rho
\end{equation}
 where $m$ is the same of the Laughlin state term and $a_\mu$ is an emergent $U(1)$ gauge field whose conserved current are quasi-particles. The peculiarity of such action is that when it is put in a confined geometry it is no more gauge invariant due to boundary terms. The easiest way to save the theory is by the introduction of a new degree of freedom which lives on the boundary and cancel the gauge dependent term after a gauge transformation. This is achieved by a field $\phi$ whose action is:
\begin{equation}\label{eq:1_Floreanini_action}
    S = -\frac{m}{4\pi} \int_{y=0} dxdt \; \partial_t\phi \partial_x \phi +v(\partial_x\phi)^2
\end{equation}
The velocity $v$ is a free parameter and cannot be fixed at the effective theory level. The equation of motion are:
\begin{equation}
    \partial_t\partial_x \phi + v\partial_x^2 \phi = 0 
\end{equation}
A new field can be defined as the space derivative of $\phi$ so that its equation of motion acquire an evident physical meaning:
\begin{equation}
\rho = \frac{1}{2\pi}\partial_x \phi \qquad \longrightarrow \qquad \partial_t \rho(x,t) +v \partial_x\rho(x,t) = 0
\end{equation}
This is the same chiral wave equation we wrote in the precedent section \ref{sec:1_hydrodinamical}. Solutions are of the form $\rho(x+vt)$, so that only right propagating waves are present and $v$ is actually the velocity of such waves.
Then in order to check that this field $\rho$ is actually the charge density one need to look at the coupling to the electromagnetic field in the effective theory and then on the boundary of it. This indeed turns out to be what one expects, for example the coupling to a vector potential with components $A_t$ and $A_x$ is:
\begin{equation}
    S_J = \int_{y=0}dxdt \; (A_t+vA_x) \frac{1}{2\pi}\partial_x \phi
\end{equation}
The coupling to $A_t$ tells us that $\rho$ is a charge density and the coupling to $A_x$ tells us that the current along $x$ is $v\rho$. Everything is consistent with what we found from the hydrodynamical approach, a last piece missing is the Hamiltonian which can be easily derived from the action in equation \ref{eq:1_Floreanini_action} and coincide with the Hamiltonian we saw above: 
\begin{equation}
    H = \frac{\pi v }{\nu}\int dx \rho^2(x) 
\end{equation}
The new ingredient here is that we are starting from a new field $\phi$ to which we still need to give a physical interpretation. By taking the difference between the phase field at two point we are computing the integral of the charge density:
\begin{equation}
    Q =\int_{a}^b dx \; \rho =     \int_{a}^b dx \; \frac{1}{2\pi}\partial_x\phi = \frac{1}{2\pi} (\phi(b)-\phi(a))
\end{equation}
In a closed geometry with a single edge the charge on the edge must be an integer, so that the difference $\phi(L)-\phi(0)$ must be $2\pi q$ with $q$ the total charge on the boundary. However if we have periodic boundary condition this is a bit odd, shouldn't we have $\phi(L)=\phi(0)$ ? The answer is no because $\phi$ is actually a scalar field with period $2\pi$\footnote{Sometimes in the literature a different normalisation is used so that the action becomes independent of $m$ at the price of having a periodicity of $2\pi \sqrt{m}$}. For this reason $\phi$ is often called the \textit{phase} field or a \textit{compact} boson. One could then ask what is observable, evidently $\phi$ alone is not but for example $\partial_x \phi$ or $\phi(a)-\phi(b)$ must be and indeed they are related to the charge excess on the boundary. There is another function that seems to satisfy this \textit{condition} for being an observable, the exponential $e^{i\phi}$. In the next chapter we will see how this will play the role of the quasi-particle field in the edge theory. \\
The quantisation of the theory follows in the same way as we did from the hydrodynamical approach. The theory in momentum space is diagonalized by plasmonic excitations which carry a momentum $k$ and an energy $vk$. Again we must be careful with the UV sector where we we know our effective theory will break down and so we put an exponential cutoff in the momentum component which we denote in the sum as a $\sum{}^{'}$. The field $\phi$ can also be written in terms of the the normalised ladder operators of equation \ref{eq:1_ladder_a} as the charge density field:
\begin{align}
    \phi(x) &= \sum_{k>0}{}^{'} f_k a_k e^{ikx} + f_k^* a^\dagger_k e^{-ikx} \qquad\qquad \text{with} \qquad f_k = -i \sqrt{\frac{2\pi \nu}{k L}}\\
    \rho(x) &= \sum_{k>0}{}^{'} r_k  a_k e^{ikx} + r_k a^\dagger_k e^{-ikx} \qquad \qquad \text{with} \qquad r_k =\sqrt{\frac{\nu k}{2\pi L}}
\end{align}
By using the commutation relation $[a_k,a_{k'}^\dagger]=1$ one can derive the equal time commutation relation for the fields $\phi(x)$ and $\rho(x)$. 
We start by the commutation of the field $\phi$ and then by deriving we will obtain those with $\rho$. 
\begin{align}
    [\phi(x),\phi(y)] =& \Bigl[\sum_{k>0} {}^{'}f_k a_k e^{ikx} + h.c.,\sum_{k'>0} {}^{'} f_{k'} a_{k'} e^{ik'y} + h.c.\Bigl] = \\
   =&\sum_{k,k'>0}{}^{'}|f_k|^2 (e^{ik(x-y)}[a_k,a^\dagger_{k'}] +e^{-ik(x-y)}[a_k^\dagger,a_{k'}]) = \\
    =& \frac{2\pi \nu}{ L}2\,Im\Bigl( \sum_{k>0} \frac{1}{k} \; e^{ik(x-y +ia)} \Bigl)= \\
    =& \frac{2\pi \nu}{ L}2\,Im\Bigl( \frac{L}{2\pi}\sum_{n_k=1}^{\infty} \frac{1}{n_k} \;( e^{i(x-y +ia)2\pi/L})^{n_k} \Bigl)= \\
    =& -2\nu  \,Im\Bigl( \ln(1-e^{i(x-y +ia)2\pi/L})\Bigl)= \\
    =& -2\nu \,Im\Bigl( \ln(i(x-y +ia)2\pi/L)\Bigl)= \\
    =& -2\nu \,\text{atan}(- \frac{x-y}{a})= \\
    =& \frac{\pi i}{m} \text{sign}(x-y) \\
\end{align}
Where we work in the limit $a\ll|x-y|\ll L$ and we used the Taylor expansion of the logarithm. The other commutation relations are easily obtained by differentiating:
\begin{align}\label{eq:1_commutation_fields1}
    &[\phi(x),\phi(y)]= \frac{\pi i}{m} \text{sign}(x-y)\\\label{eq:1_commutation_fields2} &[\rho(x),\phi(y)]= \frac{i}{m} \delta(x-y) \\\label{eq:1_commutation_fields3}
    &[\rho(x),\rho(y)]= -\frac{ i}{2\pi m} \partial_x\delta(x-y)
\end{align}
Note the highly non-local commutation relation of the field $\phi$ and that it is the conjugate field of the charge density becouse of equation \ref{eq:1_commutation_fields2}. We remind that these commutation relations are valid as long as $|x-y|$ is bigger then the cutoff $a$.
\subsection{Quasi-particles at the edge} \label{sec:1_qp_p}
We are now ready to attack the question on where are the quasi-particles in the edge theory. A particle of c
harge $q$ in position $x$ must create a spike in the density. From the commutation relations  in \ref{eq:1_commutation_fields2} we know that density and phase are conjugate fields, then to add a charge in $x$ we need apply a displacement operator of the form $e^{-i\phi(x)}$. To check if this operator has the correct property of adding a charge we compute its commutation relation with the charge field:
\begin{equation}
    [\rho(x), e^{-i\phi(y)}] = [\rho(x),\phi(y)] (-i )e^{i\phi(y)} = \frac{1}{m} e^{-i\phi(y)}
\end{equation}
This means that acting with $e^{-i\phi(y)}$ on a state we will create a localized charge of $1/m$, exactly the charge of our quasi-particles in the Laughlin state. We can also compute the statistics of this excitations:
\begin{align}
    e^{-i\phi(x)}e^{-i\phi(y)} = & e^{-i(\phi(x)+\phi(y))}e^{-\frac{1}{2}[\phi(x),\phi(y)]} = \\
    = & e^{-i\phi(y)}e^{-i\phi(x)}e^{\frac{1}{2}[\phi(y),\phi(x)]}e^{-\frac{1}{2}[\phi(x),\phi(y)]}= \\
    = & e^{-i\phi(y)}e^{-i\phi(x)} e^{i\frac{\pi}{m} \text{sign}(x-y)}
\end{align}
where we've use the Baker-Campbell-Hausdorff formula $e^Ae^B = e^{A+B}\, e^{1/2[A,B]}$ valid when $[A,B]$ commute with both $A$ and $B$. The statistical phase is $\alpha=\pm 1/m$ as for the quasi-particles in the bulk! Since we have $PBC$ on $x$ the difference $x-y$ could be either positive or negative depending of where we fix the origin but we should not worry about this. The sign ambiguity exist also in the bulk from the two possible position exchange in a clockwise or anti-clockwise manner. We are happy here that when we consider $m=1$ we actually get fermions with unit charge, there should not be any exotic excitations in the IQH. We then conclude that the quasi-particle creation operator on the edge must be something like:
\begin{equation}
    \Psi_{qp}^\dagger(x) = e^{-i\phi(x)}
\end{equation}
There is a subtle issue with the normalization of this operator that we will fix later. At this point we can go on and define also the particle operator for the edge. By following what we just did for the quasi-particle operator is easy to verify that the particle operator is
\begin{equation}
    \Psi_{p}^\dagger(x) = e^{-im\phi(x)}
\end{equation}
which satisfies:
\begin{equation}
    [\rho(x),\Psi_p(y)]= \Psi_p(y)\qquad\qquad\Psi_p(x)\Psi_p(y) = (-1)^m \Psi_p(y)\Psi_p(x)
\end{equation}
It carries charge 1 and is a fermion (boson) when $m$ is odd (even). Our edge theory for the Laughlin knows what is happening in the bulk, whether the particles are fermions or bosons and which statistics the quasi-particles should have. 
\subsection{Correlation functions}\label{sec:1_corr_funct}
One important piece of the edge theory that we can derive without too much effort are the correlation functions at zero temperature. The fields are chirally moving along the edge so that we can take them to be function of a single variable $\phi(x,t)=\phi(x+vt)$. The correlation function of the phase field is:
\begin{align}
    \langle\phi(x)\phi(y)\rangle =& \bra{0}\sum_{k'>0}{}^{'} f_{k'} a_{k'}e^{ik'x}+h.c.\sum_{k>0}{}^{'} f_{k} a_{k}e^{iky}+h.c\ket{0}=\\
    =& \sum_{k>0}{}^{'} |f_k|^2 e^{ik(x-y)} = \\
    =& \frac{2\pi \nu}{ L}\frac{L}{2\pi}\sum_{n_k=1}  \frac{1}{n_k} \Bigl(e^{i (x-y+i a)2\pi/L}\Bigl)^{n_k} =\\
    =& -\frac{1}{m}\ln( 1- e^{i (x-y+i a)2\pi/L} )  =\\
    =& \frac{1}{m}\ln( x-y) + const
\end{align}
where we used the Taylor expansion of the logarithm and considered $a\ll x\ll L$. The correlation function of the density is obtain by differentiating with respect to $x$ and $y$:
\begin{equation}
    \langle\rho(x)\rho(y)\rangle = - \frac{1}{4\pi^2 m} \frac{1}{(x-y)^2}
\end{equation}
Then we can also compute propagators for the charged excitations we found, particles and quasi-particles. These are a little bit more subtle because we have to deal with exponentials. To carry out the calculations it is better to divide the phase field in creating and annihilating  part: 
\begin{equation}
    \phi^{-(+)} = \sum_{k>0}{}^{'} f_{k}^{(*)} a_{k}^{(\dagger)}e^{+(-) ikx} \qquad [\phi^+(x),\phi^-(y)]= -\frac{1}{m}\ln(i(x-y+ia)\frac{2\pi}{L} )
\end{equation}
where in the commutation relation we are taking only the limit of $a,x<<L$. Now using multiple times the Baker-Campbell-Hausdorff formula to commute the exponential we get:
\begin{align}
\langle\Psi_{qp}(x)&\Psi_{qp}^\dagger(y)\rangle =  \bra{0}e^{-i(\phi^+(x)+\phi^-(x))}e^{i(\phi^+(y)+\phi^-(y))}\ket{0}= \\
= & \bra{0}e^{-i\phi^+(x)}e^{-i\phi^-(x)}e^{i\phi^+(y)}e^{i\phi^-(y)}\ket{0} e^{1/2[\phi^+(x),\phi^-(x)]+1/2[\phi^+(y),\phi^-(y)]} = \\
= & \bra{0}e^{-i\phi^+(x)}e^{i\phi^+(y)}e^{-i\phi^-(x)}e^{i\phi^-(y)}\ket{0} e^{[\phi^+(0),\phi^-(0)]} e^{-[\phi^-(x),\phi^+(y)} = \\
= & \Bigl(\frac{L}{2\pi a}\Bigl)^{1/m} \Bigl(\frac{L}{2\pi }\frac{1}{i(x-y)-a }\Bigl)^{1/m} 
\end{align}
This correlation function has a divergent prefactor when $a<<L$, however this fact is linked with the definition we gave for the quasi-particle operator. This divergence comes from the fact that we are using a non-normal ordered operator and can be fixed by just using the normal-ordered version. Alternatively one can just forget those prefactors and look at the  physical part which is the non-trivial power law dependence on $x-y$. The power law is $1/m$ and for $m\neq 1$ the range of the correlations is infinite! Also the particle correlations follow an interesting power law. Following the same calculations of the quasi-particle correlations we have:
\begin{equation}
    \langle\Psi_{p}(x)\Psi_{p}^\dagger(y)\rangle\;\simeq\; \frac{1}{(x-y)^m}
\end{equation}
We remark here that in the case $m=1$ particles and quasi-particles coincide and the propagators are those of a Fermi Liquid. 
\\
\\
In this chapter we have introduced the main physical properties of Quantum Hall systems and we gave a specific framework for the $\chi$LL as the effective theory of the edge states. However we have not yet moved away from the nice picture of a completely solvable theory. In the next chapter we will see how scattering of excitations within a $\chi$LL is a non trivial problem which generates completely different behaviours with respect to a non-interacting particle picture. 
\chapter{Tunneling in chiral Luttinger liquids}\label{chap:2}
\begin{figure}[!h]
    \centering
    \includegraphics[width=10cm]{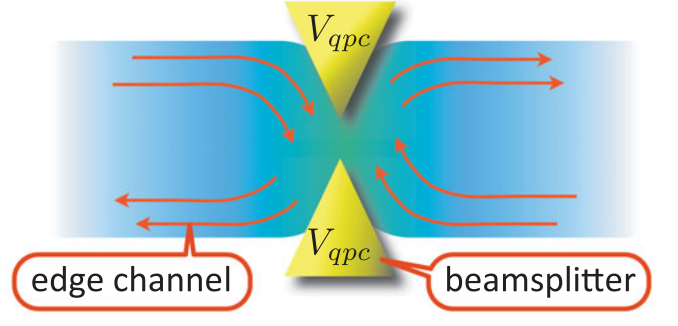}
    \caption{Schematic picture of a quantum point contact (QPC) in a Quantum Hall sample taken from \cite{EQO_review2014}. The side gate voltage $V_{qpc}$ governs the strength of the tunneling between the chiral edge state by controlling their spatial separation and acting as a beamsplitter for the incoming top left into the top right/bottom left chiral edge state or for the bottom right into the bottom left/top right chiral edge state. }
    \label{fig:2_QPC_general}
\end{figure}

We concluded the previous chapter with a review of the theoretical framework in which the edge of a FQH system is usually described. In this chapter we will introduce the phenomena which is the central topic of this thesis, namely the scattering of excitations in chiral Luttinger Liquids ($\chi$LL). This problem has been among the first ones to be analyzed in the context of $\chi$LL and, as we will see, it gives a very rich and non-trivial phenomenology. Also from the experimental point of view the geometry that one needs to realize is quite simple, one just needs to have a constriction where two edges with opposite chiralities come close enough to one another, usually referred as a Quantum Point Contact (QPC). \\
An idea of such configuration is shown in figure \ref{fig:2_QPC_general} and consist of two gates parallel to the 2DEG with a voltage bias $V_{qpc}$ that deplete the density beneath them and controls the distance between the edge states. \\
Another typical probe through which one can analyze the $\chi$LL behaviour of FQH edges is the tunneling of electrons from external sources, such as nearby metallic electrons ( \cite{chiLL_extelectron_tunn} ) or, as more recently proposed, quantum dots ( \cite{FQH_quantumdot} ). In this thesis we will not discuss this case and we will rather focus on the simple configuration which give tunneling between two FQH edges.
\section{Tunneling Hamiltonian}\label{sec:2_tunn_ham}
\begin{figure}[!h]
    \centering
    \includegraphics[width=\linewidth]{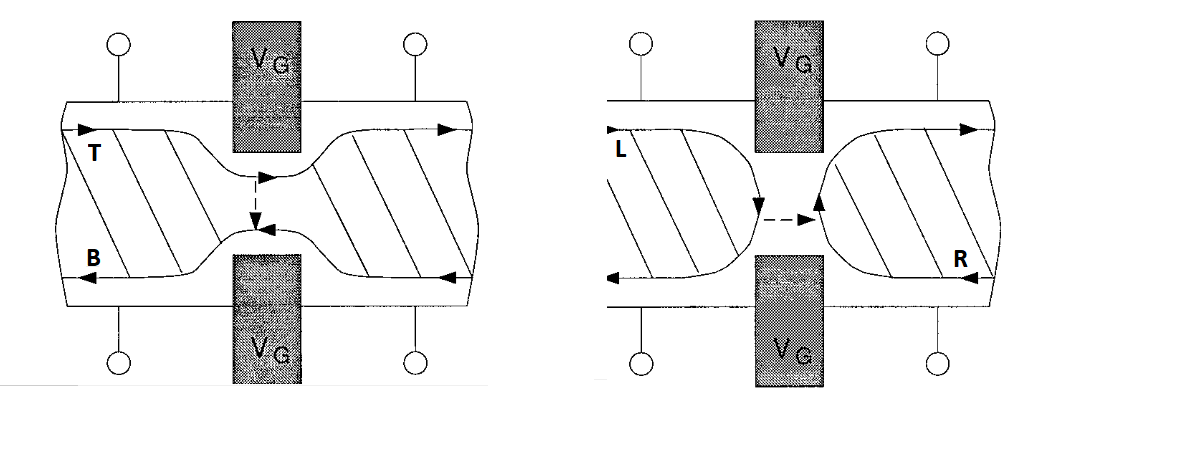}
    \caption{Different interpretations of the QPC geometry, on the left we have tunneling between a top and bottom edge while on the right we have tunneling between a left and right edge. Adapted from \cite{Chamon_sn_duality} }
    \label{fig:2_duality}
\end{figure}

The Hamiltonian of the free system, no QPC, is easily written in terms of the Luttinger phonons as we saw in chapter \ref{chap:1}. In this section we will discuss the different terms that one can add to the Hamiltonian to model the tunneling process at the QPC as shown in figure \ref{fig:2_QPC_general}. We already saw in section \ref{sec:1_qp_p} that there are different charged operators in a $\chi$LL and in principle all tunnelings terms of the type:
\begin{equation}\label{eq:2_Htun_general}
    H_T = \Gamma \psi (0) \psi^\dagger (l) + h.c
\end{equation}
could be added to the theory. Here $0$ and $l$ identify the position of the constriction and, if the two points belong to two different edges, also in which edge the operator is acting. Now the QPC geometry can be interpreted in two possible ways as depicted in figure \ref{fig:2_duality}: one where the QPC is almost open and we have a top and bottom edge connected by tunneling and the second where the QPC is almost closed and we have a right and left edge connected by tunneling. However it is problematic to transfer a fractional charge from two distinct droplets as the configuration on the right of figure \ref{fig:2_duality} suggests since it would lead to two distinct systems with a fractional charge. In this case it indeed seems more natural to use a particle tunneling with $\psi_p(x) = e^{im\phi(x)}$. For the configuration on the left this reasoning is absent and there is no real problem in transferring a fractional charge from one side to the other since they both belong to the same droplet. An heuristic argument which comes from a microscopic point of view then suggests us to use quasi-particles and not particles in this case. Moving a fraction of the unit charge instead of a full charge is expected to be more convenient and then have bigger matrix elements which results in greater coupling constants in the effective edge theory. We will then say that the best choice for the tunneling Hamiltonian in this configuration edge theory is that of quasi-particles $\psi_{qp}= e^{i\phi(x)}$ but a more convincing argument will be given later. There is an apparent duality between the two situations, the single droplet with an almost open constriction can be viewed as two distinct droplets with a strong tunneling between them and vice versa. \\
\\
Of course there could be other couplings between the two point of the constriction such as a density density interaction of the type $\rho(0)\rho(l)$ but this will not give any charge tunneling and, for this reason, it is often discarded. 
\\ 
\subsection{Renormalization group analysis}\label{sec:2_RG}
In order to analyze which tunneling term in the effective theory is the most relevant one of the first approaches have been to study each tunneling action from a renormalization group (RG) point of view. This approach has been developed in \cite{Kane1992_RG} and here we will limit ourselves to summarize some basic concepts and results. The idea is to start from the free theory with the addition of the tunneling term as perturbations and study whether these are relevant or irrelevant operators in the RG sense. To do so we need the scaling dimensions of the particle and quasi-particle fields in the free theory which can be deduced from the correlators we derived in section \ref{sec:1_corr_funct} and are:
\begin{equation}
    \langle \psi_p(0) \psi_p(t)\rangle \propto \frac{1}{t^{m}} \qquad     \langle \psi_{qp}(0) \psi_{qp}(t)\rangle \propto \frac{1}{t^{1/m}} 
\end{equation}
The scaling dimensions are then $\Delta_p= m/2$ and $\Delta_{qp} = 1/2m$. The tunneling action is of the form:
\begin{equation}
    S_{T;p/qp}[\psi] = \Gamma_{p/qp} \int dt \; \psi_{p/qp}(0) \psi_{p/qp}^\dagger(l) +h.c
\end{equation}
from which we conclude that the scaling dimension of the tunneling coupling is $\Delta_{\Gamma;p/qp} = 1 -\Delta_{p/qp}$. This gives two very different results in the FQH case $m>1$:
\begin{align}
    \Delta_{\Gamma;p} =   1- m <0 \qquad \rightarrow \qquad \text{irrelevant}\\
        \Delta_{\Gamma;qp} = 1 - \frac{1}{m} >0 \qquad \rightarrow \qquad \text{relevant}
\end{align}
The fact that the particle tunneling is an irrelevant perturbation means that as long as its coupling constant $\Gamma_{p}$ is small, perturbation theory will in general work. On the other hand there is a problem with applying perturbation theory at low energies for the quasi-particle tunneling since it is relevant and the effective coupling $\Gamma_{qp}$ explodes. However the RG flow of the quasi-particle coupling will stop at a cutoff energy scale set by $\epsilon$ which can be the temperature or the finite voltage bias. When one of these scales is high enough, then the renormalization of $\Gamma_{qp}$ does not grow too much and perturbation theory can be applied if the bare coupling allows it. We will see in the next section what actually happens in perturbation theory when the rescaling of $\Gamma_{qp}$ explodes.
\subsection{Resonant Tunneling}
The clean tunneling described by equation \ref{eq:2_Htun_general} is of course the first and easiest step but in real systems localized states may exist in the vicinity of the QPC. Such states are very frequent in electronic systems, as we discussed in chapter \ref{chap:1}, and give rise to the so-called resonant tunneling effect. We will briefly discuss this effect since it has been important in the history of FQH edges and then we will return to the clean tunneling Hamiltonian of equation \ref{eq:2_Htun_general}. \\
\begin{figure}[!h]
    \centering
    \includegraphics[width= \linewidth]{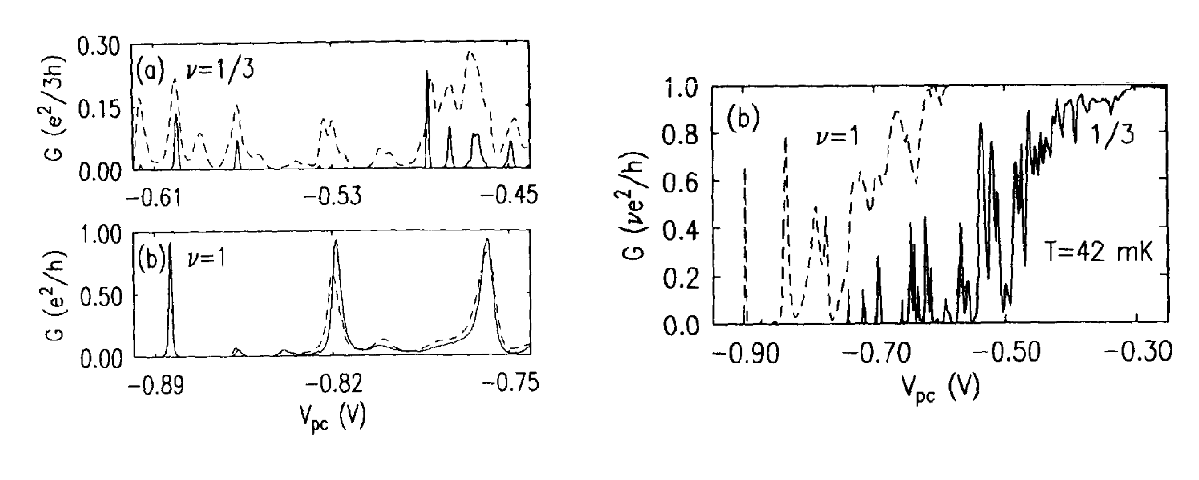}
    \caption{On the left: Resonance behaviour of the zero bias conductance $G = \frac{I}{V}|_{V=0}$ through the QPC as a function of the point contact potential $V_{pc}$ for different temperatures in the FQH (top) and IQH (bottom) case. Dashed lines refers to a higher temperature with respect to solid lines. For the shown values of $V_{pc}$ the QPC is almost pinched off and the conductance is zero when off resonance and is non-zero only on the resonances. \\
    On the right: Conductance for general strength of the QPC. At lower $V_{pc}$ the QPC is pinched off and the conductance is generally 0 except for the resonances, at higher $V_{pc}$ the QPC is open and the conductance takes its quantised value  $G= \nu e^2/2\pi\hbar$ as if the QPC was not present. \\
    Adapted from \cite{Milliken1996_resonances} } 
    \label{fig:2_res_tunn}
\end{figure}\\
An example of the phenomenology of resonant tunneling is shown in figure \ref{fig:2_res_tunn} for the conductance of a QPC geometry. In a free particle picture for the edge (IQH) the shape of those resonances is a Lorentzian with a width and weight that at low temperatures go to constants while for a $\chi$LL (FQH) the resonance is non-Lorentzian with width and weight that shrink to zero in the zero temperature limit (\cite{resonances_theory93} ). Moreover the $\chi$LL predicts particular exponents with which this happens. The resonances are usually observed in the zero bias conductance ($y$ axis of all three panels in figure \ref{fig:2_res_tunn}) by changing the gate voltage potential (see figure \ref{fig:2_res_tunn}) and so the relative energy between localized states and edges. The behaviour of such resonances have been one of the first experimental indication of the $\chi$LL behaviour of FQH edge states (\cite{Milliken1996_resonances}).  \\
During the rest of the thesis we will usually avoid these resonances and we will focus on the tunneling Hamiltonian of equation \ref{eq:2_Htun_general}.
\section{Conductance}
As we somehow already mentioned in the previous section, an important observable is the response of the system to a constant voltage difference or, equivalently, an incoming constant current on the constriction. To clarify the notation we will always refer to interpretation of the QPC as the quasi-particle tunneling configuration (left in figure \ref{fig:2_duality}) so that $i_0$ arrive in the top channel, $i_t$ is the current after the QPC in the top channel and $i_b$ is the current backscattered in the bottom channel. This response is measured via the the conductance:
\begin{equation}
    G = \frac{i_t}{V} = G_0 \frac{i_t}{i_0} \qquad G_0 = \frac{e^2}{2\pi\hbar} \frac{1}{m}
\end{equation}
where $G_0$ is the quantised conductance without tunneling, $i_t$ is the transmitted current and $i_0$ is the incident current generated by the voltage difference $V$. The tunneling current will be equal to the backscattered current $i_b$. Then we will usually refer to $t=i_t/i_0$ as the transmittivity of the QPC. The conductance is typically looked in the zero bias limit ($i_0\rightarrow 0$); otherwise one can look at the differential conductance $G_{diff} = G_0\;\partial i_t / \partial i_0 $ or directly at the characteristics $i_t(i_0)$ or $i_b(i_0)$. 
\subsection{Perturbation theory}\label{sec:2_perturbation_theory}
The most straightforward way to include the tunneling term in the Hamiltonian is through perturbation theory and also historically this has been the first tool to be applied ( \cite{Wen_pert_theory} ). Considering the two dual limits of figure \ref{fig:2_duality} one can apply perturbation theory for the quasi-particle tunneling when $t\simeq 1$ and particle tunneling when $t\simeq 0$. The results for the perturbation theory for particle tunneling ($t\simeq 0$) are:
\begin{align}
    i_t \propto \Gamma_{p}^2 i_0^{2m-1}  \qquad \text{for}\;\; k_B T \ll i_0 /G_0\\
    i_t \propto \Gamma_{p}^2 i_0 T^{2m-2}  \qquad \text{for}\;\; k_B T\gg i_0 /G_0
\end{align}
Note that at low temperatures the behaviour is non-linear for $m\neq 1$ and hence deviates from an Ohmic behaviour. Instead at high enough temperature the behaviour is linear yet with a non-trivial power law dependence of conductance on the temperature. \\
For the quasi-particle tunneling near $t\simeq 1$, the results are:
\begin{align}
    i_b \propto \Gamma_{p}^2 i_0^{-(m-2)/m}  \qquad \text{for}\;\; k_B T \ll i_0 /G_0\\
    i_b \propto \Gamma_{p}^2 i_0 T^{-2(m-1)/m}  \qquad \text{for}\;\; k_B T \gg i_0 /G_0
\end{align}
Note that these are the same of the particle provided we exchange $m\rightarrow1/m$ and $i_t \rightarrow i_b$.
In the high temperature limit we will then have a finite correction to the zero bias conductance $G_0$ which depend on the temperature with a non-trivial power law. In particular $m>1$ the power law is negative and hence as the temperature is increased the constriction becomes effectively weaker. At high incoming currents we have a clear power law for the backscattered current which is negative for $m>2$ and hence imply that at higher currents again the constriction becomes effectively weaker. This result of having a stronger effective backscattering at lower energy scales was actually already anticipated in the RG analysis of section \ref{sec:2_RG}. Following that reasoning the RG analysis also anticipated problems that in the small $i_0$ and $T$ region a perturbative theory could not have worked because of the perturbation being a relevant operator. Indeed in the $T\rightarrow0$ and $i_0 \rightarrow 0 $ limits the perturbation theory gives a divergent result for the backscattered current which of course is in contrast with its assumption $i_b/i_0<<1$. Within RG one could also recover the perturbative regime scalings as done in \cite{Kane1992_RG}. \\
Perturbation theory allows us to solve the problems in the two opposite limits of open and closed QPC but at this point the duality between the two pictures of particle and quasi-particle tunneling is an assumption. We are not guaranteed that the strong coupling limit of one tunneling Hamiltonian recovers the weak tunneling regime of the other. To verify it we in principle need to access the strong coupling regime and this cannot be done by definition in perturbation theory.  

\subsection{Exact results}\label{sec:2_exact}
An exact solution to the problem of a QPC contact with quasi-particle tunneling at constant voltage bias $V$ has been achieved using non-perturbative methods based on the thermodynamic Bethe ansatz ( \cite{Exact_Fendley1,Exact_Fendley2} ). We will not go into the detail of such solutions and we will limit to give some basic ideas. The big difference is that these methods use a new interacting electrons basis where the incident current is treated as a regular flow of kinks and antikinks of the phase field $\phi(x)$, carrying $\pm 1$ charge. These scatter off the constriction following a factorized $S$ matrix and can either be transmitted or backscattered. The backscattered current at zero temperature is then expressed as:
\begin{equation}
    i_b(V,T_\Gamma ) = e v \int_{-\infty}^{A(V)}d\alpha\; d_+(\alpha) |S_{+-}(\alpha-\alpha_\Gamma )|^2
\end{equation}
where $T_\Gamma$ is an energy scale set by the coupling $\Gamma$, $d_+(\alpha)$ is the density of incoming kinks at an energy parametrized by $e^\alpha$. At finite temperature one can also show that the differential conductance can be expressed as:
\begin{equation}
    G(T,V) = G_0 f\Bigl( \frac{T}{T_\Gamma}, \frac{eV}{2\pi k_B T}\Bigl)
\end{equation}
with $f$ an universal function which depend only on $m$ and can be fixed by $G(T,V=0)$. This formalism recovers the two limits we found in perturbation theory and explains the whole crossover between the two regimes. In particular it gives a justification for the duality hypothesis we only sketched before. Indeed the strong coupling regime of the quasi-particle tunneling is captured by the weak coupling regime of the dual picture with a particle tunneling and vice-versa. If one wants to be more accurate, for example as discussed in \cite{Exact_reviewSaleur2002}, the duality holds up to density-density coupling terms which however do not influence the DC properties as they do not transfer current\footnote{We will see in chapter \ref{chap:5} that in the strong coupling dynamics of the quasi-particle tunneling we do have an effective density-density coupling.}. The duality we discussed is then an exact mathematical result and not only a conjecture.
\section{Shot Noise}\label{sec:2_shotnoise}
Another important probe of the physics of these QPC are the noise properties of the tunneling current. Indeed it was thanks to these type of measurements that the fractional charge of Laughlin excitations have been detected for the first time (\cite{DePicciotto1997}).\\
If the tunneling of the charge carriers in a QPC is uncorrelated, the shot noise can be used to determine their charge. To understand how this works, let us imagine a classical picture where the incident current $i_0$ consists of charges $q$ which backscatter independently at the QPC with a probability $p$ so that the backscattered current is $i_b= p i_0$. We then define the rate at which the carriers arrive at the constriction as $\lambda_q = i_0/q$ and we note that the backscattered charge $Q_{\Delta T}$ in a time window $\Delta T$ is described by a Poisson process with rate $\lambda_q p$ where average and variance are:
\begin{equation}
   \langle Q_{\Delta T} \rangle= p\lambda_q \Delta T \, q =  i_b \Delta T \qquad \langle Q_{\Delta T}^2\rangle - \langle Q_{\Delta T} \rangle^2 =   p\lambda_q \Delta T\, q^2 = q i_b \Delta T
\end{equation}
As a result, the backscattered current fluctuations carry information about the charge of the individual carrier. In our $\chi$LL we know that both fractional charges and particles exist but in general their scattering of a QPC is not uncorrelated as in the simple classical picture. However there is a limit where we can treat the tunneling events as basically uncorrelated, namely the weak backscattering limit for the quasi-particles near total transmittivity $t\simeq 1$ and the weak transmittivity limit for the particles near $t\simeq 0$. The shot noise is usually measured in the backscattered current at zero frequency. It is defined as\footnote{Beware that sometimes in the literature (e.g \cite{DePicciotto1997}) a definition with a prefactor $2/\Delta T$ is used instead of $1/\Delta T$.}:
\begin{align}
    S_{sn} =& \frac{1}{\Delta T} \int_0^{\Delta T} dt\int_0^{\Delta T} dt'\, \langle i_b(t)i_b(t')\rangle - \langle i_b(t)\rangle \langle i_b(t')\rangle =\\
    =&\frac{1}{\Delta T} ( \langle Q_{\Delta T}^2 \rangle - \langle Q_{\Delta T} \rangle^2 )
\end{align}
and can be computed in perturbation theory for the quasi-particle and particle tunneling (\cite{Chamon_sn_duality}). At zero temperature one finds that:
\begin{align}
    &S_{sn}^{qp}(i_b) = \frac{e}{m} i_b \qquad \text{for}\;\; i_b \ll i_0 \\
    &S_{sn}^{p}(i_b) = e (i_0-i_b) \qquad \text{for}\;\; i_0-i_b\ll i_0
\end{align}
which corresponds to the classical picture of fluctuations for fractional charges in the quasi-particle case and for integer charges in the dual picture. These results can be recovered as the two opposite limits of weak and strong backscattering in the exact solution ( \cite{Exact_Fendley2} ) we discussed in section \ref{sec:2_exact} and a crossover between this two limits is recovered where the effective charge $q$ that can be measured goes continuously from $e/m$ to $e$. However, in the intermediate regime the independent particle scattering picture described above fails since correlations between tunneling events are important. Finite temperature corrections can also be calculated and are particularly relevant for comparison with experiments. The analytical expressions which works in the weak quasi-particle tunneling regime $t\simeq 1 $ and in the weak particle tunneling regime $t\ll1$ are:
\begin{align}
    &S^{qp}_{sn}(i_b,T) = \frac{e}{m} i_b t\Bigl[ \coth\Bigl(\frac{1}{m}\frac{i_0}{G_0 2 k_B T} \Bigl)-m\frac{G_0 2 k_B T}{i_0}\Bigl]+ 2 k_B T G_0\\
    &S^{p}_{sn}(i_b,T) = e (i_0-i_b) (1-t)\Bigl[ \coth\Bigl(\frac{i_0}{G_0 2 k_B T} \Bigl)-\frac{G_0 2 k_B T}{i_0}\Bigl]+ 2 k_BT G_0
\end{align}
The two temperature regimes gives for the quasi-particle case $t\simeq1$:
\begin{align}
    &S^{qp}_{sn}(i_b,T) = \frac{e}{m} i_b t + 2k_B T G_0 \qquad \text{for}\;\; k_B T \ll i_0/G_0\\
    &S^{qp}_{sn}(i_b,T) =  2k_B T G_0 \qquad\qquad\text{for}\;\; k_B T \gg i_0/G_0
\end{align}
and similarly for the particle case:
\begin{align}
    &S^{p}_{sn}(i_b,T) = e (i_0-i_b) (1-t) + 2k_B T G_0 \qquad \text{for}\;\; k_B T \ll i_0/G_0\\
    &S^{p}_{sn}(i_b,T) =  2k_B T G_0 \qquad\qquad\text{for}\;\; k_B T \gg i_0/G_0
\end{align}
At low temperature we recover the linearity between $S$ and $i_b$ ($i_0-i_b$) but with the addition of a constant noise which goes to 0 with $T\rightarrow 0$. On the other hand at high temperatures the noise is entirely dominated by thermal fluctuations and the signature of the charge $e$ or $e/m$ is absent.
\\

\section{Anyon detection}
\begin{figure}[!h]
    \centering
    \includegraphics[width=\linewidth]{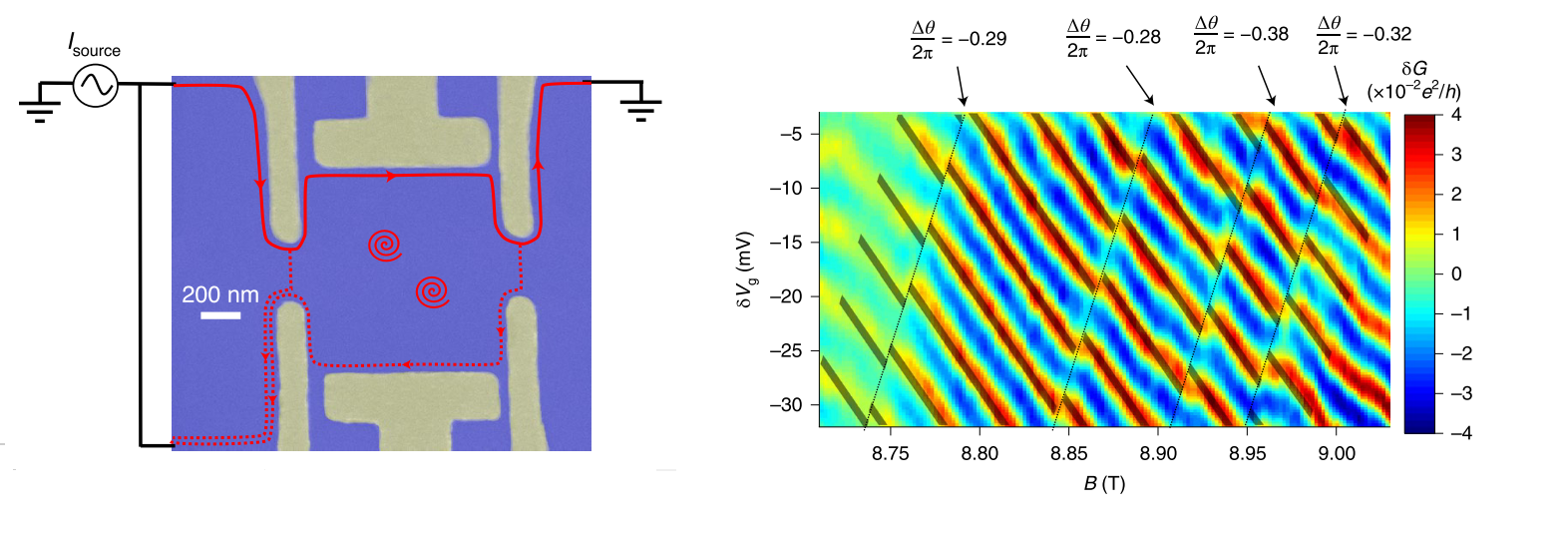}
    \caption{On the left: typical geometry of a Fabry-Perot interferometer. Red lines indicate the chiral current flows and dashed red lines indicate the tunneling paths. Vortexes represent localized quasi-particles in the bulk.\\
    On the right: Conductance oscillations in the $B$-$\delta V_g$ plane. The lines where the Aharonov-Bohm oscillations change in phase correspond to the introduction of quasi-particles in the interferometer area. \\
    Adapted from \cite{Nakamura2020}}
    \label{fig:2_FP_interferometer}
\end{figure}

The other property which we would like to probe is the anyonic statistics of the quasi-particles. Again tunneling experiments through edges of FQH liquids have been the main tools for the experimental verification of the exotic statistics of its quasi-particles. Even though the fractional charge have been measured thanks to shot noise measurements in the early day of the FQH, a convincing experimental verification of the braiding phase $e^{i\pi/m}$ only came in 2020 (\cite{Nakamura2020}). The same year has actually witnessed two different experimental verification of such effect, a more direct one using a Fabry-Perot interferometer for quasi-particles  (\cite{Nakamura2020}) and a more indirect one which use a sort of anyon collider (\cite{AnyonCollider_exp} ) following the theoretical proposal in \cite{AnyonCollider_theory}.\\
The latter experiment uses two QPC in the weak backscattering regime to generate two distinct dilute beams of quasi-particles that are then sent towards a third QPC giving rise to an "anyonic" collider. Since the noise properties in the outgoing channels of such collider depend on the statistics of the incident particles an indirect measurement of the braiding statistics can be obtained. We mention here also another proposal in \cite{AnyonCollider_bulk} for the detection of anyonic braiding effects through the scattering of anyonic bulk molecules, similar in spirit to the anyon collider on the edges we just described. However this proposal treats only bulk quasi-particles and is then best suited for synthetic materials such as cold atoms rather then electronic systems where bulk exitations live in a much more complicated world.\\
We now instead spend a little bit more words for the anyonic interferometer experiment. The idea of using an interferometer to measure a phase is probably the most intuitive one and indeed it has been the first to be proposed ( \cite{Wen_interferometer}). The geometry of such Fabry-Perot interferometer is shown in figure \ref{fig:2_FP_interferometer} and consists of two QPC placed one after the other so to enclose an area $A$. 
These devices usually operate with both constrictions in the weak backscattering regime, so that most of the current passes through each constriction and only a small fraction is backscattered. The two tunneling Hamiltonian that one should use in the description must have a phase difference: quasi-particles that encircle the interferometer area will get a phase $\Phi$ resulting from an Aharonov-Bohm contribution due to the magnetic field and a braiding phase due to quasi-particles which may be localized in the bulk:
\begin{equation}
    \Phi = \Phi_{AB} + \Phi_{s} = \frac{2\pi}{m}\frac{AB}{\phi_0} + \frac{2\pi}{m} N_{qp}
\end{equation}
where $\phi_0$ is the flux quantum and $N_{qp}$ is the number of quasi-particles localized in the bulk. The Aharonov-Bohm phase shows a so-called super-periodicity: because the charge of the quasi-particle is $e/m$, as we increase the magnetic field we need to add $m$ flux quantum to return to the same initial phase. The statistical phase is $2\pi/m$ as a full round trip around another quasi-particles exchanges twice the position and then provide twice the phase $\pi /m$. The Hamiltonian will then look as 
\begin{equation}
    H_T = \Gamma_1 \cos( \Delta \phi_1 ) + \Gamma_2 \cos( \Delta \phi_2 + \Phi )
\end{equation}
where the phase difference in the two tunneling terms summarizes the two contributions just described.\\
The backscattered current from the two QPC then interferes and shows oscillations as $\Phi$ varies from 0 to $2\pi$. In real devices there are two external handles through which it is possible to control the phase $\Phi$ and these are the magnetic field $B$ and the area of the interferometer that can be controlled by side gate voltages $V_s$. The area in principle depend also on $B$ but it is in general only a slow dependence which can be neglected. These have the the effect of changing the AB contribution continuously and can induce a change in the population of the quasi-particle localized state in the bulk at some sample dependent values of $B$ and $V_{s}$. A typical 2D colormap of the conductance oscillations is shown in figure \ref{fig:2_FP_interferometer} where one clearly see the lines where the AB oscillations abruptly change by a phase which exactly fits the introduction of one quasi-particle in the bulk. 
The physics of such interferometers is not at all trivial in particular because of the long range part of the Coulomb interaction that screens the dependence on $N_{qp}$ of $\Phi$ by changing the interferometer area whenever a quasi-particle enters it(see for example \cite{Interferometers_lasttheory2020}). The experiment in \cite{Nakamura2020} used a particular configuration with two screening wells parallel to the 2DEG that successfully decoupled the edges and bulk of the interferometer. We refer to the literature for a complete discussion of such topics ( \cite{Interferometers_review2011,Review_FeldmanHalperin2021}  ). \\
\section{Time dependent properties}\label{sec:2_timedep}
\begin{figure}[!h]
    \centering
    \includegraphics[width=\linewidth]{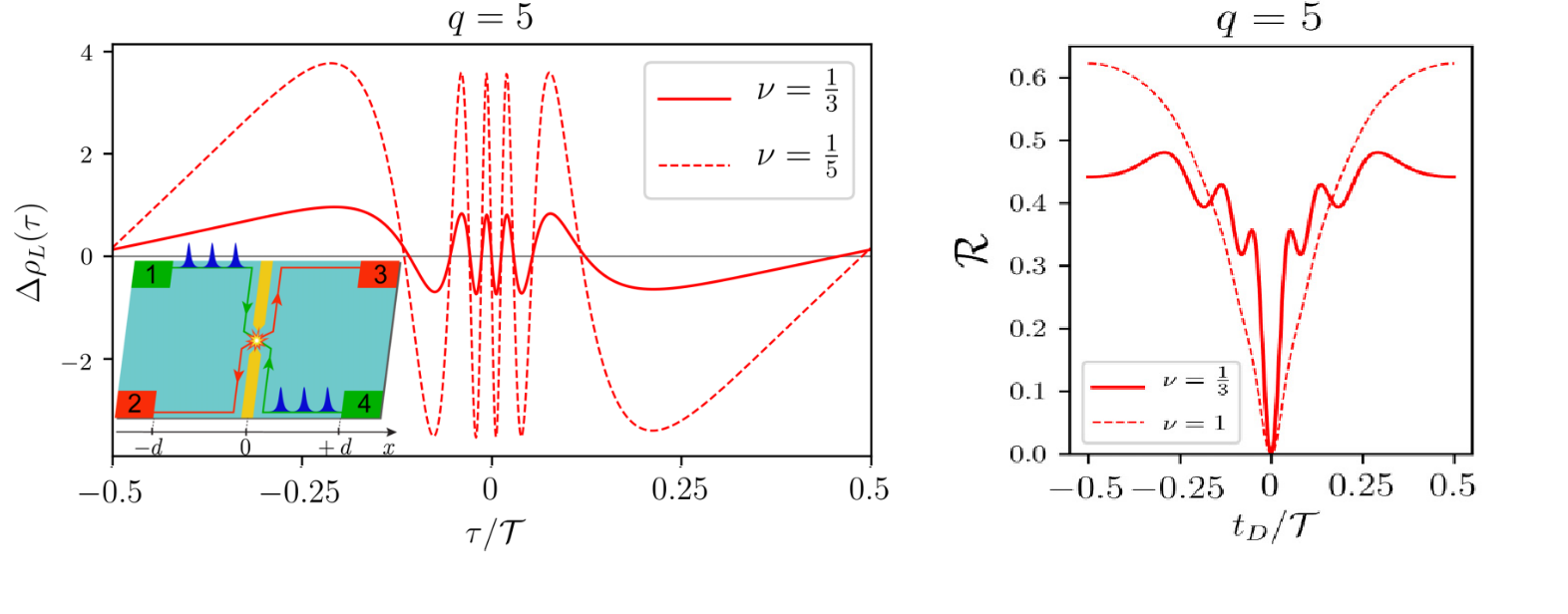}
    \caption{On the left: Crystallization effect in the backscattered current $\Delta \rho_L$ as a function of time $\tau$ generated by a $q$-leviton scattering against a QPC. The number of peaks is equal to the charge $q$. The inset shows a pictorial representation of the QPC geometry with two trains of levitons scattering off the QPC. The signature in the backscattered current refers to a configuration where only one of two such trains is present and $\mathcal{T}$ refers to the period of the leviton sequence. Each leviton carries $q=5$ and indeed 5 peaks are observed\\
    On the right: Excess noise signature of the collision between two trains of levitons with a time delay $t_D$. The noise $\mathcal{R}$ is normalized with the value of the Hanbury-Brown and Twiss configuration where levitons are injected from only one side of the QPC. \\
    Adapted from \cite{Cristallization}}
    \label{fig:2_cristallization}
\end{figure}

Everything we saw in the last sections is done at constant voltage bias. If on one side the the DC properties of scattering in $\chi$LL has been well studied both from a theoretical and from an experimental point of view not as much has been done at finite frequencies and with time dependent drives. The time dependent control of excitations at the edge of FQH systems is still an open problem.\\
In the last years a growing attention has been devoted to this, in particular having in mind the huge achievements in the so called Electron Quantum Optics (EQO) community with IQH edge states ( \cite{EQO_review2014} ). The EQO use the coherent transport of electrons in integer quantum hall edges and has succesfully implemented Hong-Ou-Mandel experiments as well as time dependent sources of coherent electronic states ( \cite{EQO_coherence_SingleElectronSource}). Within the context of time dependent excitations a particular role is played by \textit{levitons}, spatially Lorentzian shaped wavepackets which live above the Fermi sea (\cite{EQO_levitonsReview2017} ). Levitons has been widely use in EQO and in the last years they have been extended also to the context of FQH edges (\cite{Levitons_fqh}) where a different phenomenology arise because of the $\chi$LL nature of the modes. For the FQH case in \cite{Cristallization} the scattering of levitons with an integer charge $q$ at a QPC has been analyzed by using perturbation theory in the quasi-particle tunneling amplitude  and the phenomenon of \textit{crystallization} have been predicted. The name "crystallization" comes from the particular spatial shape that the bakscattered excitation gets, as it is composed by a sequence of peaks and valleys in a number exactly equal to $q$ (see left in figure \ref{fig:2_cristallization}). In the same context a possible noise signature has been proposed for the collision of two such $q$-levitons sent towards a QPC with a time delay $t_d$  (see right in figure \ref{fig:2_cristallization}). \\
 Still no experimental verification of such phenomena have been claimed.\\
On a more general ground an interesting question is whether it is possible to build a reliable time resolved quasi-particle source in FQH edges as those existing for electrons in the IQH case of EQO. There have been a couple of proposals which involve on one side an externally driven antidot in the bulk weakly coupled to the edges (\cite{QuasiparticleEmitter_Ferraro2015}) and on the other levitons scattering off a QPC (\cite{EQO_levitonsReview2017}). However again no experimental implementation has been tested to our knowledge. A clearer understanding of the real time dynamics in $\chi$LL is in our opinion needed. Its non-trivial correlations drastically change the free electron picture of IQH systems which have driven all the discoveries in the EQO context. As we will see in the following chapters a different perspective on the crystallization effect and more generally on time dependent dynamics is one of the main result of this thesis. \\
For finite frequency drives, few results have been obtained at the experimental level in FQH edges and in particular they treat AC voltages at a fixed frequency. Recently in \cite{PASN_fqh} the so called photo-assisted shot noise (PASN) have been characterized for two different FQH edges $m=3$ and $m=5$ confirming with a new tool the existence of fractional charges $e/m$. The theoretical description of such effects has been described a long time before such experiment (\cite{Chamon_sn_duality,PASN_fqh_theory1} ) but as stressed in \cite{PASN_fqh} the experimental combination of microwave irradiation ($>10 \,GHz$) for the AC voltage, ultralow temperatures for the FQH state realization ($\simeq 20\,mK$) and sub-$fA/Hz^{1/2}$ current noise measurements has revealed highly challenging. This tells us that complications in passing from IQH to FQH edges is not only in the theoretical description but the experimental realization itself is a factor to not forget. \\
In this context the realization of FQH states in synthetic systems (e.g. cold atoms \cite{Cooper_topoatoms}, superconducting circuits \cite{Carusotto_PhotonicMaterials2020} and Rydberg polartions \cite{Simon2020_Rydberg_fqh} ) could give a lot of physically valuable information. Here the physical system is much more tunable and new type of measurement, time dependent ones for example, becomes easily accessible. Their problem, in some platforms more than in others, remain the scalability and whether a thermodynamic limit where the $\chi$LL applies can be reached, as in the case of electronic systems.
\\
\\
\newline
In this chapter we gave a broad overview on the problem of scattering in $\chi$LL. We discussed how it has been the tool for important discoveries, from the fractional charge to the quantum statistics of the FQH quasi-particles. In the end we introduced the problem of time dependent situations which have been less discussed in the literature but seems to show a non-trivial phenomenology. We also pointed out how a great push to perform time dependent measurements comes from the field of EQO where the theoretical framework is simpler because it is a free particle problem. The next chapters will discuss the original results of the thesis which aim at giving a different and clearer perspective on scattering in $\chi$LL.

\chapter{Classical theory}\label{chap:3}
\begin{figure}
    \centering
    \includegraphics[width=10cm]{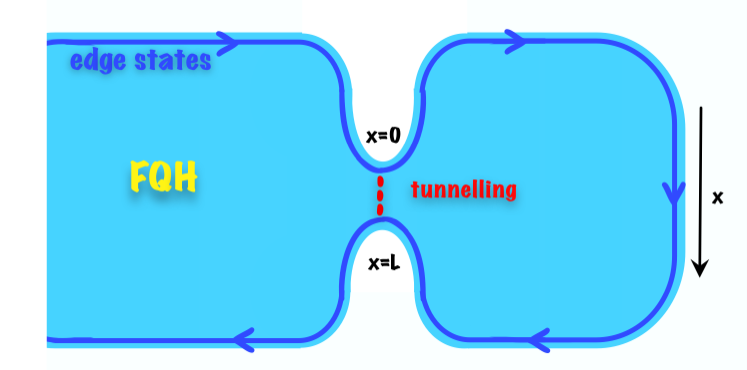}
    \caption{Schematic picture of the geometry under consideration}
    \label{fig:3_geometry}
\end{figure}

After reviewing the basics of QH systems and presenting state-of-the-art and on-going experimental challenges, in this and the following chapters we will discuss the results we obtained during the thesis work. \\
In this chapter we will give a look at the classical trajectories which arise from the chiral Luttinger Liquid theory. This is the easiest starting point for a discussion of the dynamics of excitations which scatter on a constriction. As we saw in the previous chapter a lot has been done for the problem of constant voltage bias but the physics of time dependent excitations has not been fully investigated. This approach a priori does not tell anything about the quantum evolution of the system but as we will see in chapters \ref{chap:4} and \ref{chap:5} we are actually not too far. We remind that indeed when linearity is present the expectation values of observables follow in general the classical trajectory.\\
The geometry that we will look at is that of a disk with a constriction where excitations can scatter from one side of the disk to another. The coordinate describing the edge is $x\in[0,L]$ and we will consider the constriction happening between $x=0$ and $x=l$ in the geometry depicted in figure \ref{fig:3_geometry}.
The action for the free phase field is the Floreanini-Jackiw as we saw in section  \ref{sec:1_effective_theory} for right moving excitations:
\begin{equation}
    S[\phi] = -\frac{m}{4\pi} \int dxdt \,\partial_t\phi \partial_x\phi + v (\partial_x \phi)^2
\end{equation}
which describe a right moving chiral boson. The closed geometry we are considering has the advantage of needing a single edge, we only need one phase field. This imply that even when we allow scattering in some points the total charge on the edge will be a conserved quantity. As we discussed in section \ref{sec:1_effective_theory} the periodic condition on $\phi$ are:
\begin{equation}
    \phi(L) = \phi(0) + 2\pi Q
\end{equation}
where $Q$ is the total charge on the edge, the zero mode. In our treatment we will always consider $Q=0$, whenever a localized charge is consider we can always think about having the opposite charge far away and forget it when talking about scattering at the constriction. 
The equation of motion is obtained by differentiating with respect to $\delta\phi(x,t)$ and is:
\begin{equation}
    \partial_t\partial_x \phi+v\partial_x^2 \phi = 0
\end{equation}
When written in terms of the density field we have the the chiral wave equation:
 \begin{equation}
     \partial_t \rho(x,t) +v\partial_x \rho(x,t) = 0
 \end{equation}
 In the next section we will derive the equation of motion for different couplings in the constriction.

 \section{Equation of motion}
 We will look at 3 different possibilities: quasi-particles, particles and density scattering. Because of the linear dispersion we remind that  space and time have the same physical scales just multiplied by a prefactor so talking about one or the other is basically the same.
 \subsection{Quasi-particle and particle tunneling}
 As we already saw at the quantum level in section \ref{sec:1_qp_p} the quasi-particles and particles excitation are governed by fields which are written as exponentials of the phase field. The coupling which describe the tunneling of a quasi-particle from one point to the other has been introduced in section \ref{sec:2_tunn_ham} an the correspoding action term will then be:
 \begin{equation}
     S_T^{qp} = \frac{m}{4}\Gamma \int dt \;e^{i\phi(0,t)}e^{-i\phi(l,t)}\; + \;h.c.
 \end{equation}
 where $\Gamma$ is a coupling constant and $e^{i\phi}$ is the quasi-particle field. The coupling for the tunneling of particles will be the same but with the particle field:
 \begin{equation}
          S_T^{p} = \frac{1}{4}\Gamma \int dt \;e^{im\phi(0,t)}e^{-im\phi(l,t)}\; + \;h.c.
 \end{equation}
 The normalisations are chosen so to get only $\Gamma$ in the equation of motion.
 We will first discuss the quasi-particle case and then the particle will follow easily. The equation of motion are computed by minimising with respect to the field $\phi$:
 \begin{align}
     \frac{\delta S_T^{qp}}{\delta \phi(x,t)} = \Gamma \int dt' \;\frac{\delta \,\cos(\phi(l,t')-\phi(0,t') )}{\delta \phi(x,t)} = \\ \Bigl( \delta(x-l) - \delta(x) \Bigl) \sin(\phi(l,t)-\phi(0,t)) 
 \end{align}
 The phase field difference in the $\sin$ has the clear interpretation as the charge excess in $x\in[0,l]$:
 \begin{equation}
     \phi(l,t)-\phi(0,t)= 2\pi \int_0^l dx \; \rho(x,t) = 2\pi q(t)
 \end{equation}
 The quasi-particle coupling can then be interpreted as a cosine potential with periodicity 1 for the charge $q$ enclosed in the region which we will call \textit{cavity}. In writing the full equation of motion we must be careful to the sign of the free action and its normalization. The result is:
 \begin{equation}\label{eq:3_eqmot_qp}
     \partial_t \rho(x,t) = -v \partial_x \rho(x,t) - \Gamma (\delta(x)-\delta(x-l) ) \sin( 2\pi q(t))
 \end{equation}
 This describes a chiral propagation of the density field plus two source terms which respect the total conservation of current at the constriction point. The source terms introduce a discontinuity in the density field at the contact points that can be calculated from equation \ref{eq:3_eqmot_qp} by integrating in a small window around the constriction point at $x=0$ and $x=l$. For example around $x=0$ we have:
 \begin{align}
     \partial_t \int_{-\epsilon}^\epsilon dx\;\rho(x,t) = - v&\int_{-\epsilon}^\epsilon dx \,\partial_x \rho(x,t) +\\
     &- \Gamma \sin(2\pi q(t)) \int_{-\epsilon}^\epsilon dx\, (\delta(x)-\delta(x-l) )
 \end{align}
 where the r.h.s is of order $\epsilon$, the first term in the r.h.s. gives the difference $\rho(0^+)-\rho(0^-)$ and the last term gives the discontinuity. The results on $x=0$ and $x=l$ are:
 \begin{align}\label{eq:3_qp_tcurrent}
     \rho(0^+,t) &= \rho(0^-,t) -\frac{\Gamma}{v}  \sin(2\pi q(t))\\
     \rho(l^+,t) &= \rho(l^-,t) + \frac{\Gamma}{v} \sin(2\pi q(t)) \label{eq:3_qp_bcurrent}
 \end{align}
 Given this conditions and the chiral propagation, the equation of motion for $\rho(x,t)$ can actually be written in terms of the variable $q(t)$. This is achieved by integrating the equation \ref{eq:3_eqmot_qp} over $x$ from $0^+$ to $l^-$:
 \begin{align}
     &\int_{0+}^{l^-} dx\;\partial_t \rho(x,t) = - v\int_{0^+}^{l^-} dx\;\partial_x \rho(x,t) - \int_{0^+}^{l^-} dx\;\Gamma (\delta(x)-\delta(x-l) ) \sin( 2\pi q(t))\\
     &\partial_t q(t) = -v\rho(l^-,t) +v \rho(0^+,t) \\
     &\partial_t q(t) = v\rho(0^-,t) - v\rho(0^+,t-l/v) - \Gamma\sin( 2\pi q(t))\\
     &\partial_t q(t) = v\rho(0^-,t) - v\rho(0^-,t-l/v) - \Gamma\sin( 2\pi q(t)) + \Gamma\sin( 2\pi q(t-l)) \label{eq:3_eqmot_q_qp}
 \end{align}
 where we used the chiral constraint $\rho(x,t) =\rho(0,t-x/v)$ valid away from the constriction. In this way we managed to write an equation of motion for $q(t)$ only in terms of the incoming current $\rho(0^-,t)$. Equation \ref{eq:3_eqmot_q_qp} is a first order ordinary differential equation non local in time that can be numerically solved quite easily for $q$ given the forcing term $\rho(0^-,t)$. The physical meaning of the terms is clear, the forcing $\rho$ represents the current entering the cavity at time $t$ and the escape of the current that had entered the cavity at $t-l$. Analogously the last two terms in $\Gamma$ are the tunneling current entering at time $t$ and the tunneling current which had entered at $t-l$ escaping the cavity. Once we know $q(t)$ we can determine the charge density at any point, of particular interest is the outgoing current
 \begin{equation}
     \rho(l^+,t) = \rho(0^-,t-l) -\frac{\Gamma}{v} \sin(2\pi q(t-l)) + \frac{\Gamma}{v}  \sin(2\pi q(t) ) 
 \end{equation}
 where the interpretation of each term is what we described above for the equation of motion. \\
 The equation of motion for the particle tunneling are very similar, we only have a different periodicity in the sine: 
 \begin{equation}\label{eq:3_eqmot_p}
     \partial_t \rho(x,t) = - v\partial_x \rho(x,t) - \frac{\Gamma}{v}  (\delta(x)-\delta(x-l) ) \sin( 2\pi m  q)
 \end{equation}
 and for the charge in the cavity:
 \begin{equation}
          \partial_t q(t) = \rho(0^-,t) - \rho(0^-,t-l) - \frac{\Gamma}{v} \sin( 2\pi mq(t)) + \frac{\Gamma}{v} \sin( 2\pi m q(t-l))
 \end{equation}
 \subsection{Density-density coupling}
 Another possible interaction in the constriction is of density-density type. The action term which correspond to such an interaction is:
 \begin{equation}
     S_{d}[\phi] = \frac{m}{2}\gamma \int dt \int dx_1dx_2\; \delta_\Delta(x_1)\delta_\Delta(x_2-l)\rho(x_1,t) \rho(x_2,t) 
 \end{equation}
 where now we use a regularized delta over $\Delta$ which has the physical meaning of giving a scale to the constriction. 
 The equation of motion are again found by minimising with respect to $\phi(x,t)$:
 \begin{align}
     \frac{\delta S_{d}}{\delta \phi(x,t)} =& -\frac{m}{8\pi^2}\gamma \int dt' \int dx_1dx_2\; \delta'_\Delta(x_1)\delta_\Delta(x_2-l)\frac{\delta \phi(x_1,t')}{\delta \phi(x,t)}\partial_x\phi(x_2,t')+\\
     &\;\;+\delta_\Delta(x_1)\delta'_\Delta(x_2-l)\partial_x\phi(0,t')\frac{\delta \phi(x_2,t')}{\delta \phi(x,t)}\\
     =& -\frac{m}{4\pi}\gamma  \delta'_\Delta(x)\int dy\; \delta_\Delta(y-l)\rho(y,t)+\int dy\;\delta_\Delta(y)\delta'_\Delta(x-l)\rho(y,t)
 \end{align}
 So that we have:
 \begin{align} \label{eq:3_eqmotion_density}
     \partial_t \rho(x,t) = -v \partial_x \rho(x,t)& - \gamma \delta'_\Delta(x) \int dy\;\delta_\Delta (y-l)\rho(y,t) +\\
     &-\gamma \delta'_\Delta(x-l) \int dy \;\delta_\Delta (y)\rho(y,t)
 \end{align}
 The physical intuition behind a density-density coupling is probably more straightforward than the cosine coupling we saw above that comes from the counter intuitive form of the particles fields. Whenever an excess charge is present on the boundary it will affect the charge distribution nearby since it disturbs the equilibrium.

\section{Wavepacket scattering}
As a first application we will now look at what happens at the classical level to a charge wavepacket sent towards the constriction point using the different couplings we discussed in the previous section. Throughout the section we will mainly use gaussian wavepackets characterized by two quantities, the width $\sigma$ and the charge $Q$:
\begin{equation}
    \rho_{wp}(x) = Q \frac{1}{\sqrt{\pi\sigma^2}}e^{-x^2/\sigma^2}
\end{equation}
We remind that away from the constriction the chiral propagation $\rho(x,t) = \rho(x+vt,0)$ is valid. As we anticipated above in order to justify a generic charge $Q$ we would also need a charge $-Q$ somewhere on the edge, we consider it to be far away from the constriction so that it does not influence the dynamics of the single wavepacket. The different time/space scales present in the problem are the length of the cavity $l$, the inverse coupling constant $\Gamma^{-1}$ or $\gamma^{-1}$ and now also the wavepacket size $\sigma$. Again we remark that time and space differ just by a proportionality constant $v$. 
\subsection{Quasi-particle and particle tunneling}\label{sec:3_qp_p_tunneling}
\begin{figure}
\centering
\begin{subfigure}{.5\textwidth}
  \centering
  \includegraphics[width=7cm]{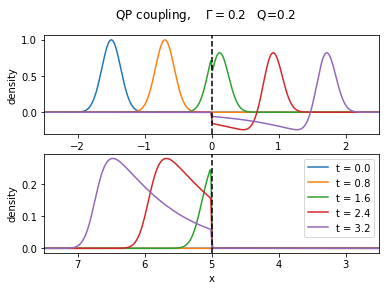}
  \caption{$Q<0.5$ case}
  \label{fig:fig:3_qp_weak_02}
\end{subfigure}%
\begin{subfigure}{.5\textwidth}
  \centering
  \includegraphics[width=7cm]{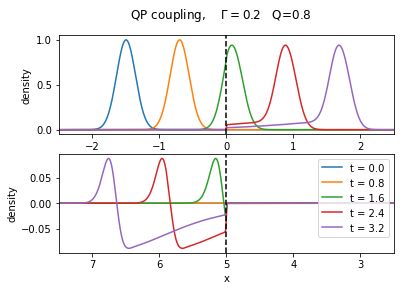}
  \caption{$Q>0.5$ case}
  \label{fig:fig:3_qp_weak_08}
\end{subfigure}
\caption{Weak tunneling of wavepackets with a different amount of charge. Both left and right figure show a density normalized so that the maximum height is 1. The upper panels and lower panels should be viewed as a single representation of the x-axis with the right part of panels identified (see figure \ref{fig:3_geometry}). The constriction is at $x=0$ and $x=5$. The wavepacket width is $\sigma=0.2$, the velocity is $v=1$ and the cavity length $l=5$}
\label{fig:3_qp_weak}
\end{figure}

\begin{figure}
    \centering
    \includegraphics[width=8cm]{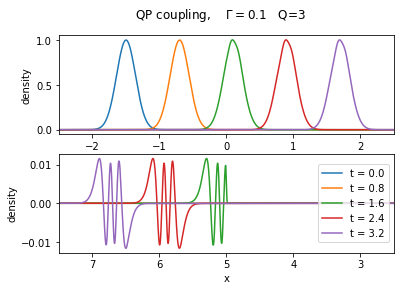}
    \caption{Crystallization of a wavepacket of charge $Q=3$ that generates an excitation with 3 distinct peaks and valleys.  The density normalized so that the maximum height is 1. The upper panels and lower panels should be viewed as a single representation of the x-axis with the right part of panels identified (see figure \ref{fig:3_geometry}). The constriction is at $x=0$ and $x=5$. The wavepacket width is $\sigma=0.2$}
    \label{fig:3_cristallization}
\end{figure}

The differential equations we need to solve are equation \ref{eq:3_eqmot_qp} and \ref{eq:3_eqmot_p} . Again since they differ just by a renormalisation of the charge we will look at the quasi-particle coupling and then draw parallelism with the particle one. Let us consider the case of short wavepackets $\sigma \ll l$. At times $t<l$ the memory terms are zero and the equation for $q$ describe exactly the overdamped dynamics of pendulum with a forcing torque. The equation for an overdamped and forced pendulum is:
\begin{equation}
b \dot{\theta}(t) =F(t)- mgl \,\sin(\theta(t)) 
\end{equation}
where $b$,$F$,$m$,$g$ and $l$ are respectively the friction coefficient, the forcing torque, the mass, the gravitational acceleration and the length of the pendulum. It is then easy to think about our scattering dynamics in term of this physical simple example with the following identifications:
\begin{align}
    q &\leftrightarrow \frac{\theta}{2\pi}\\
    \frac{\Gamma}{v}  &\leftrightarrow 2\pi\frac{mgl}{b} \\
    \rho(0^-,t) &\leftrightarrow 2\pi\frac{F}{b}
\end{align}
The solutions without a forcing term are just exponentially fast relaxations towards the minimum at $q=0,\pm1,\pm2 ...$ as the pendulum relax towards $\theta=0,\pm2\pi,\pm 2\pi ... $. There are then unstable minima at $q=\pm1/2,\pm 3/2 ... $ as the pendulum is unstable when $\theta = \pm \pi,\pm 3\pi...$ . The typical time of relaxation is set by the coupling $\Gamma^{-1}$ as for the pendulum by the friction coefficient. We will now look at what happens in different regimes of $\Gamma \sigma$
\paragraph{Weak tunneling: $\frac{\Gamma \sigma}{v}   << 1$}
In the pendulum analogy this correspond to a fast angular momentum transfer to the pendulum so that the dissipation is negligible. From the point of view of the charge excitation dynamics it is the limit where the wavepacket passes the constriction almost undisturbed, so all the initial charge $Q$ is transmitted. After the wavepacket has passed the charge in the cavity $q$ wants to reach one of the minima at $q=0,\pm1,\pm2 ...$ in the same way the pendulum relaxes towards its equilibrium point. In our physical system it is interesting to note the non-local nature of our coupling, even when the wavepacket is deep inside the cavity away from the constriction there is some tunneling current at the constriction which is bringing the cavity charge $q$ towards an integer. There is even a more puzzling aspect about this dynamics, it highly depends on the wavepacket charge $Q$. Take for the moment $0<Q<1$, there are two possible direction where $q$ can go, to $q=0$ and $q=1$ depending only on the wavepacket charge $Q$. In the pendulum language this correspond to two situations where with a kick we do one full orbit or the kick is not strong enough to overcome the $\theta=\pi$ ($q=0.5$) point. All this dynamics of $q$ is reflected in the current coming out of the constriction which is given by equation \ref{eq:3_qp_bcurrent} and in our case is just:
\begin{equation}
    \rho(l^+,t) = \frac{\Gamma}{v}  \sin(2\pi q(t) ) 
\end{equation}
where the $\rho(l^-,t)$ term is zero since we consider $t<l$ so no excitations had the time to travel the whole cavity. The outgoing current is then proportional to $\frac{\Gamma}{v} $ and is exponentially decaying in time. The clearer effect is however that it will be positive when $0<q<1/2$ and negative when $1/2<q<1$. This means that the when a wavepacket of charge $1/2<Q<1$ hits the constriction the quasi-particle coupling starts to eject negative current out of the constriction. Examples of such dynamics are shown in figure \ref{fig:3_qp_weak}. We remark here the non-trivial dynamics that the quasi-particle coupling is generating, the amount of charge present in the wavepacket decides between two possible outcomes, a very clear feature. \\
Another very interesting phenomena is the scattering of an excitation with a charge which is bigger than $Q>1$. From the point of view of the pendulum analogy is nothing more than doing more than one orbit before relaxing to the minimum but in the our system this means that the tunneling current of equation \ref{eq:3_qp_bcurrent} is changing sign every time $q(t)$ passes through half integers $q=\pm0.5,\pm1.5 ...$. This is a clear signature in the outgoing current which will oscillates from $\Gamma$ to $-\Gamma$ as many times as the closest integer to $Q$ which is the minima where the trajectory will relax. Examples of this are shown in figure \ref{fig:3_cristallization} . 
This shape of the outgoing current should closely remind the crystallization of levitons (\cite{Cristallization}) we anticipated in section \ref{sec:2_timedep}. The classical trajectories surprisingly reproduce this behaviour, although with some differences, for example the height of the excitation in the classical theory is proportional to the coupling $\Gamma$ while in the perturbative quantum theory is quadratic in $\Gamma$.\\ 

\begin{figure}
    \centering
    \includegraphics[width=8cm]{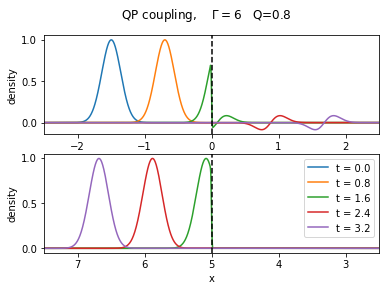}
    \caption{Strong tunneling regime for the scattering of a wavepacket. The wavepacket is completely reflected and a neutral excitation with the shape of the derivative of the wavepacket tunnels inside the cavity. The density normalized so that the maximum height is 1. The upper panels and lower panels should be viewed as a single representation of the x-axis with the right part of panels identified (see figure \ref{fig:3_geometry}). The constriction is at $x=0$ and $x=5$. The wavepacket width is $\sigma=0.2$ and the velocity $v=1$}
    \label{fig:3_qp_strong}
\end{figure}

\paragraph{Strong tunneling: $\frac{\Gamma\sigma}{v} \gg 1$}
Again it useful to think about the dynamics in term of our pendulum analogy. The limit of strong tunneling correspond to the limit where the restoring force is much greater then the kick we are giving to the pendulum. In this limit any excitation which is sent to the constriction bounces back in the only possible channel which is at the other point of the constriction $x=l^+$\footnote{Because of chirality we cannot have excitations going back from $x=0$}. The idea is that in order for $q$ to remain near 0 there must be a tunneling current entering the cavity exactly cancelling the incoming wavepacket, thus giving rise to an excitation going out of the cavity which correspond to the incoming wavepacket. There are however corrections to this picture coming from finite $\Gamma\sigma$ values. To understand these we approximate the sine in the equation of motion as its argument which is equivalent to approximate the cosine potential of the pendulum as a parabola. This is justified as long as $q<<1$. Now the differential equation for $q$ is linear and it is useful to go to Fourier space:
\begin{equation}
    i\omega\Tilde{q}(\omega) = v\Tilde{\rho}(0^-,\omega) -\Gamma 2\pi \Tilde{q}(\omega) \;\;\; \longrightarrow \;\;\;\Tilde{q}(\omega) = v\Tilde{\rho}(0^-,\omega) \frac{1}{i\omega +2\pi\Gamma}
\end{equation}
Now the density just after the constriction can be found from equation \ref{eq:3_qp_tcurrent} in Fourier space:
\begin{equation}
    \Tilde{\rho}(0^+,\omega) = \Tilde{\rho}(0^-,\omega) -\frac{\Gamma}{v}  2\pi \Tilde{q}(\omega) = \Tilde{\rho}(0^-,\omega) \Bigl(1-\frac{1}{i\omega +2\pi \Gamma}\Bigl)
\end{equation}
Since we have wavepackets of size $\sigma$ the frequencies we are interested in are of order $v\sigma^{-1}$ and then in the limit $\Gamma \sigma/v >>1$ we can use $\Gamma>>\omega$ and get:
\begin{equation}
    \Tilde{\rho}(0^+,\omega) = \frac{i\omega}{2\pi \Gamma}\Tilde{\rho}(0^-,t)
\end{equation}
Or in real space:
\begin{equation}
    \rho(0^+,t) = \frac{1}{2\pi\Gamma} \partial_t \rho(0^-,t)
\end{equation}
Equivalently we can derive a closed expression for the outgoing current $\rho(l^+,t)$ in terms of the incoming $\rho(0^-,t)$ in Fourier space:
\begin{equation}
    \Tilde{\rho}(l^+,\omega) = \Gamma 2\pi \Tilde{q}(\omega) = \Tilde{\rho}(0^-,\omega) \frac{\Gamma 2\pi}{i\omega +2\pi\Gamma} \simeq  \Tilde{\rho}(0^-,\omega) (1 - \frac{1}{\Gamma 2\pi} i\omega) 
\end{equation}
and in real space:
\begin{equation}
    \rho(l^+,t) = \rho(0^-,t) - \frac{1}{\Gamma 2\pi } \partial_t \rho(0^-,t) \simeq \rho(0^-,t-\frac{1}{\Gamma 2\pi}
\end{equation}
The correction to the bouncing of the excitations off the constriction is an excitation which has the shape of the derivative of the incoming wavepacket and is proportional to $\Gamma^{-1}$ as shown in figure \ref{fig:3_qp_strong}. This indeed correspond to an a small shift in time of the original excitation.
\\
The intermediate regime $\Gamma\sigma/v \simeq 1$ is an interpolation between the two regimes where no clear features are present. The physical intuition of the overdamped pendulum is still useful and one can get all the situation we have not treated so far. \\
The particle coupling as we said at the beginning gives the same results, just with the periodicity of the sine reduced by a factor $1/m$. This is particularly relevant in the weak tunneling regime $\Gamma \sigma/v <<1$. The classical trajectory seems to predict the same phenomena of crystallization of the excitations but now the number of peaks in the backscattered excitation is equal to the nearest integer to $Q/m$ where $Q$ is the charge of the incoming wavepacket. 
\paragraph{long time limit: $t>l$}
Once we clarified the dynamics at times where the excitations do not see the finite size of the cavity $l$ we can look at what the memory terms add. The pendulum parallelism is now not that useful, there are no such memory terms. \\
\begin{figure}
\begin{subfigure}{.5\textwidth}
  \centering
  \includegraphics[width=7cm]{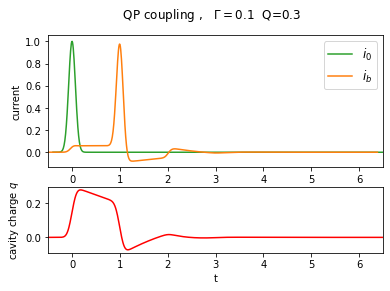}
  \caption{Weak tunneling}
  \label{fig:3_qp_longtime_weak}
\end{subfigure}
\begin{subfigure}{.5\textwidth}
  \centering
  \includegraphics[width=7cm]{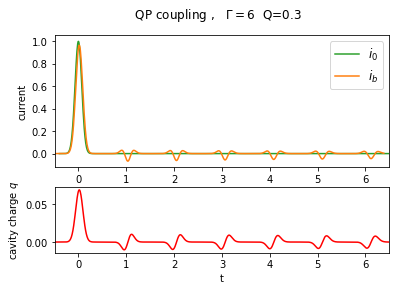}
  \caption{Strong tunneling}
  \label{fig:3_qp_longtime_strong}
\end{subfigure}
\newline
\begin{subfigure}{.5\textwidth}
    \centering
  \includegraphics[width=7cm]{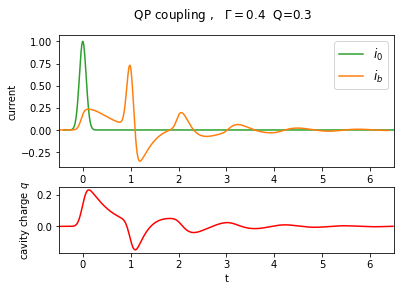}
  \caption{Intermediate tunneling}
  \label{fig:3_qp_longtime_intermediate}
\end{subfigure}
\begin{subfigure}{.5\textwidth}
    \centering
  \includegraphics[width=5cm]{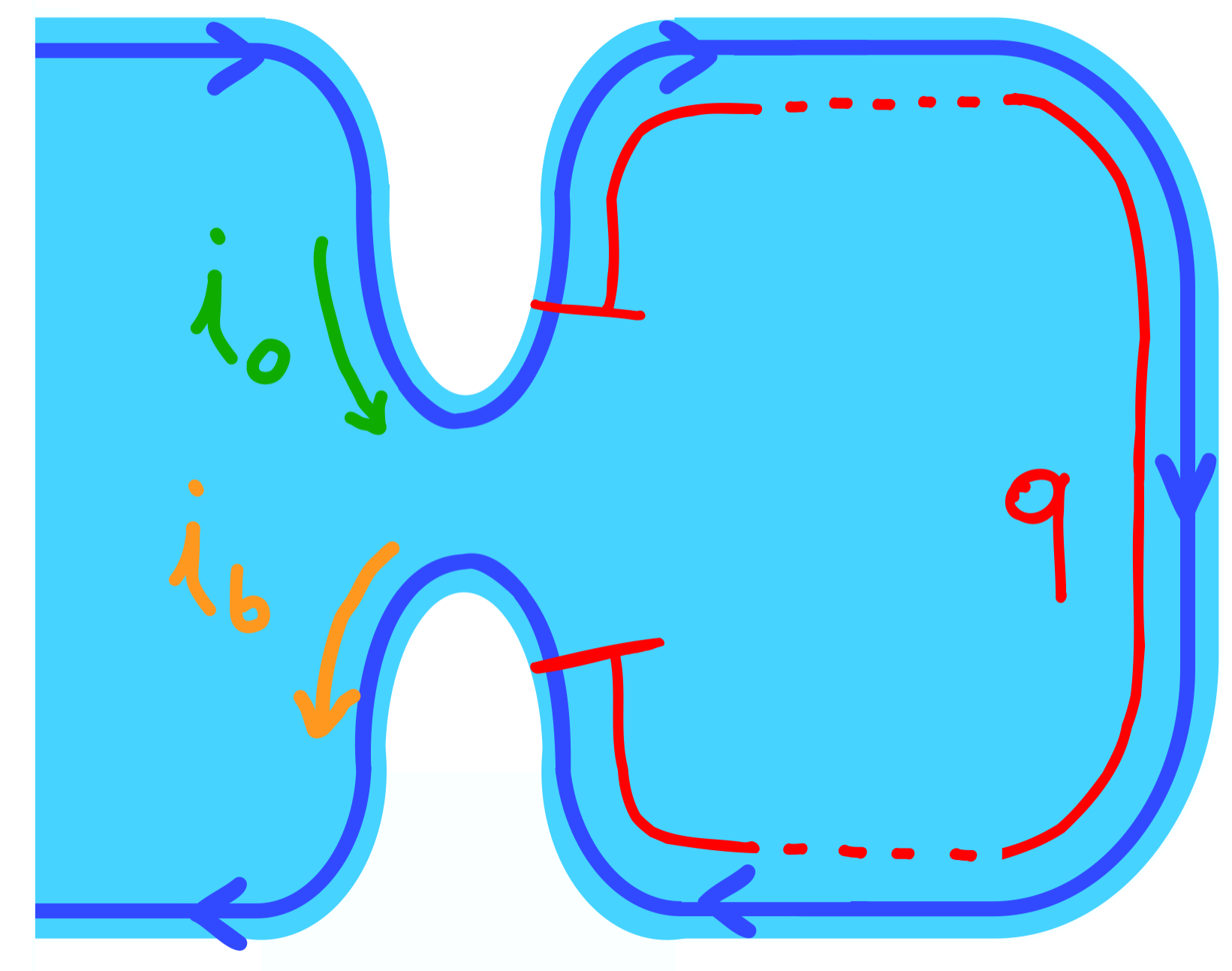}
  \caption{Geometry under consideration}
  \label{fig:3_qp_longtime_geometry}
\end{subfigure}
    \caption{Long time behaviour of the wavepacket scattering on a cavity of length $l=1$ in the weak/intermediate/strong tunneling regimes. The incoming current $i_0$ is the current at $x=0^-$ before the constriction while the outgoing current $i_b$ is the current at $x=l^+$.  The density is normalized so that the maximum height is 1. The wavepacket has size $\sigma=0.1$ and charge $Q=0.3$, the velocity is $v=1$}
    \label{fig:3_qp_longtime}
\end{figure}
 In the weak tunneling case where we have a relaxation dynamics of $q$ towards the closest integer to $Q$ the relaxation is messed up by the fact that the initial wavepacket after a time $l$ escapes the cavity and the charge $q$ drops by a quantity close to $Q$. For example if we have an initial wavepacket of charge $Q=0.3$ at times $t< l$ we will have $q$ which starts at $q(0)\simeq0.3$ and decreases towards 0 exponentially in a time $\Gamma^{-1}$. As soon as the wavepacket escape the cavity at $t=l^+$ the charge $q$ drops to a value which is smaller than zero, depending on how much of the initial charge has been compensated by the current at the constriction. The charge $q$ will then oscillate around zero for some roundtrip times until the charge $q$ is rest at 0 as shown in figure \ref{fig:3_qp_longtime_intermediate} and \ref{fig:3_qp_longtime_weak}. \\
 In the strong backscattering limit, shown in figure \ref{fig:3_qp_longtime_strong}, the excitations present inside the cavity is what we just saw and once the excitation hits again the constriction it remains trapped inside the cavity and a secondary excitation is transmitted outside just as the first wavepacket had created an excitation travelling inside the cavity. The shape of the outgoing secondary excitation is the second derivative of the initial wavepacket. Note also that there is a shift in time of the outgoing excitation with respect to $t= \frac{l}{v}$ since every time the excitation in the cavity hits the constriction it gets slightly delayed as we saw previously for the strong backscattering regime. 
 At strong tunneling instead we will have a completely backscattered wavepacket and an excitation with the shape of the first derivative of the wavepacket inside the cavity. Every time that this excitation bounces back on the constriction it will generate an outgoing excitation with the shape of the second derivative of the initial wavepacket as you can see in figure \ref{fig:3_qp_longtime_strong}.\\
\\

It is a good point now to stop and think about what it is happening. We are describing the classical trajectories of a quantum theory and there is at this point no evidence that these should be somehow meaningful. Indeed if we remember that for $m=1$ the theory must describe free electrons it is really difficult to convince ourselves that the trajectories seen so far are of any meaning. For free electrons one can also set up an exact ab initio calculation as in \cite{Nardin2020}. A. Nardin actually implemented the constriction geometry for an IQH and the effect of the constriction is just to scatter particles from one edge to the other giving the equivalent in the effective theory of $\rho(l^+,t) = R \rho(0^-,t)$ with $R$ a reflection coefficient. These are reasonable doubts and we will resolve them in the next chapter, up to now let's continue our characterization of the classical theory.

\subsection{Density-density coupling }\label{sec:3_dd_scattering}
\begin{figure}
    \centering
    \includegraphics[width=8cm]{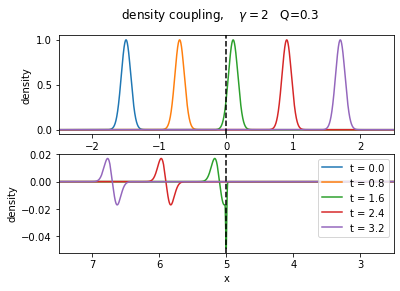}
    \caption{scattering of a wavepacket with the density coupling. The upper panels and lower panels should be viewed in as a single representation of the x-axis with the constriction at $x=0$ and $x=5$. $\sigma=0.2$ and $v=1$}
    \label{fig:3_density}
\end{figure}

The density-density coupling is described by the equation of motion in equation \ref{eq:3_eqmotion_density}. Solving this type of differential equation with derivatives of delta functions in multiplicative terms equation is a bit odd and tricky and one should always work with the regularized version we showed. Once the differential equation is regularized it can be easily solved numerically since it is first order both in time and space, e.g. with the Euler method. The solutions to the scattering of a wavepackets are neutral excitations, differently from the quasi-particle and particle coupling the density is not able to transfer charge from one point to the other. The shape of excitations that comes out of the constriction in the scattering process is the derivative of the incoming wavepacket. This is shown in figure \ref{fig:3_density} for a gaussian wavepacket. 
\\
This phenomenology closely remind that of the strong tunneling limit for the quasi-particle coupling. The mapping between the two situations tells us that in the strong tunneling regime for quasi-particles the constriction becomes basically closed so that one can see the system as a two distinct edges coupled by a weak density-density interaction. This should closely reminds us of the discussion on the duality of chapter \ref{chap:2}. Indeed there we discussed how in the strong coupling limit of the quasi-particle tunneling we should recover the dual picture with weak particle tunneling up to density-density coupling. Here we find that at the classical level the strong coupling of the quasi-particle tunneling give rise to density-density couplings however no weak particle tunneling seems to be present. Of course we cannot ask too much from the classical theory as quantum fluctuations are completely absent but it is nonetheless curios that already something non-trivial is present at the classical level.
\\
\\
\\
In this chapter we characterized the classical solutions to the scattering of wavepackets in a chiral Luttinger Liquid. The conclusion is that even the classical theory is rich of interesting behaviours but not all of this  seems to be plausible in all regimes. What is surely missed by the classical trajectories is the in principle easy limit of free particles at $m=1$. However away from $m=1$ we are able to recover the crystallization phenomena of \cite{Cristallization} valid for $m>2$ then something must change in the quantum theory while we change $m$ that we remind does not appear in the evolution equation in the quasi-particle case. At a naive level we can say that at higher $m$ the charge quantum fluctuations are given by the quasi-particle charge and then are reduced by a factor $1/m$ with respect to the IQH case. Reduced quantum fluctuations hints at a more \textit{classical} behaviour of the theory. In the following chapter we will try to make this argument more quantitative. 
\chapter{Truncated Wigner approach}\label{chap:4}
In the context of quantum optics many theoretical tools have been developed for the inclusion of quantum fluctuations on top of a classical theory. The Quantum Hall edges we described in the first two chapters in terms of Luttinger Liquids seems to be well suited for such approaches. The free theory is indeed a collection of bosonic modes with linear dispersion which closely reminds photons. The "quantumness" of the theory is important away from the free theory when some non-linearity is added. The non-linearity we will have to deal with is the tunneling Hamiltonian for particles and quasi-particles which at the classical level gives rise to the nice and interesting dynamics we saw in the previous chapter. The density-density coupling instead is quadratic and then the classical trajectories will survive as quantum expectation values.\\ 
In this chapter we will use the Wigner representation which describes a quantum system in terms of a quasi-probability distribution over the phase space. The time evolution of the Wigner distribution is a complicated object when non-linearities are present, such as the tunneling Hamiltonian in our system. We will approximate such time evolution with a truncated Wigner approximation (TWA) that will allow us to recover a Fokker-Planck equation for the quasi-probability distribution with no diffusive term that we will solve by integrating a Langevin-like equation. In the end in order to test our approximation we will apply this new approach in situations where the tunneling in Luttinger Liquids has been solved via other methods proving the efficacy of our approach and testing its regime of validity. \\
We must also mention here that Wigner functions have been already used in the context of IQH edges. The difference is that in the IQH case the free particle basis makes the tunneling problem trivial and the Wigner function is defined differently from what we will do below (see for example \cite{EQO_wigner_interf_2017} ) using the free particle operator. Our approach instead will use the basis of the Luttinger "phonons" which is free at every $m$ but only without tunneling and use the TWA to solve the tunneling problem. The two uses are completely different as our Wigner function will be used to solve a full many-body problem while in the other case the Wigner function is just used as a tool to describe the single-particle space. 
\section{Wigner quasi-probability evolution}
The Wigner quasi-probability distribution, or Wigner function,  has been introduced in the context of quantum correction to statistical mechanics almost 90 years ago (\cite{Wigner1932} ) and has been successfully and widely used in a variety of different context (\cite{Wigner_review}). For example in the quantum optics community it has been successfully introduced for the first time 30 years ago in the description of non-linear bistable cavities \cite{Vogel1} \cite{Vogel2}. Here we will just give a brief introduction taking the angle we need for the thesis; for a more detailed and comprehensive discussion we refer to textbooks such as  \cite{QObook_WM} . \\
\newline
The general idea is that instead of working with a waveunction or the Fock space one can look at quantum expectation values as stochastic averages over some classical probability distribution over the phase space. Our phase space can be described in term of independent bosonic modes for each momentum $k$ and so via coherent states:
\begin{equation}
    \ket{\{\alpha_k\}} =  \prod_{k>0} e^{-\frac{1}{2}|\alpha_k|^2}e^{ \alpha_k \hat{a}_k^\dagger} \ket{0}
\end{equation}
We remind here the properties of coherent states:
\begin{align}
    a_k \ket{\{\alpha_k\}} = \alpha_k \ket{\{\alpha_k\}} \qquad \braket{\{\alpha_k\}|\{\beta_k\}} = \prod_k \exp\Bigl(-\frac{1}{2}(|\alpha_k|^2 + |\beta_k|^2 -2\alpha^*_k\beta_k)\Bigl)
\end{align} 
A basic ingredient for the definition of the Wigner function is the displacement operator $D(\{\xi_k\})$. With it we can define the characteristic function $\chi(\{\xi_l\})$ of a quantum state as:
\begin{equation}
    D(\{\xi_k\})= \prod_k\exp(\xi_k a_k^\dagger -\xi^* a_k) \qquad \chi(\{\xi_k\}) = Tr[\rho_M D(\{\xi_k\})]
\end{equation}
where $\rho_M$ is the density matrix of the state. Note the subscript $M$ to distinguish it from the charge density $\rho$. The Wigner function is then defined as a Fourier transform of the characteristic function:
\begin{equation}
    W(\{\alpha_k\}) = \int \prod_k \frac{d^2\xi_k}{\pi^2}\;  \chi(\{\xi_k\}) \exp\Bigl(\sum_k\alpha_k\xi_k^*-\alpha_k^*\xi_k\Bigl)
\end{equation}
The Wigner function is normalized to 1:
\begin{equation}
    \int\prod_k d^2 \alpha_k\; W(\{\alpha_k\} = 1
\end{equation}
but in general it will not satisfy the non-negativity condition of a probability distribution. 
However we can use it as a quasi-probability distribution and use it to compute averages. To show this we note that we can write any symmetrically ordered\footnote{For symmetrically ordered operator we mean for example $:a^\dagger a:_s = 1/2(a^\dagger a + a a^\dagger$)} product of ladder operators by taking the derivative of the displacement operator:
\begin{equation}
    \prod_k:(a_k^\dagger)^{n_k} a_k^{m_k} :_s = \prod_k\frac{\partial^{n_k}}{\partial \xi_k^{n_k}}\frac{\partial^{m_k}}{\partial {\xi_k^*}^{m_k}}\;\;D(\{\xi_k\})|_{\xi_k=\xi_k^*=0 \,\forall \,k}
\end{equation}
Now expectation values are computed as trace with the density matrix so:
\begin{align}
    \langle \prod_k:(a_k^\dagger)^{n_k} a_k^{m_k} :_s \rangle = Tr\Bigl[\rho_M \prod_k:{a_k^\dagger}^{n_k} a_k^{m_k} :_s\Bigl] =\\
    = Tr\Bigl[\rho_M\prod_k\frac{\partial^{n_k}}{\partial \xi_k^{n_k}}\frac{\partial^{m_k}}{\partial {\xi_k^*}^{m_k}}\;D(\{\xi_k\})|_{\xi_k=\xi_k^*=0 \,\forall \,k}\Bigl] = \\
    = \prod_k\frac{\partial^{n_k}}{\partial \xi_k^{n_k}}\frac{\partial^{m_k}}{\partial {\xi_k^*} ^{m_k}}\;\chi(\{\xi_k\})|_{\xi_k=\xi_k^*=0 \,\forall \,k} = \\
    = \int \prod_k d^2 \alpha_k\; \prod_k\alpha_k^{n_k}{\alpha_k^*}^{m_k}\; W(\{\alpha_k\})
\end{align}
which shows that the Wigner function has the interpretation of a quasi-probability.\\
We stop here to remark that the symmetric order is particularly well suited for the calculation of the quantities we are interested in. For example let's consider the expectation value of the density field written in terms of $a_k$:
\begin{equation}
    \langle\rho(x)\rangle = \sum_{k>0}' r_k \langle a_k\rangle e^{ikx} +h.c.
\end{equation}
where the primed sum $\sum'$ reminds us of the cutoff (see section \ref{sec:1_hydrodinamical}). We can compute this expectation value from the Wigner quasi-probability distribution. Then when we would like also to deal with correlations $\langle \rho(x)\rho(y)\rangle $ and indeed we have:
\begin{align}
    \langle\rho(x)\rho(y)\rangle = \langle \Bigl(\sum_{k>0}' r_k a_ke^{ikx} +h.c.\Bigl)\Bigl(\sum_{k'>0}' r_k a_{k'}e^{ik^{'}y} +h.c.\Bigl) \rangle =\\ \sum_{k,k'>0}' r_kr_{k'}\langle a_k a_{k'}\rangle e^{i(kx+k'y)}+  r_kr^*_{k'}\langle a_k a^\dagger_{k'}\rangle e^{i(kx-k'y)}+\\
    +r^*_kr_{k'}\langle a^\dagger_k a_{k'}\rangle e^{i(-kx+k'y)}+ r^*_kr^*_{k'}\langle a^\dagger_k a^\dagger_{k'}\rangle e^{-i(kx+k^{'}y)}
\end{align}
Products of two creation or two annihilation are already in symmetric order. For $k\neq k^{'}$ the ladder operators commutes so all product of two operator with $k\neq k^{'}$ can also be put in symmetric order. The remaining term is:
\begin{align}
    &\sum_{k>0}' |r_k|^2\langle a_k a^\dagger_{k}\rangle e^{ik(x-y)} + |r_k|^2\langle a^\dagger_k a_{k}\rangle e^{-ik(x-y)} = \\
    =&\sum_{k>0}' |r_k|^2 \langle : a_k a^\dagger_{k} :_s\rangle \cos(k(x-y)) +  i\sum_{k>0}'|r_k|^2 \sin(k(x-y))  =\\
    = &\sum_{k>0}' |r_k|^2 \langle : a_k a^\dagger_{k} :_s\rangle \cos(k(x-y)) +  i \langle[\rho(x),\rho(y)]\rangle
\end{align}
The first part is a symmetric ordered expectation value and the second term is the commutator of the density field at different points which disappear if we consider the symmetrized product $\rho(x)\rho(y)+\rho(y)\rho(x)$ in the correlator or if we consider $x\neq y$.\\
\subsection{Time evolution}
The question now is what to do with this formalism. In principle we can access the Wigner quasi-probability only at times where we know the density matrix $\rho_M$, however it is possible to write down a time evolution differential equation for $W$. We define a more general Wigner function where we replace the density matrix with a general operator $A$:
\begin{align}
    &\chi(\{\xi_k\};A) = Tr\Bigl[A D(\{\xi_k\})\Bigl] \\ &W(\{\alpha_k\};A) = \int \prod_k \frac{d^2\xi_k}{\pi^2}\;  \chi(\{\xi_k\};A) \exp\Bigl(\sum_k\alpha_k\xi_k^*-\alpha_k^*\xi_k\Bigl)
\end{align}
Of course this function does not necessarily have the meaning of a quasi-probability distribution when $A\neq\rho_M$. We can in any case make some formal manipulation using the following properties of the displacement operator:
\begin{align}
    \frac{\partial D(\{\xi_k\})}{\partial \xi_k} = \Bigl(-\frac{\xi^*_k}{2}+ a^\dagger\Bigl) D(\{\xi_k\}) =  D(\{\xi_k\})\Bigl(\frac{\xi^*_k}{2}+ a^\dagger\Bigl) \\
    \frac{\partial D(\{\xi_k\})}{\partial \xi^*_k} = \Bigl(\frac{\xi_k}{2}+ a^\dagger\Bigl) D(\{\xi_k\}) =  D(\{\xi_k\})\Bigl(-\frac{\xi_k}{2}+ a^\dagger \Bigl)
\end{align}
and arrive at the rules of table \ref{tab:4_wigner_rules}.
\begin{table}[h!]
    \centering
    \begin{tabular}{c|c|c }
       Operator  & Characteristic function & Quasi-probability  \\ \hline
        $A$ & $\chi(\{\xi_k\; A\})$ & $W(\{\alpha_k\};A)$ \\
        $\lambda A + \mu B$ &$\lambda  \chi(\{\xi_k\; A\}) + \mu \chi(\{\xi_k\; B\})$ & $\lambda W(\{\alpha_k\};A) + \mu W(\{\alpha_k\};B)$ \\
        $a_k A$ & $\Bigl(-\frac{\partial}{\partial \xi_k^*}-\frac{1}{2}\xi_k\Bigl)\chi(\{\xi_k\; A\})$ & $\Bigl( \alpha_k + \frac{1}{2}\frac{\partial}{\partial \alpha_k^*}\Bigl) W(\{\alpha_k\};A)$ \\
        $a_k^\dagger A$ & $\Bigl(\frac{\partial}{\partial \xi_k}-\frac{1}{2}\xi_k^*\Bigl)\chi(\{\xi_k\; A\})$ & $\Bigl( \alpha^*_k -\frac{1}{2}\frac{\partial}{\partial \alpha_k}\Bigl) W(\{\alpha_k\};A)$ \\
        $ A a_k$ & $\Bigl(-\frac{\partial}{\partial \xi_k^*}+\frac{1}{2}\xi_k\Bigl)\chi(\{\xi_k\; A\})$ & $\Bigl( \alpha_k - \frac{1}{2}\frac{\partial}{\partial \alpha_k^*}\Bigl) W(\{\alpha_k\};A)$ \\
        $A a_k^\dagger $ & $\Bigl(\frac{\partial}{\partial \xi_k}+\frac{1}{2}\xi_k^*\Bigl)\chi(\{\xi_k\; A\})$ & $\Bigl( \alpha^*_k +\frac{1}{2}\frac{\partial}{\partial \alpha_k}\Bigl) W(\{\alpha_k\};A)$ \\
    \end{tabular}
    \caption{Characteristic function and Wigner function manipulation rules.}
    \label{tab:4_wigner_rules}
\end{table}\\
These becomes useful when we want to write $W(\{\xi_k\};A\rho_M)$ in therms of $W(\{\xi_k\};\rho_M)$ where $A$ is some operator function $a$ and $a^\dagger$. Given an Hamiltonian $H$ this happens for example when we want to solve the time evolution equation $\dot{\rho}_M(t)= -i/\hbar [H,\rho_M(t)]$. Using the linearity in table \ref{tab:4_wigner_rules} the time evolution of the Wigner quasi-probability is:
\begin{align}\label{eq:4_Wigner_ev_general}
    \partial_t W(\{\alpha_k\}; \rho_M(t) ) = -i /\hbar\Bigl( W(\{\alpha_k\}; H\rho_M(t) ) -W(\{\alpha_k\}; \rho_M(t)H) \Bigl)
\end{align}
At this point we need to specify the Hamiltonian of our system which we introduced in the last chapters (see figure \ref{fig:3_geometry} and section \ref{sec:2_tunn_ham}). It is composed by two parts, the free propagation $H_0$ and the tunneling $H_T$:
\begin{equation}
    H = H_0 + H_T \qquad H_0 = \sum_{k>0}' \hbar vk \; a_k^\dagger a_k \;\;\;H_T = \Gamma \cos( \phi(l) - \phi(0) ) 
\end{equation}
where we are using the quasi-particle tunneling and passing to the particle tunneling just requires the addition of a factor $m$ inside the cosine.\\
Starting from equation \ref{eq:4_Wigner_ev_general} in general we will end up with a differential equation for $W(\{\alpha_k\};\rho_M(t))$ in the variables $\{\alpha_k\}$ with derivatives of any order. If the maximum order of the derivatives is the second then the equation would be a pseudo Fokker-Planck equation where first order derivatives correspond to the deterministic drift and the second order to the diffusion\footnote{The diffusion matrix may come up non-positive definite and then the "diffusion" would have a counterintuitive role.}. Then if the evolution for $W$ is a Fokker-Planck equation it is straightforward to sample its evolution by means of a Langevin-like stochastic differential equation. We will now see if this is true for our system under some approximation. From now on we will simplify the notation and stop to write the dependence on $\alpha_k$ every time $W(\{\alpha_k\};A) = W(A)$.
\paragraph{Free Hamiltonian}
The free Hamiltonian is quadratic and then the classical solutions for the time evolution of the charge density are the actual quantum trajectory of its expectation value. We then expect to have only a drift term in the time evolution of $W$ which correspond to the classical evolution.\\
We use again the linearity of table \ref{tab:4_wigner_rules} and we worry about one $k$ mode at a time. It is straightforward to find:
\begin{align}
    W\Bigl( [a^\dagger a ,\rho_M(t)] \Bigl)&=\Bigl[ \Bigl(\alpha^* -\frac{1}{2}\frac{\partial}{\partial \alpha}\Bigl)\Bigl(\alpha +\frac{1}{2}\frac{\partial}{\partial \alpha^*}\Bigl) + \\
    &\;\;-\Bigl(\alpha -\frac{1}{2}\frac{\partial}{\partial \alpha^*}\Bigl)\Bigl(\alpha^* +\frac{1}{2}\frac{\partial}{\partial \alpha}\Bigl)\Bigl] W\Bigl(\rho_M(t) \Bigl) =\\
    &= -\Bigl(\frac{\partial}{\partial \alpha} \alpha - \frac{\partial}{\partial \alpha^*}\alpha^*\Bigl)W\Bigl(\rho_M(t)\Bigl)
\end{align}
As expected we have only first order derivatives, the Fokker-Planck equation will then only have a drift term. Substituting what we just found for the the single mode in the equation with all modes we have:
\begin{equation}
    -\frac{i}{\hbar}W\Bigl([H_0,\rho_M(t)]\Bigl) = \Bigl[\sum_{k>0}'\frac{\partial }{\partial \alpha_k} (i v  k\;\alpha_k ) + \frac{\partial }{\partial \alpha_k^*} (-iv k\;\alpha_k^* ) \Bigl]\;W\Bigl(\rho_M(t)\Bigl)
\end{equation}
So the Fokker-Planck equation for the free propagation is:
\begin{equation}
    \frac{\partial}{\partial t}W\Bigl(\rho_M(t)\Bigl)  = \Bigl[\sum_{k>0}'\frac{\partial }{\partial \alpha_k} (iv k\;\alpha_k ) + \frac{\partial }{\partial \alpha_k^*} (-iv k\;\alpha_k^* ) \Bigl]\;W\Bigl(\rho_M(t)\Bigl)
\end{equation}
We can then write the Langevin equation for the "stochastic" motion of the $\alpha_k$ which will just be:
\begin{align}
    &\partial_t \alpha_k = -ivk \;\alpha_k\qquad\longrightarrow\qquad \alpha_k(t) = \alpha_k(0)\,e^{-iv kt} \\
    &\partial_t \alpha_k = ivk\alpha_k^* \qquad\longrightarrow\qquad \alpha^*_k(t) = \alpha^*_k(0)\,e^{iv kt} 
\end{align}
Otherwise we add up the $\alpha_k$ to obtain a stochastic charge density $\rho_s(x,t)$ defined just as the charge density but with $\alpha_k$ instead of ladder operators:
\begin{equation}
    \rho_s(x,t) = \sum_{k>0}^{'} r_k \alpha_k(0) e^{-ik(x-vt)} + h.c
\end{equation}
The time dependence give just a chiral propagation $x-vt$ of the stochastic charge density. We can also write the corresponding Langevin equation for $\rho_s(x,t)$:
\begin{equation}
    \partial_t\rho_s(x,t) = - \partial_x \rho_s(x,t) 
\end{equation}
which is exactly the classical chiral wave equation. The free Hamiltonian term is quadratic and quantum fluctuations do not modify the dynamics of the expectation values, indeed we can take the stochastic average of the above equation and since it is linear in $\rho_s$ it simply becomes:
\begin{equation}
    \partial_t\langle \rho(x,t)\rangle= -\partial_x \langle \rho(x,t) \rangle
\end{equation}
with $\langle \rho(x,t)_s \rangle_s = \langle \hat{\rho}(x,t) \rangle$
\paragraph{Tunneling Hamiltonian}
We will now treat the tunneling term which is non-linear in $a$ and $a^\dagger$. Let us consider the quasi-particle tunneling, the calculations for the particle tunneling will be the same. Since we know how to deal with product of ladder operators we expand the cosine to all orders in Taylor series so that the tunneling Hamiltonian is written as:
\begin{equation}
    H_T = \Gamma \cos(\Delta \phi) =  \Gamma \sum_{n=0}^\infty \frac{(-1)^n}{(2n)!}  \Delta \phi ^{2n}
\end{equation}
Where we call $\Delta \phi =\phi(l/2)-\phi(-l/2)$. Notice that for these calculations we will shift the constriction in order to have it symmetrical with respect to $x=0$ and make calculations clearer. Using the linearity of table \ref{tab:4_wigner_rules} we can treat each power separately. Now we write the phase difference in terms of the ladders operators as:
\begin{align}
    \phi(l/2)-\phi(-l/2) &= \sum_{k>0}' f_k a_k (e^{ikl/2}-e^{-ikl/2}) + h.c. = \\
    &=\sum_{k>0}' 2i f_k a_k e^{-ikl/2} sin(kl/2) + h.c. =\\
    &=\sum_{k>0} F_k a_k + h.c.
\end{align}
where $F_k = 2i f_k = 2 \sqrt{2\pi\nu/kL}\sin(kl/2)$ is a real quantity. Then we treat each $\Delta \phi$ of the $2n$ that compose each term consequently and we have:
\begin{align}
    W\Bigl( (\sum_{k>0} F_k a_k +h.c.)A \Bigl) = \Bigl( \sum_{k>0} F_k (\alpha_k + \frac{1}{2}\frac{\partial}{\partial \alpha_k^*} ) + F_k (\alpha_k^*-\frac{1}{2}\frac{\partial}{\partial \alpha_k} )\Bigl)W( A )\\
    W\Bigl( A (\sum_{k>0} F_k a_k +h.c.)\Bigl) = \Bigl( \sum_{k>0} F_k (\alpha_k - \frac{1}{2}\frac{\partial}{\partial \alpha_k^*} ) + F_k (\alpha_k^*+\frac{1}{2}\frac{\partial}{\partial \alpha_k} )\Bigl)W( A )
\end{align}
It it is better now to work with real and imaginary part of $\alpha_k = \alpha_k'+i\alpha_k''$ and $\frac{\partial}{\partial \alpha_k}=\frac{\partial}{\partial \alpha_k'} -i\frac{\partial}{\partial \alpha_k} $ so that the $2n^{th}$ power terms gives:
\begin{align}
    W\Bigl( (\sum_{k>0} F_k a_k +h.c.)^{2n}\rho_M \Bigl) = \Bigl( \sum_{k>0} F_k(2\alpha_k' + i \frac{\partial}{\partial \alpha_k''} ) \Bigl)^{2n}W( \rho_M )\\
    W\Bigl( \rho_M (\sum_{k>0} F_k a_k +h.c.)^{2n}\Bigl) = \Bigl( \sum_{k>0} F_k(2\alpha_k' - i \frac{\partial}{\partial \alpha_k''})\Bigl)^{2n}W( \rho_M )
\end{align}
And so the commutator for the $2n^{th}$ term is:
\begin{align}
    W\Bigl([\Delta \phi^{2n},\rho_M]\Bigl) = \Bigl[ \Bigl( \sum_{k>0}^{'} F_k (2\alpha_k' + i \frac{\partial}{\partial \alpha_k''} )\Bigl)^{2n} - \Bigl(\sum_{k>0}' F_k (2\alpha_k' - i \frac{\partial}{\partial \alpha_k''}) \Bigl)^{2n} \Bigl] W(\rho_M)
\end{align}
We then re-sum all the expansion of the cosine:
\begin{align}
   \sum_{n=0}^\infty \frac{(-1)^{2n}}{(2n)!} W\Bigl([\Delta \phi^{2n},\rho_M]\Bigl) = \Bigl[ \cos&\Bigl( \sum_{k>0}' F_k(2\alpha_k' + i \frac{\partial}{\partial \alpha_k''} )\Bigl) +\\
   -&\cos\Bigl( \sum_{k>0}' F_k(2\alpha_k' - i \frac{\partial}{\partial \alpha_k''} )\Bigl)  W(\rho_M)
\end{align}
Now note that the derivatives are only with respect to the imaginary part of the $\alpha_k$ and only the real part of $\alpha_k$ appear. This means that we can safely treat the argument of the cosine as a number, everything commutes. We use the expansion of the cosine in term of its real and imaginary part $\cos(a+ib) = \cos(a)\cosh(b) - i \sin(a)\sinh(b)$ to get:
\begin{equation}
    W\Bigl([\cos(\Delta \phi),\rho_M]\Bigl) = -2i\sin(\sum_{k>0}' 2F_k \alpha_k') \sinh( \sum_{k>0} F_k \frac{\partial}{\partial \alpha_k''} ) W(  \rho_M )
\end{equation}
The tunneling Hamiltonian then contributes to the evolution of the Wigner quasi-probability as:
\begin{align}
    -\frac{i}{\hbar}W\Bigl( &[H_T,\rho_M(t)]\Bigl) = - 2 \Gamma \sinh\Bigl( \sum_{k>0} ' F_k \frac{\partial}{\partial \alpha_k''}  \Bigl)\sin\Bigl(\sum_{k>0}' 2F_k \alpha_k'\Bigl) W(  \rho_M (t)) =\\
 &= - 2 \Gamma \sinh\Bigl( \sum_{k>0}' i\frac{F_k}{2}( \frac{\partial}{\partial \alpha_k} -  \frac{\partial}{\partial \alpha_k^{*}})  \Bigl)\sin\Bigl(\Delta \phi_s(\{\alpha_k\})\Bigl) W(  \rho_M (t)) 
\end{align}
where we define the stochastic $\Delta \phi_s$ as the operator $\Delta \phi$ with the replacement of ladder operators with the coordinates $\alpha_k$. The full time evolution equation for the Wigner function is then:
\begin{align} \label{eq:4_Wigner_ev_allder}
    \frac{\partial}{\partial t}&W(\rho_M(t))  = \Bigl[\sum_{k>0}'\frac{\partial }{\partial \alpha_k} (iv k\;\alpha_k ) + \frac{\partial }{\partial \alpha_k^*} (-iv k\;\alpha_k^* ) \Bigl]\;W(\rho_M(t)) + \\
    &- 2 \Gamma \sinh\Bigl( \sum_{k>0}' i\frac{F_k}{2}( \frac{\partial}{\partial \alpha_k} -  \frac{\partial}{\partial \alpha_k^{*}})  \Bigl)\sin\Bigl(\Delta \phi_s(\{\alpha_k\}\Bigl) W(  \rho_M (t)) 
\end{align}

The evolution equation takes derivatives at all orders with respect to the coordinates $\alpha_k$ and is then nearly impossible to solve it exactly. Note that the particle tunneling evolution can be obtained by just adding a prefactor $m$ in the hyperbolic sine and in the sine. Up to this point we have not done any approximation and this is a good point where to do one. 
\subsection{Truncation }
The hyperbolic sine is nicely linear when its argument is close to $0$. The approximation that we would like to do is:
\begin{equation}\label{eq:4_wigner_truncation}
        \sinh\Bigl( \sum_{k>0}' \frac{F_k}{2} (\frac{\partial}{\partial \alpha_k}-\frac{\partial}{\partial \alpha_k^*} ) \Bigl) \simeq  \sum_{k>0}' \frac{F_k}{2} (\frac{\partial}{\partial \alpha_k}-\frac{\partial}{\partial \alpha_k^*}) 
\end{equation}
In this way also the tunneling Hamiltonian would just contribute with a drift term to the Fokker-Planck equation:
\begin{equation}
    -\frac{i}{\hbar}W\Bigl( [H_T,\rho_M(t)]\Bigl) \simeq - \Gamma \Bigl(\sum_{k>0} iF_k( \frac{\partial}{\partial \alpha_k} -  \frac{\partial}{\partial \alpha_k^{*}})  \Bigl)\sin\Bigl(\sum_{k>0}^{'} F_k (\alpha_k+\alpha_k^*)\Bigl) W(  \rho_M (t)) 
\end{equation}
making the evolution easily solvable with a Langevin-like stochastic differential equation. This approximation is non-perturbative in $\Gamma$ and depends in principle on the function $W$. There is however a parameter we can vary for which the approximation becomes more and more accurate, namely the FQH filling fraction $1/m$. In fact each $F_k$ brings a factor $\frac{1}{\sqrt{m}}$ so that the $2n+1$ term in the hyperbolic sine expansion is reduced by a factor $1/m^n$ with respect to the linear term, then in the $m\rightarrow \infty$ limit the approximation is exact. \\
Now imagine we want to do the same approximation for the particle tunneling Hamiltonian:
\begin{equation}
    H_{T} = \Gamma \cos( m \Delta \phi )
\end{equation}
The calculations are the same but now we would have a prefactor $m$ before the argument of hyperbolic sine in equation \ref{eq:4_wigner_truncation} and hence the factor with which every term in the expansion scale is $m^n$ instead of $1/m^n$. Clearly the limit of big $m$ here work against our approximation and we must then not expect much from the truncation in the particle tunneling case.\\
Let us then focus on the quasi-particle case where we would also like to work at finite $m$ and we can try to be a bit more specific on the terms we are neglecting. An exact calculation of such terms is complicated and depend of the Wigner function, but nonetheless we can give a rough estimate based on the typical shape of $W$. In the free theory, when $\Gamma=0$, the Wigner function is a gaussian distribution on every variable $\alpha_k$:
\begin{equation}\label{eq:4_Wigner_gaussian}
    W(\{\alpha_k\};\rho_M) = \prod_{k>0} \frac{1}{\pi} \exp(- |\alpha_k|^2/ 2\sigma^2_k ) \qquad \text{with} \qquad \sigma_k^2(\beta) = \frac{1}{2}+n_k(\beta)
\end{equation}
where $n_k = 1/(e^{\beta \hbar v k }-1)$ is the occupation of the bosonic mode $k$. One can get convinced that this is the correct answer since it gives the correct expectation value for the expectation value $\langle :a_k a^\dagger_k :_s\rangle$
\begin{align}
    &\langle :a_k a_k^\dagger:_s \rangle = \frac{1}{2} ( 1+ 2 \langle a_k^\dagger a_k \rangle ) = \frac{1}{2}+ n_k(\beta) \\
   &\langle |\alpha_k|^2 \rangle_s = \sigma_k^2(\beta) = \frac{1}{2}+ n_k(\beta) 
\end{align}
If the Wigner distribution is a gaussian function then we can replace $\partial/\partial \alpha_k''\simeq \alpha_k'' / \sigma_k(\beta)$ so that what we need to be small is something like:
\begin{align}
    Q(\{\alpha_k\}) = \sum_{k>0}' \frac{F_k}{2\sigma^2_k(\beta)} \alpha_k'' 
\end{align}
Every term in the expansion of the hyperbolic sine will then be smaller with respect to the precedent by a factor $Q^2(\{\alpha_k\})/(2n+1)(2n)$. This in principle depends on the point in the phase space $\{\alpha_k\}$ where we look but what we are interested in is an estimate of the error we are doing on \textit{average} in the space of the $\{\alpha_k\}$. In the free theory there are no correlations in the phase space for different $\alpha_k$ and hence we can write our small quantity as:
\begin{align} \label{eq:4_wigner_small_quantity}
    \Tilde{Q}^2 \simeq \int \prod \frac{d\alpha_kd\alpha_k^*}{\pi^2}W(\{\alpha_k\}) Q(\{\alpha_k\})^2 \simeq \sum_{k>0} \frac{F_k^2}{4\sigma_k^2(\beta)} 
\end{align}
There is, as already mentioned, a dependence with $m$ as $\Tilde{Q}^2\propto 1/m$. Now we can also look at what the $k$ sum is doing and we start to look at zero temperature where $\sigma_k^2=1/2$:
\begin{equation}
    \Tilde{Q}^2(\beta=\infty)  = \frac{1}{m}\sum_{k>0}'4\pi\frac{ \sin^2(kl/2)}{kL} = \frac{1}{m} \int_0^{\infty} dk \,e^{-ka} \frac{\sin(kl/2)}{k} 
\end{equation}
This integral seems problematic since it diverges when the cutoff goes to zero $a\rightarrow 0$. However we know that $a$ is a finite physical scale and hence we must live with the fact that $\Tilde{Q}^2$ depends on it. For a physically acceptable $l/a \simeq 10^2$ we have:
\begin{equation}
    \Tilde{Q}^2(\beta=\infty)  \simeq \frac{1}{m} \log( l/a ) \simeq \frac{1}{m} 4.6
\end{equation}
In the hyperbolic sine expansion this means that:
\begin{equation}
    \sinh \Bigl( \sum_{k}' \frac{F_k}{2} (\frac{\partial}{\partial \alpha_k}-\frac{\partial}{\partial \alpha_k^*} ) \Bigl) \simeq O(\sqrt{\Tilde{Q}^2}) ( 1 + \underbrace{\sum_{n=1}^\infty \frac{1}{(2n+1)!} O(\Tilde{Q}^2)^{2n}}_{\text{truncation}} ) 
\end{equation}
The approximation seems problematic when we stay at $m<5$ but as we will later see still a lot physics is captured even at $m=1$. There are other two ingredients we are missing here and are the temperature and the coupling $\Gamma$. The role of the former is intuitive, it increases the fluctuations and hence the variance of the Wigner function which, as can be seen from equation \ref{eq:4_wigner_small_quantity}, goes in the direction of reducing $\Tilde{Q}^2$ and making our approximation more legitimate. The role of $\Gamma$ instead is not obvious from our treatment since it changes the Wigner function. We can try however to give a physical interpretation to what happens. The reason for which we called the small parameter $\Tilde{Q}^2$ is that indeed it has a similar expression with respect to the fluctuations of the charge $q$ in the cavity region $l$:
\begin{equation}
    \langle q^2\rangle = \sum_{k>0} F_k^2 \sigma_k{}
\end{equation}
These fluctuations are expected to be reduced as a restoring force proportional to $\Gamma$ is introduced. Then by comparing with equation $\ref{eq:4_wigner_small_quantity}$ we can say that $\Gamma$ will probably work against our approximation in a non-trivial way. In any case as we will see later the fluctuations of the charge $q$ within our approximation for strong values of $\Gamma$ will not change drastically (see section \ref{sec:4_equilibrium}). 
\subsection{Langevin equation}
We can then write an approximated time evolution equation for $W$ which only includes first derivative contrary to equation \ref{eq:4_Wigner_ev_allder} which has derivatives of all orders and is of little use. The approximated evolution equation is a Fokker-Plank equation with only drift terms:
\begin{align} \label{eq:4_Wigner_evolution_full}
    \frac{\partial}{\partial t}&W(\rho_M(t))  = \Bigl[\sum_{k}'\frac{\partial }{\partial \alpha_k} (iv k\;\alpha_k ) + \frac{\partial }{\partial \alpha_k^*} (-iv k\;\alpha_k^* ) \Bigl]\;W(\rho_M(t)) + \\
    &-  \Gamma \Bigl( \sum_k' i F_k( \frac{\partial}{\partial \alpha_k} -  \frac{\partial}{\partial \alpha_k^{*}})  \Bigl)\sin\Bigl(\Delta \phi_s(\{\alpha_k\}\Bigl) W(  \rho_M (t)) 
\end{align}
The corresponding Langevin equations for the coordinates $\alpha_k$ are:
\begin{align}
    &\partial_t \alpha_k = -i vk \, \alpha_k -i\Gamma F_k \sin(\sum_{k'>0}F_k(\alpha_{k'}+\alpha_{k'}^* ) ) \\
    &\partial_t \alpha_k^* = i vk \, \alpha_k^*+ i\Gamma F_k \sin(\sum_{k'>0}F_k(\alpha_{k'}+\alpha_k^* ) ) 
\end{align}
We can also write the Langevin equation for the stochastic charge density $\rho_s$ defined as the charge density but with the $\alpha_k$ instead of ladder operators: 
\begin{align}
    \partial_t \rho_s(x,t)&= \sum_{k>0} r_k  \partial_t \alpha_k e^{-ikx}+h.c. = \\
    &= \sum_{k>0}' r_k (-ivk) \alpha_k + h.c. \sum_{k>0}' -i \Gamma r_k F_k \sin( \Delta \phi_s ) + h.c. = \\
    &= -v \partial_x \rho_s(x,t) -\Gamma \sin( \Delta \phi_s ) \sum_{k>0} i r_k F_k e^{ikx}+h.c 
\end{align}
where $\Delta \phi_s$ is the stochastic phase difference. The sum on $k$ can be carried out and is:
\begin{align}
     \sum_{k>0} i r_k F_k e^{ikx}+h.c =&\sum_{k>0}  2\sin(kl/2) \sqrt{\frac{2\pi\nu}{ kL}}  \sqrt{\frac{ \nu k}{2\pi L}} ( i e^{ikx} - i e^{-ikx} )  =\\ 
     =&-2\nu i\frac{L}{2\pi}\frac{1}{L}\int dk \, \sin(kl/2)e^{ikx} \\
     =&  \nu (\delta(x-l/2)-\delta(x+l/2)) 
\end{align}
valid in the limit $L\gg l\gg a$. We can absorb the prefactor $\nu$ into the definition of the coupling $\Gamma$. The Langevin equation is then:
\begin{equation}
    \partial_t \rho_s(x,t)= -v \rho_s(x,t) -\Gamma \sin( \Delta \phi_s ) (\delta(x-l/2) -\delta(x+l) ) 
\end{equation}
This is exactly the classical equation of motion! It is a completely deterministic equation and the stochasticity comes only from the initial condition $\rho_s(x,t=0)$. This indeed must reproduce the Wigner function at the initial state which, in the free case $\Gamma=0$, is a simple gaussian distribution as we already mentioned in the previous section in equation \ref{eq:4_Wigner_gaussian}.\\
To solve the Langevin equation we can thus sample the initial gaussian variable $\alpha_k(t=0)$ and evolve the corresponding stochastic charge density. Note however that in order to reach the an initial state at finite $\Gamma$ we will need to wait for the relaxation to a new equilibrium state as we will discuss in section \ref{sec:4_equilibrium}. Before doing so let us give a look at the numerical integration of the Langevin equation for the $\rho_s(x,t)$ which we will use in the remaining part of the thesis. 
 \subsection{Numerical integration}
 As for the classical equation of motion we can use the chiral propagation away from the constriction and discuss the time evolution of the stochastic charge $q_s(t)$ governed by the same equation as the classical one (equation \ref{eq:3_eqmot_q_qp}) which is:
 \begin{equation}\label{eq:4_langevin_q_qp}
     \partial_t q_s(t) = v\rho_s(0^-,t) - v\rho_s(0^-,t-l/v) - \Gamma\sin( 2\pi q_s(t)) + \Gamma\sin( 2\pi q_s(t-l/v))
\end{equation}
Then we can compute the stochastic charge density using:
\begin{align}
    &\rho_s(0^+,t) = \rho(0^-,t) - \frac{\Gamma}{v} \sin(2 \pi q(t) ) \\
    &\rho_s(l^+,t) = \rho(0^-,t-l/v) + \frac{\Gamma}{v} \sin(2 \pi q(t) ) - \frac{\Gamma}{v} \sin(2 \pi q(t-l/v) )
\end{align}
and the density in all other points by adding the chiral propagation $\rho_s(x,t)= \rho_s(x-vt,0)$ away from the constriction. 
Here the initial condition $\rho(x,t=0)$ clearly appear as a stochastic noise with non-trivial time correlations which are exactly the correlations present in the $\chi$LL. The result is then a first order stochastic differential equation with memory terms and correlated noise.\\
The first step to solve it is to discretize time in small steps which we take as $dt=0.001$ and hence the unit length will be $dx=v dt$. Note that due to the memory terms when integrating the equation we will need to store at least $O(lv/dt)$ numbers. Then we also need to specify the cutoff length which we will take in general as $a=0.01/v$, note that this scale must be greater than the time discretization and that contrary to the time step the cutoff is a physical scale. Another scale to which we must pay attention is the length of the system $L = N_{L} dx$ where $N_{L}$ is the number of discrete steps. Usually $N_{L}$ sets also the total number of time steps for which we integrate the Langevin equation as after that excitations have travelled along all the sample. Then we have the coupling time scale set by $\Gamma^{-1}$, the temperature time scale set by $\hbar \beta  $ and the cavity roundtrip time $l/v$. Of all the physical time scales $ \hbar \beta $, $\Gamma^{-1}$,$l/v$ and $a/v$ the latter should always be much smaller than the others since beyond that we are in the UV region where the $\chi$LL fails. Then we need to generate the free system fluctuations $\alpha_k$ as gaussian random variables from which we can obtain the charge density $\rho(0^-,t)$ using fast Fourier transform algorithms, for a numerical cost $O(N_{L})$. The integration of a first order differential equation as equation \ref{eq:4_langevin_q_qp} is cheap and using an integration method such as the Euler method it costs O($N_{L}$). This means that the integration algorithm will be relatively fast.

\section{Equilibrium}\label{sec:4_equilibrium}
\begin{figure}[h!]
    \centering
    \includegraphics[width=9cm]{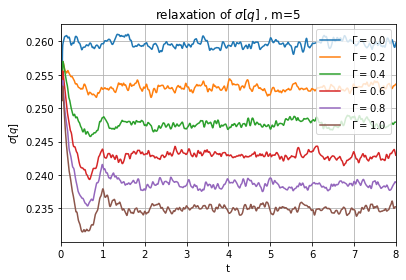}
    \caption{Relaxation of the cavity charge fluctuations $\sqrt{\langle q^2(t)\rangle}$ in the zero temperature limit after the tunneling quench to the weak coupling regime for an $m=5$ system. The relaxations occurs within a cople of roundtrip times $l/v$. $l=1$, $v=1$ and $t_0=0.01$}
    \label{fig:4_relaxation_sigmaq_wt}
\end{figure}

\begin{figure}[h!]
    \centering
    \includegraphics[width=9cm]{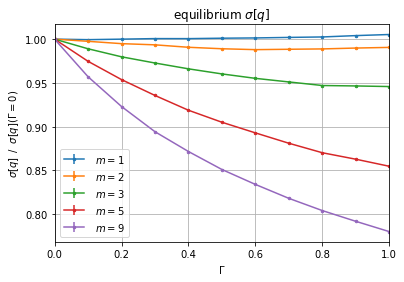}
    \caption{Equilibrium asymptotic value of the cavity charge fluctuations $\sqrt{\langle q^2(t)\rangle}$ relative to the $\Gamma=0$ case. Increasing $\Gamma$ in the regime of weak coupling $\Gamma$ reduces the fluctuations, more efficiently for higher $m$ where the fluctuations at $\Gamma=0$ are already smaller (not shown in figure). The figure refers to the zero temperature limit with $l=1$ and $v=1$ }
    \label{fig:4_equilibrium_sigmaq_wt}
\end{figure}

\begin{figure}[h!]
    \centering
    \includegraphics[width=10cm]{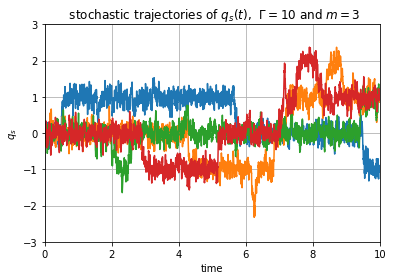}
    \caption{Stochastic trajectories of $q_s(t)$ in a system with $\Gamma = 5$ and $m=3$. $v=1$ and $l=1$}
    \label{fig:4_trajectories_q_st}
\end{figure}

\begin{figure}[h!]
    \centering
    \includegraphics[width=10cm]{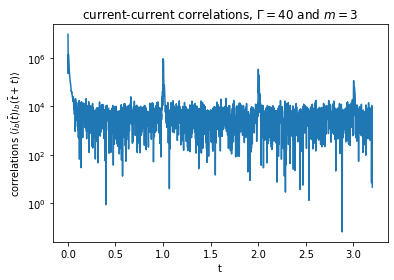}
    \caption{Equilibrium time correlations in the backscattered current $|\langle i_b(\Bar{t})i_b(\Bar{t}+t)\rangle|$ for $m=3$ and $\Gamma=40$. Peaks at multiples of the roundtrip time $l/v=1$ are clearly visible over the noisy background which hides the power law correlation of the free Luttinger liquid $-1/t^2$}
    \label{fig:4_equilibrium_correlation_l}
\end{figure}

\begin{figure}[h!]
    \centering
    \includegraphics[width=10cm]{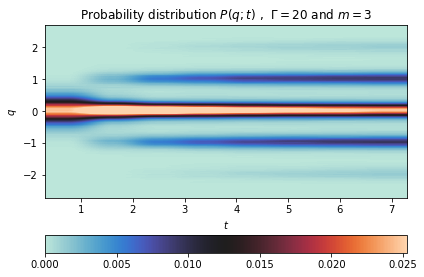}
    \caption{Time evolution of the probability distribution $P(q;t)$ in the zero temperature limit for $m=3$ and $\Gamma = 20$. The population around side minima of the cosine potential grows in time while the width of the peaks fast reaches an equilibrium value. $l=1$, $v=1$ and $t_0=2 $}
    \label{fig:4_relaxation_Pq_st}
\end{figure}

\begin{figure}[h!]
\begin{subfigure}{.5\textwidth}
  \centering
  \includegraphics[width=7cm]{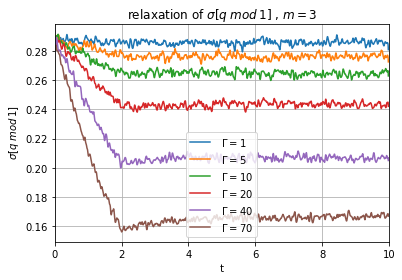}
  \caption{ $m=3$}
  \label{fig:4_relaxation_sigmaqd_st_fixm}
\end{subfigure}
\begin{subfigure}{.5\textwidth}
  \centering
  \includegraphics[width=7cm]{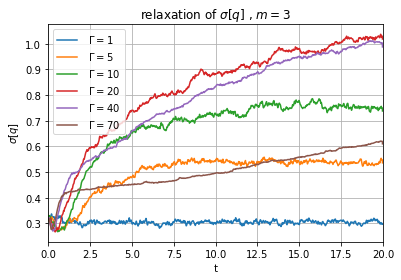}
  \caption{ $m=3$}
  \label{fig:4_relaxation_sigmaq_st_fixm}
\end{subfigure}
\newline
\newline
\begin{subfigure}{.5\textwidth}
  \centering
  \includegraphics[width=7cm]{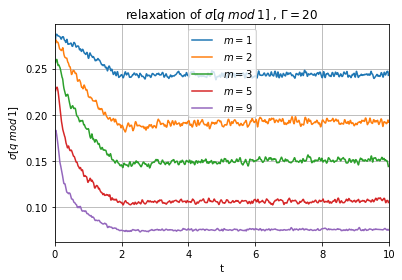}
  \caption{ $\Gamma=20$}
  \label{fig:4_relaxation_sigmaqd_st_fixGamma}
\end{subfigure}
\begin{subfigure}{.5\textwidth}
  \centering
  \includegraphics[width=7cm]{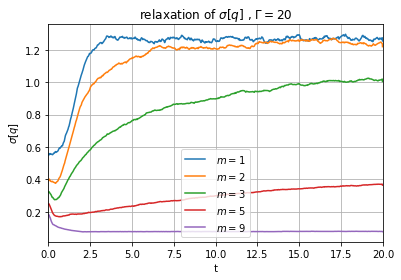}
  \caption{ $\Gamma=20$}
  \label{fig:4_relaxation_sigmaq_st_fixGamma}
\end{subfigure}
    \caption{Relaxation of the cavity charge fluctuations $\sqrt{\langle q^2(t)\rangle}$ (right) and of the modulus of cavity charge fluctuations around the cosine potential minima $\sqrt{\langle (q(t) \;\mod \,1)^2\rangle}$ (left) in the zero temperature limit after the tunneling quench to the strong coupling regime. $l=1$, $v=1$ and $t_0=2 \,l/v$ }
    \label{fig:4_relaxation_st}
\end{figure}

\begin{figure}[!h]
    \centering
    \includegraphics[width=10cm]{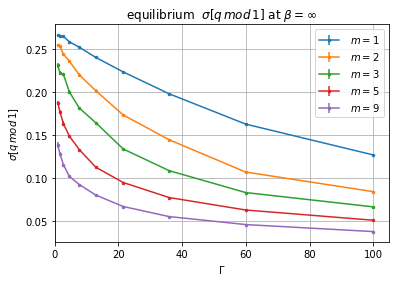}
    \caption{Equilibrium asymptotic fluctuations around the cosine potential minima $\sqrt{\langle (q \;\mod \,1)^2(t)\rangle}$ as a function of $\Gamma$ for different $m$. }
    \label{fig:4_equilibrium_sigmaqd_st}
\end{figure}

In this section we will look at the equilibrium properties of our system at $\Gamma >0$. This is of course a preliminary step for what will come later but it is important also to test our approximation. As we anticipated above we can only start from $\Gamma =0$ where the system is free and we know its quantum state. We will use a ramp protocol for the switching on of the tunneling: 
\begin{equation}\label{eq:4_ramp_equilibrium}
       \Gamma(t) = \begin{cases} 0 \qquad\qquad \text{for}\;\;t<0\\
       \Gamma\; t/t_0\qquad \text{for}\;\;0\leq t< t_0 \\
       \Gamma \qquad \qquad\text{for}\;\; t\geq t_t0 \end{cases}
\end{equation}
where $t_0$ is the switching time. \\
In this section we will usually set $l=1$ and $v=1$. The Langevin equation \ref{eq:4_langevin_q_qp} has memory terms and we should add the time dependence also to the $\Gamma$ so that we do not need $q$ before the ramp because it does not enter the equation. We instead need $\rho(0^-,t)$ also at times $-l/v<t<0$. To check whether the system has reached a new equilibrium we should look at quantities such as the variance of $q$ and at its distribution.\\
In order to have a clearer interpretation we first work in the zero temperature limit $\hbar \beta \gg l/v, \Gamma^{-1}$. We can then distinguish two regimes of weak and strong coupling by comparing $\Gamma^{-1}$ and $l/v$. In the weak tunneling regime $\Gamma l/v<1$ we expect the new equilibrium to be only perturbatively distant from the free state. On the contrary the strong tunneling regime $\Gamma l/v>1 $ will probably show a different equilibrium state.
\paragraph{Weak coupling $\Gamma l /v < 1$}
In figure \ref{fig:4_relaxation_sigmaq_wt} we show the fluctuations of the charge $q_s$ as a function of time for different couplings $\Gamma$ in the weak coupling regime.
\\
The relaxation is fast and within a time $l/v$ this observable has reached the equilibrium. In general we see that at weak coupling the effect of $\Gamma$ is to shrink the distribution of $q$ around $q=0$ reducing its quantum fluctuations around it. In figure \ref{fig:4_equilibrium_sigmaq_wt} we plot the new equilibrium values $\sqrt{\langle q^2 \rangle}$ as a function of the coupling for different values of $m$. 
\\
In the weak coupling regime the corrections to the equilibrium distribution $P(q_s)$ are small and the distribution essentially remain a gaussian with a slightly reduced width.
\paragraph{Strong coupling $\Gamma l /v < 1$}
The strong tunneling regime shows a peculiar new behaviour. In this regime the coupling is strong enough that the minima of cosine potential are able to trap the stochastic trajectories of $q_s$, we show some examples in figure \ref{fig:4_trajectories_q_st}.
On top of this we must add a random tunneling to neighboring minima. This jumps of $\Delta q_s = \pm 1$ will generate excitations of charge $\pm1$ in the outgoing current at $x=l^+$ and the conjugate excitation entering the cavity at $x=0^+$ which after a time $l/v$ will hit the constriction again. At this point the charge $q$ can jump back to its original value with the excitation escaping the cavity or remain at the new minimum with the excitation remaining inside the cavity. Of course this process may overlap to other processes of the same kind and the correlations between these is non-trivial. In any case there will be a clear signature in the time correlations of the outgoing charge density $\langle\rho(l^+,\Bar{t})\rho(l^+,\Bar{t}+t)\rangle$ as shown in figure \ref{fig:4_equilibrium_correlation_l}. On top of the free Luttinger correlations $-1/t^2$ we now also have a strong anticorrelation at integer multiples of the roundtrip time $t=n l/v$.
\\
There are then two physical quantities that need to reach an equilibrium, the distribution among neighboring minima which reflects on the fluctuations of the cavity charge $\langle q^2 \rangle$ and the fluctuations within the minima of the potential which can be kept under control by looking at $\langle ( q\; \mod \,1 )^2 \rangle$. Note that the latter is actually what matters for the outgoing current where the charge $q$ enters in a periodic function as $\sin(2\pi q)$. However working in a finite window means that even with a uniformly spread distribution the variance is finite and is $\sigma_{ma}[q\, mod\, 1] = \sqrt{1/12}\simeq0.29$. To show these two processes in figure \ref{fig:4_relaxation_Pq_st} we plot how the stochastic probability distribution $P(q)$ evolve in time, as you can see the width of every peak relaxes almost immediately to a constant value while the population around side minima (e.g $q=\pm1$) grows in time. 
\\
In figure \ref{fig:4_relaxation_st} we show the behaviour in time of fluctuations around the minima and the total variance of $q$. 
\\
The former relaxes quite fast while the latter takes longer times sometimes non accessible to the simulation. For example in the top right panel at the highest shown values of $\Gamma$ for $m=3$ the fluctuations $\sigma[q]$ drastically slow their relaxation time (brown and purple lines). Longer simulations confirm that the asymptotic value increase with $\Gamma$.  In figure \ref{fig:4_equilibrium_sigmaqd_st} we show the dependence of the fluctuations around the minima equilibrium quantities for different values of $m$ as a function of $\Gamma$. \\
\subsection{Temperature}
\begin{figure}[!h]
\begin{subfigure}{.5\textwidth}
   \centering
    \includegraphics[width=7cm]{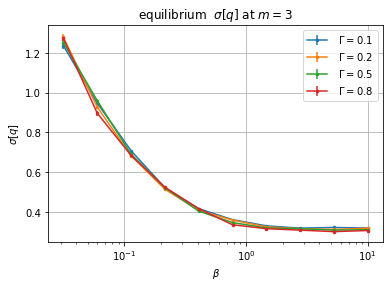}
    \caption{Weak tunneling at $m=3$}
    \label{fig:4_equilibrium_sigmaq_wt_T_m3}
\end{subfigure}
\begin{subfigure}{.5\textwidth}
  \centering
    \includegraphics[width=7cm]{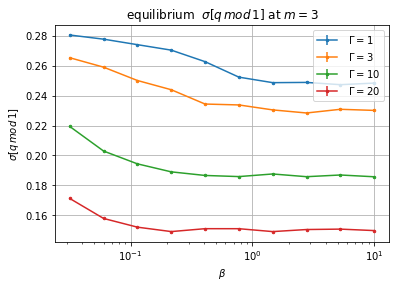}
    \caption{Strong tunneling at $m=3$}
    \label{fig:4_equilibrium_sigmaqd_st_T_m3}
\end{subfigure}
    \caption{Finite temperature dependence of the equilibrium asymptotic fluctuations at $m=3$. For weak tunneling (left) the plot shows $\sqrt{\langle q ^2\rangle_{\beta}}$  while for strong tunneling (right) the plot shows $\sqrt{\langle (q \;\mod \,1)^2\rangle_{\beta}}$. In both cases the zero temperature regime is achieved as $\hbar \beta \gtrsim l/v=1$. $t_0=1$, $v=1$ and $l=1$}
    \label{fig:4_equilibrium_T_m3}
\end{figure}

The addition of the temperature does not change too much the picture we have in mind, in the stochastic problem is enters just as an increase of the free fluctuations of the $\alpha_k$. There are other two scales to which it must be compared, the cavity length $l$ and the coupling $\Gamma$, in particular we will recover the limit of zero temperature whenever we have $\hbar \beta \gg \Gamma^{-1}$ or $\hbar \beta \gg l/v$. In the weak $\Gamma$ regime we again have a slight modification of the equilibrium distribution, as shown in figure  \ref{fig:4_equilibrium_sigmaq_wt_T_m3} .
\\
Also in the strong coupling regime where the minima of the cosine potential are able to trap the stochastic trajectories of $q$ the qualitative behaviour is similar. The quantitative change is in the size of the fluctuations and as we already argued it will be seen only when $\hbar \beta \ll \Gamma^{-1}$. In figure \ref{fig:4_equilibrium_sigmaqd_st_T_m3} for example we show the behaviour of the equilibrium fluctuations of $q$ around its minima for different $\Gamma$ at $m=3$ as a function of $\beta$ relative to the zero temperature limit $\beta=\infty$. Notice that for great values of $\Gamma$ the change in the fluctuations is less evident as the coupling time scale dominates the temperature time scale. 
\begin{figure}[!h]
    
\end{figure}
\section{DC transport properties}
\begin{figure}[!h]
    \centering
    \includegraphics[width=12cm]{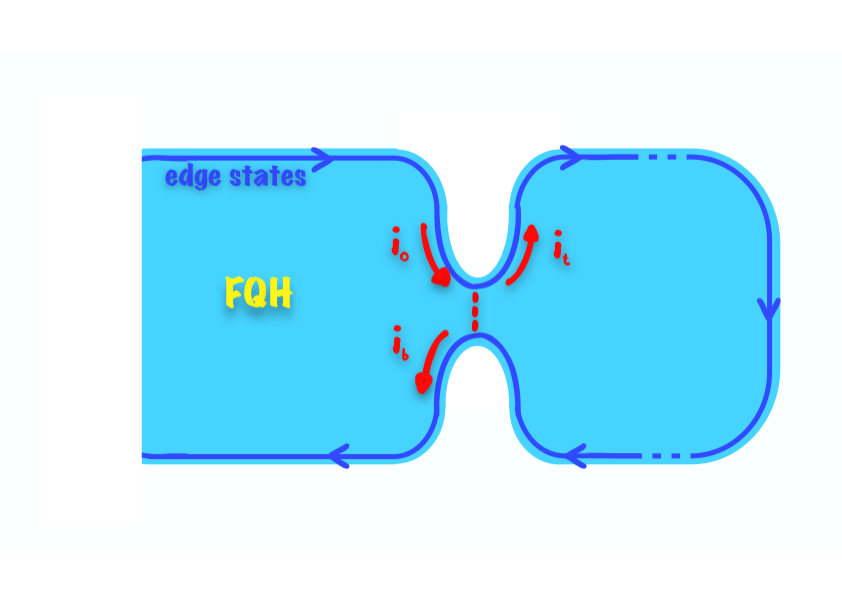}
    \caption{Configuration for the calculation of DC properties within our approximation. Dashed lines for the edge states means that the cavity is long enough that the current $i_t$ does not have the time to perform a full roundtrip.}
    \label{fig:4_geometry_ss}
\end{figure}

In this section we will go out of equilibrium and we will perturb our system with a constant current bias $i_0$ incident on the constriction. This can be easily added to our evolution equation by changing the mean values of the $\alpha_k$ in the initial distribution or in real space to add a constant to $\rho_s(0^-,t)$ so that:
\begin{equation}
    \rho_s(0^-,t) =i_0 f_{t_0}(t-t_i) + \sum_k' r_k \alpha_k e^{ikx}+h.c.
\end{equation}
where $f_{t_0}(t-t_i)$ is a ramp function as that of equation \ref{eq:4_ramp_equilibrium} with $t_i$ the time where the ramp starts. We will usually consider $t_0=2$ both for the coupling ramp and for the incident current $i_0$ with the latter delayed by a $t_i=2$ which in the equilibrium calculations we saw was enough for the fluctuations around each minima to reach the $i_0=0$ equilibrium.\\
In order to look at the steady state properties of the constriction we will look at very large $l$ so that a steady state is reached before the injected current has made its way to other point of the constriction, hence time will always be less than the roundtrip time $t<l/v$. A schematic picture of the configuration in shown in figure \ref{fig:4_geometry_ss}.
The observable we will mainly look at is the backscattered current $i_b$:
\begin{equation}
    i_b(t) = \langle \rho(l^+,t) \rangle =\underbrace{\langle \rho(l^-,t) \rangle}_{=0} + \langle \Gamma \sin(2\pi q(t) ) \rangle 
\end{equation}
where the time dependence will disappear in the steady state. Note also that we can avoid to average over the current at $x=l^-$ since we stay at times $t<l$ and the average charge density in the vacuum is $0$. \\
Remember that due to chirality the backscattering is happening in the only possible channel at the other point in the constriction. Because of charge conservation we then have also the transmitted current $i_t= i_0 - i_b$ as:
\begin{equation}
    i_t = \langle \rho(0^+) \rangle = \langle \rho(0^-) \rangle - \langle \Gamma \sin(2\pi q ) \rangle = i_0 - i_b
\end{equation}
We also argue that the relaxation to a steady state for $i_b$ is fast in all regimes since it is a function of $\sin(2\pi \,q)$ that as we saw in the precedent section relaxes almost immediately.\\

\begin{figure}
\begin{subfigure}{.5\textwidth}
    \centering
  \includegraphics[width=7.5cm]{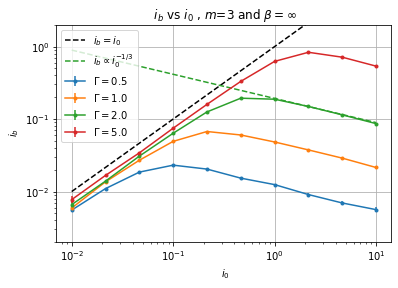}
    \caption{$\beta = \infty$ }
  \label{fig:4_ibvsi0_beta}
\end{subfigure}
\begin{subfigure}{.5\textwidth}
  \centering
  \includegraphics[width=7.5cm]{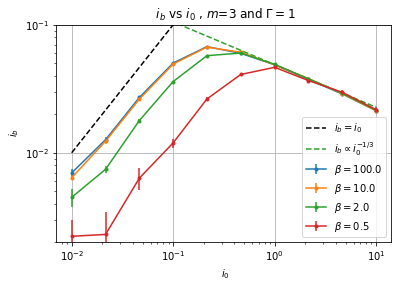}
  \caption{$\Gamma=1$}
  \label{fig:4_ibvsi0_Gamma}
\end{subfigure}
    \caption{Typical behaviour of the backscattered current $i_b$ against the incident current $i_0$ for a $m=3$ system at zero temperature for different coupling values (left) and at a fixed coupling for different temperatures (right). Dashed lines guide the interpretation of the figure, in low $i_0$ region the dependence is linear and in the high $i_0$ region a power law is present. The crossover scale depend on $\Gamma$ and $\beta$.}
    \label{fig:4_ibvsi0_typical}
\end{figure}

The interest in these observables comes from the fact that we can measure the Hall conductivity as:
\begin{equation}
    \sigma_{H} = \frac{i_t}{V_0} = \frac{1}{m}\frac{e^2}{2\pi\hbar}\;\; \frac{i_t}{i_0}
\end{equation}
where in the response to a voltage difference $V_0$ between the channel at $x=-\infty$ and $x=+\infty$ generates as a response the $i_0$ current. 
We can use our machinery of the TWA to calculate the response $i_b$ in all possible regimes varying the physical scales $\Gamma$, $\beta$, $i_0$ and the Laughlin parameter $m$ to explore different behaviours.
The typical behaviour  at zero temperature for different $\Gamma$ is shown in figure \ref{fig:4_ibvsi0_beta} while in figure \ref{fig:4_ibvsi0_Gamma}  we show the behaviour for a fixed value of $\Gamma$ and different temperature. 
It is important to have a physical idea of what is happening to the stochastic trajectories when we vary the various parameters. The incident current $i_0$ is injecting charge in the cavity, in particular if $\Gamma=0$ we would have $q=i_0 t$. Then also the tunneling term is injecting charge in the cavity for a maximum amount of $\Gamma t$. This means that when $i_0$ is greater than $\Gamma$ the charge $q$ will increase in time with no hope for the tunneling current to follow the incident current. On the other hand when $i_0$ is less than $\Gamma$ the tunneling term can be strong enough to keep $q$ in the minima of the cosine potential. To this picture we must of course add the fluctuations, for example also at $i_0\ll \Gamma$ where the constriction is in principle strong enough to keep $q$ at the cosine potential minima we may still have some current which passes through the constriction giving a finite $i_t$ and hence $i_b<i_0$. Intuitively as the fluctuations and hence temperature increase $i_t$ should increase and $i_b$ decrease, as shown at low $i_0$ in the right panel of figure \ref{fig:4_ibvsi0_typical}. In the other limit $i_0 \gg \Gamma$ instead the incident current is strong enough to pass through the constriction and only a small portion is backscattered, $i_b \ll i_0$ as can be seen in the big $i_0$ part of figure \ref{fig:4_ibvsi0_typical}. The fluctuations will clearly play a non-trivial role and we remark that the Luttinger parameter $1/m$ enters only in the size of the fluctuations. \\
 \subsection{Conductivity}
\begin{figure}

\begin{subfigure}{.5\textwidth}
    \centering
  \includegraphics[width=\textwidth]{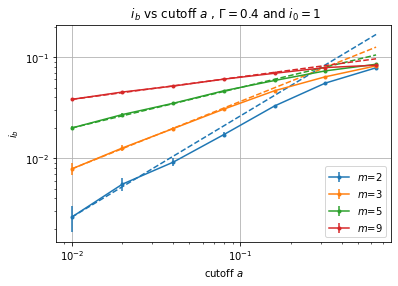}
    \caption{Cutoff $a$ dependence. $i_0 = 1$ and $\Gamma=0.4$ at $\beta = \infty$. The dashed lines are power laws $i_b\propto a^{2/m}$ }
    \label{fig:4_ibvsa}
\end{subfigure}
\begin{subfigure}{.5\textwidth}
  \centering
  \includegraphics[width=\textwidth]{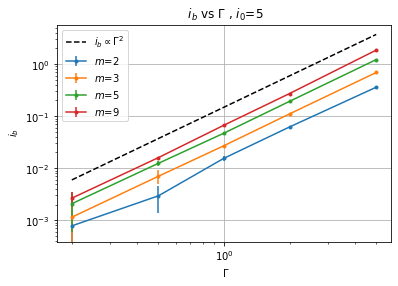}
  \caption{Coupling $\Gamma$ dependence. $i_0 =10$ and cutoff $a=0.01$}
  \label{fig:4_ibvsGamma}
\end{subfigure}
\newline
\newline
\begin{subfigure}{\textwidth}
  \centering
  \includegraphics[width=9cm]{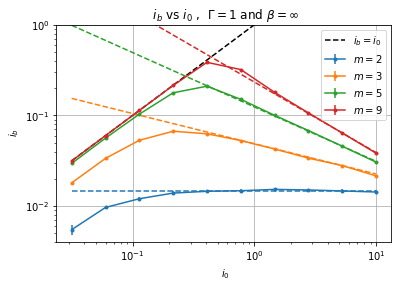}
  \caption{Incident current $i_0$ dependence. $\Gamma =1$ and cutoff $a=0.01$. The dashed lines are power laws $i_b\propto i_0^{-(m-2)/m}$.}
  \label{fig:4_ibvsi0_powerlaw}
\end{subfigure}
    \caption{Backscattered current dependence on various quantities at zero temperature $\beta=\infty$. All of them agrees with the power laws of perturbation theory }
    \label{fig:4_dc_wt}
\end{figure}

\begin{figure}[h!]
    \centering
    \includegraphics[width=10cm]{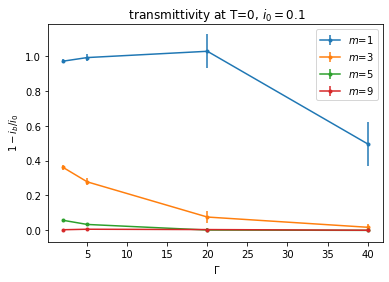}
    \caption{Transmittivity for $i_0\rightarrow 0$ at zero temperature. For FQH $m>1$ quasi-particle tunneling the transmittivity should be zero. Evidently at low values of $m$ the TWA erronously predict a finite value}
    \label{fig:4_transmittivity_T0}
\end{figure}

\begin{figure}[h!]
    \centering
    \includegraphics[width=10cm]{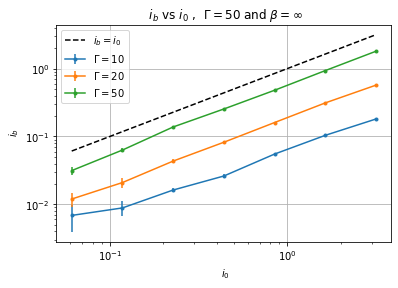}
    \caption{Backscattered current at $m=1$ and $\beta=\infty$ for different $\Gamma$. The backscattered current varies linearly with the incident current as one expect from an Ohmic picture where the carriers are free. The dashed line indiccate $i_b=i_0$}
    \label{fig:4_ibvsi0_IQH}
\end{figure}

\begin{figure}[h!]
    \centering
    \includegraphics[width=10cm]{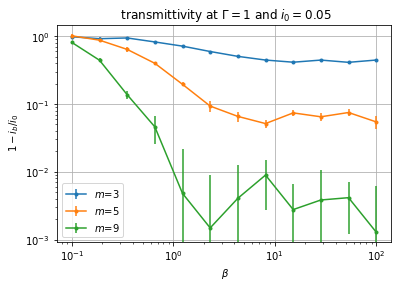}
    \caption{Transmittivity for $\Gamma=10$ as a function of the temperature. In the low temperature (high $\beta$) regime the transmittivity should go to zero but it instead goes to a finite constant (error of the TWA). At high temperatures (low $\beta$) instead the transmittivity correctly approach 1. }
    \label{fig:4_transmittivity_T}
\end{figure}

As the  first application we compute the small temperature behaviour $\beta^{-1}\ll \Gamma ,i_0$ of the conductivity or equivalently of the backscattered current $i_b$. From what we saw in chapter \ref{chap:2} in the weak tunneling regime $i_b\ll i_0$ we should have a clear power law dependence of $i_b$ to $i_0$, $\Gamma$ and the cutoff $a$ as:
\begin{equation}
    i_b \propto \Gamma^2 i_0^{-(m-2)/m)} a^{2/m}
\end{equation}
In figure \ref{fig:4_ibvsi0_powerlaw} we show the different dependence on $i_0$ for different values of $m$ and how the power law in the regime $i_b\ll i_0$ at big $i_0$ matches with the expected $-(m-2)/m$. Again for the limit of $i_b\ll i_0$ in figure \ref{fig:4_ibvsGamma} we show the quadratic dependence on $\Gamma$ and in figure \ref{fig:4_ibvsa} the power law dependence on the cutoff $a^{2/m}$.
These results matches exactly the perturbation theory in the tunneling Hamiltonian we discussed in section \ref{sec:2_perturbation_theory}. Although remember that we can go beyond perturbative results in $\Gamma$ and look at the other regime where $i_b$ is comparable to $i_0$. This is achieved for example by taking $i_0\ll \Gamma $ and it gives a linear behaviour for the dependence of $i_b$ to $i_0$ or equivalently a constant transmittivity $t=1-i_b/i_0$. This is shown in figure \ref{fig:4_transmittivity_T0} and depends both on $m$ and on $\Gamma$.
For high $m$ where the noise is lower there is a tendency to completely lock the phase $\Delta \phi$ and have a small transmitted current, the constriction becomes infinitely strong. On the other hand for lower values of $m$ the fluctuations are stronger and having a completely locked phase $\Delta \phi$ requires very strong $\Gamma$.\\
However as we discussed in section \ref{sec:2_exact} the zero bias conductance at zero temperature should be zero for a $\chi$LL with quasi-particle tunneling at $m>1$. This is not the case, in figure \ref{fig:4_transmittivity_T0} we clearly see a finite transmittivity for even at $m>1$. The most straightforward explanation is that the TWA fails to correctly predict the zero bias conductance of the zero temperature limit, in support of the approximation we however notice that the error (finite transmittivity) decreases as $m$ increases in figure \ref{fig:4_transmittivity_T0}. \\
The behaviour at $m=1$ is shown in figure \ref{fig:4_ibvsi0_IQH} . Here we see that no power law are present in the $i_b(i_0)$ characteristics and we find a simple linear behaviour which do not depend on the temperature and hence agrees with the free particle picture at $m=1$. 
\\
\paragraph{Temperature dependence}
We now put temperature into the game and for high enough temperature (small $\beta$) it is possible to observe a dependence on $\beta$ in the $i_b$ vs $i_0$ characteristics as shown in figure \ref{fig:4_ibvsi0_Gamma} for $m>1$.
The temperature is reducing the amount of backscattered current by increasing the fluctuations and making the tunneling less effective. This picture of holds when $(\hbar\beta )^{-1}$ is the dominant scale both with respect to the current $i_0$ and to the coupling $\Gamma$ otherwise we fall back on a low temperature regime as it happens for high values of $i_0$ in figure \ref{fig:4_ibvsi0_Gamma}. We can give a look at the opposite limit of high temperatures at small $i_0$ where the temperature gives still a linear behaviour $i_b \propto i_0$ but with a different coefficient and hence a different transmittivity with respect to the zero temperature case we analysed in figure \ref{fig:4_transmittivity_T0}. In figure \ref{fig:4_transmittivity_T} we show the transmittivity as a function of the inverse temperature for different values of $m$ and fixed $\Gamma$. At small temperatures (high $\beta$) the transmittivity should be zero but as we see pointed out in the zero temperature case the TWA capture this only for big $m$. At high temperatures the strength of the QPC is effectively reduced and indeed we observe the transmittivity tending to 1 at low $\beta$.
\\

\begin{figure}[!h]
    \centering
    \includegraphics[width=10cm]{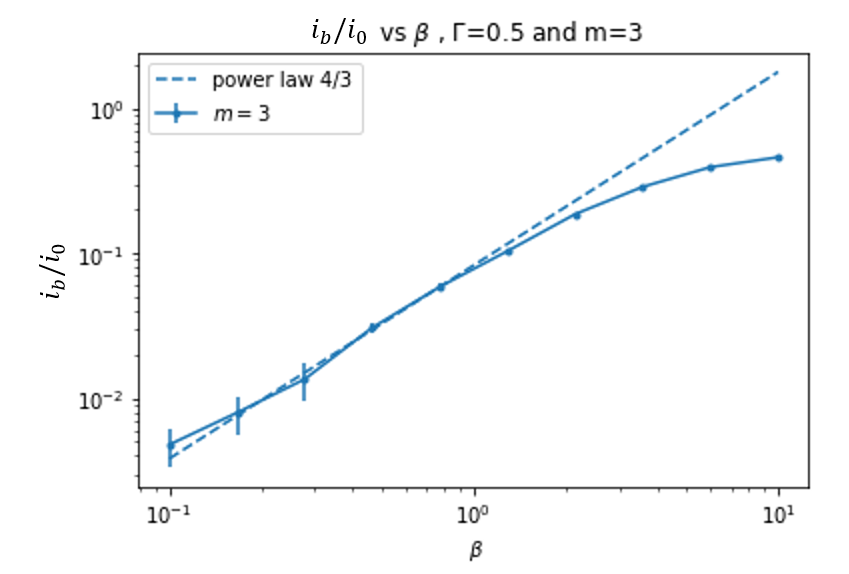}
    \caption{Correction to the zero bias transmittivity $1-t=i_b/i_0$ measured for a small enough bias $i_0=0.01$ in the linear regime as a function of the inverse temperature $\beta$ at $m=3$ and $\Gamma =0.5$. The dashed line shows the perturbation theory power law scaling $\beta^{2-2/m}$ discussed in section \ref{sec:2_perturbation_theory}. }
    \label{fig:4_Tscaling}
\end{figure}

In general the transmittivity of the quasi-particle tunneling is known to have particular temperature scalings that depend on $m$ (see section \ref{sec:2_perturbation_theory} ). As an example of that we show the high temperature (low $\beta$) scaling $\beta^{2-2/m}$ for the correction to the quantized zero bias conductance at $m=3$ in figure \ref{fig:4_Tscaling}.
\\
This results holds in the regime of small backscattering where $i_b\ll i_0$ so that at high $\beta$ when $i_b/i_0$ is comparable to one the power law stops to be valid. \\
\newline
In this subsection we saw how the TWA well predicts all the non-trivial power laws in the limit $i_b\ll i_0$, even at $m=1$ where it recovers an ohmic behaviour $i_b \propto i_0$. The problems of in the regime $i_b\simeq i_0$ have been highlighted and the investigation of such regime within the TWA will be object of future studies.   
\subsection{Shot noise}
\begin{figure}
\begin{subfigure}{.5\textwidth}
      \centering
    \includegraphics[width=7cm]{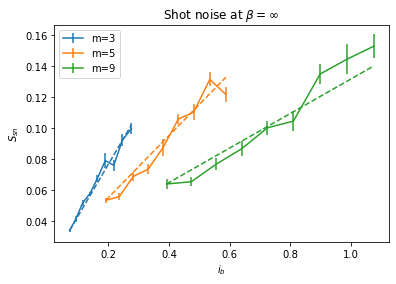}
    \caption{$i_b\ll i_0$}
    \label{fig:4_shotnoise_qp}
\end{subfigure}
\begin{subfigure}{.5\textwidth}
      \centering
    \includegraphics[width=7cm]{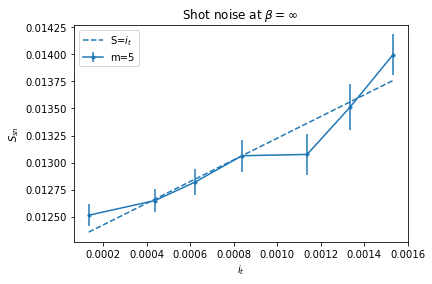}
    \caption{$i_t\ll i_0$}
    \label{fig:4_shotnoise_p}
\end{subfigure}
\caption{Shot noise in the limits $i_b\ll i_0$ (left) and $i_t\ll i_0$ (right). Dashed lines are the expected behaviour for shot noise of fractional charges $S = e/m \;i_b$ (left) and integer charges $S = e\; i_t$ (right).}
\end{figure}

Another important quantity that we can compute with this setting is the shot noise, a key quantity that has we discussed in section \ref{sec:2_shotnoise} allowed the measurement of the fractional charge of FQH excitations. This is the first quantity we compute with our TWA that involve expectation values of product of two operators. The shot noise in the backscattered channel is defined as the zero frequency component of the fluctuations in the backscattered current:
\begin{equation}
    S_{sn} = \frac{1}{T}\int_{0}^{T} dt dt' \langle \rho(l^+,t)\rho(l^+,t') \rangle - \langle \rho(l^+,t)\rangle\langle\rho(l^+,t') \rangle 
\end{equation}
with the limit $T\rightarrow\infty$ and in a time window $[0,T]$ where we are in the steady state. Instead of calculating the correlations for every point $(t,t')$ we swap integration and average to get:
\begin{equation}\label{eq:4_sn_Q}
    S_{sn} = \frac{1}{T} \langle Q_{b,T}^2 -\langle Q_{b,T}\rangle^2\rangle \qquad \qquad\text{with}\qquad Q_{b,T} = \int_0^T dt \rho(l^+,t) 
\end{equation}
Note that in the steady we have:
\begin{equation}
    \langle Q_{b,T}\rangle = i_b\;T
\end{equation}
The shot noise indeed measures the fluctuations over the mean value of the current. In order to compute equation \ref{eq:4_sn_Q} it is better to explicitly write the two term in the backscattered density and get:
\begin{align} \label{eq:4_Sn_totexact}
        S_{sn} &=   \frac{1}{T}\langle \Bigl(\int_{0}^{T} dt \rho(l^+,t)\Bigl)^2 \rangle -\frac{1}{T}\langle Q_{b,T}\rangle^2 = \\
        &= \frac{1}{T}\langle \Bigl(\int_{0}^{T} dt \rho(l^-,t)\Bigl)^2 \rangle +  \Gamma^2 \langle \Bigl( \int_0^T dt \sin(2 \pi q(t)) \Bigl)^2 \rangle + \\
        & \;\;\;+ 2\Gamma \langle \Bigl(\int_{0}^{T} dt \rho(l^-,t)\Bigl)\Bigl(\int_{0}^{T} dt'\sin(2\pi q(t')) \Bigl) \rangle -T i_b^2
\end{align}
The term first term involving only $\rho(l^-,t)$ can be calculated exactly in the free theory with discretized time $t=n_t\,dt$ :
\begin{align}
    \langle \Bigl(\int_{0}^{T} dt \rho(l^-,t)\Bigl)^2 \rangle = \int_{0}^{T} dt dt' \sum_k' |r_k|^2 \langle |\alpha_k|^2\rangle ( e^{ik(t-t')}+e^{-ik(t-t')} ) = \\
    = \sum_{k>0}' |r_k|^2(\frac{1}{2}+n_k(\beta) )2 Re \Bigl(\sum_{n_t'=0}^{n_T}\sum_{n_t=0}^{n_T} (e^{ik \,dt})^{n_t} (e^{-ik \,dt})^{n_t'}\Bigl) =\\
    =  \frac{2}{m}\sum_{k>0}' \frac{k}{2\pi L}(\frac{1}{2}+\frac{1}{e^{\beta \hbar v k}-1} )\Bigl| \sum_{n_t=0}^{n_T} (e^{ik \,dt})^{n_t}\Bigl|^2  =\\
    =  \frac{2}{m}\sum_{k>0}'\frac{k}{2\pi L}(\frac{1}{2}+\frac{1}{e^{\beta \hbar v k}-1} )\frac{| 1-e^{ik \,dt\,(N_t+1)}|^2}{|1-e^{ik\,dt}|^2}  =\\
    = \frac{2}{m}\sum_{k>0} e^{-ka} \frac{k}{2\pi L}(\frac{1}{2}+\frac{1}{e^{\beta \hbar v k}-1}) \frac{\sin^2(k(T+dt)/2)}{\sin(k\,dt/2)}
\end{align}
This sum must be carried out numerically for a general $\beta$ but does not depend on $\Gamma$ and $i_0$. Indeed this term represent the equilibrium noise of the Luttinger liquid and is a property of the free system. What we really need to compute are the other three terms of equation \ref{eq:4_Sn_totexact}. \\
As we discussed in section \ref{sec:2_shotnoise} there are two limits where the shot noise should take a precise form, $i_b\ll i_0$ ($i_t\simeq 1$) and $i_b \simeq i_0$ ($i_t<<i_0$). In this two regimes indeed we can think of the tunneling current to be generated respectively by few quasi-particles which are backscattered independently or by particles that once in a while manage to be transmitted through a very strong constriction. In our stochastic picture the shot noise in the first scenario would be suppressed as $1/m$ as the density-density stochastic correlations does and in the second scenario the integer charge would be given by the process that sees $q$ passing to different minima of the cosine potential that has unit periodicity in $q$. \\
The results for the weak backscattering $i_b \ll i_0$ with a shot noise governed by fractional charges are shown in figure \ref{fig:4_shotnoise_qp} and they match the expected results for a proportionality constant of $1/m$. For the shot noise in the limit of strong backscattering $i_t\ll i_0$ we instead have only preliminary results which at high enough $m$, such as $m=5$, seems to agree with an integer charge dominated shot noise (figure \ref{fig:4_shotnoise_p}). As we saw for the conductivity in the strong backscattering regime the TWA seems to correctly work only for big enough $m$.


In this chapter we introduced a new approach for the problem of scattering in $\chi$ LL . The TWA we used allowed us to compute many physical observables and agrees almost always to an acceptable degree with other theoretical tools which have been used in the past. We mention here that, although not treated in the thesis, the TWA can be set up also for interferometer geometries with multiple tunneling points. Up to now we only considered the steady state problems for which many know results exists. The true power of our approach is that it allows to treat time-dependent problems as the scattering of wavepackets which at the moment have not been explored in a satisfactory way. The next chapter will be devoted to this.
\\

\chapter{Real-time dynamics}\label{chap:5}
In the previous chapter we introduced the TWA and we showed that it is well behaved in steady state situations. In this chapter we will use it to compute real time dynamics. The chapter closely re-propose the situation we saw in chapter \ref{chap:3} in the classical theory for the scattering of wavepackets in the geometry of figure \ref{fig:3_geometry}. We will see how the dynamics becomes in most cases intuitive by having in mind the classical theory.\\
Again we will consider a gaussian wavepacket characterized by a charge $Q$ and a width $\sigma$:
\begin{equation}
        \rho_{wp}(x) = Q \frac{1}{\sqrt{\pi\sigma^2}}e^{-x^2/\sigma^2}
\end{equation}
The new ingredient of the TWA are stochastic fluctuations which as we already saw well mimic the quantum fluctuations for the quasi-particle tunneling Hamiltonian when $m\gtrsim 1$. This mean also that that the arrival time of the wavepacket will be chosen such that the system has reached equilibrium as discussed in section \ref{sec:4_equilibrium}. During the thesis work we mainly focused on currents as the observables of the problem and we decided to postpone the investigation of noise properties to future studies.
\section{IQH case}
\begin{figure}[!h]

\begin{subfigure}{.5\textwidth}
    \centering
  \includegraphics[width=\textwidth]{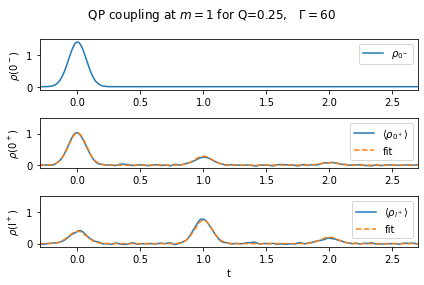}
    \caption{Q=0.25}
    \label{fig:5_IQH_q025}
\end{subfigure}
\begin{subfigure}{.5\textwidth}
  \centering
  \includegraphics[width=\textwidth]{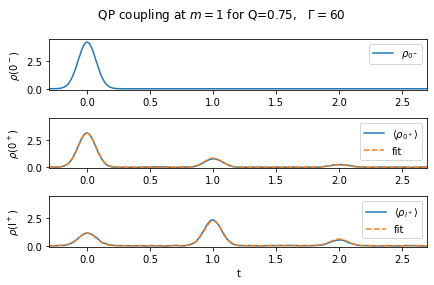}
  \caption{Q=0.75}
  \label{fig:5_IQH_q075}
\end{subfigure}
    \caption{Upper panels show the incident wavepacket, the center panels shows the transmitted current after the constriction at $x=0^+$ while the lower panel shows the backscattered current at $x=l^+$. The roundtrip time is $t_l=1$ so that in the figure also the scattering processes after one and two rountrip times are shown. Evidently the classical trajectory is absent and one recovers an Ohmic behaviour. The fit follows equation \ref{eq:5_IQH_transm} with $T= 0.72\pm1$ and $T=0.71\pm1$ for $Q=0.25$ and $Q=0.75$. $\sigma_{wp}=0.1$, $v=1$ and $\beta = \infty$ }
    \label{fig:5_IQH_nocl}
\end{figure}

In this section we will look at the quasi-particle scattering in the IQH limit with $m=1$. We know that the TWA in this limit might fail and also that the classical trajectories of section \ref{sec:3_qp_p_tunneling} did not reproduce at all the free particle behaviour we would expect. Indeed working in the basis of the Luttinger phonons is clearly not the best choice since the tunneling Hamiltonian is not quadratic as it is when written in the particle basis where the problem is a single particle one. In the weak tunneling regime at the classical level we have different behaviour for the $Q=0.75$ and $Q=0.25$ wavepackets but now using the fluctuations given by the TWA there is no difference with respect to the incident wavepacket charge as shown in figure \ref{fig:5_IQH_nocl}. \\

\begin{figure}[h!]
    \centering
    \includegraphics[width=10cm]{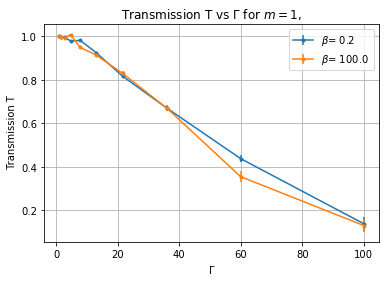}
    \caption{Transmission coefficient $T$ of the constriction in the $m=1$ case as a function of the coupling $\Gamma$ and for two values of the inverse temperature $\beta$. The transmission decrease as $\Gamma$ is decreased and it does so in a temperature independent way. Calculated by fitting equation \ref{eq:5_IQH_transm} with wavepackets of $\sigma_{wp}=0.1$ and $Q=0.5$. Cavity roundtrip time $t_l=1$}
    \label{fig:5_IQH_transm}
\end{figure}

The scattered wavepackets closely reproduce the free particle picture where:
\begin{equation}\label{eq:5_IQH_transm}
    \rho(0^+,t)= T \rho(0^-,t) \qquad \rho(l^+,t)= (1-T) \rho(0^-,t)
\end{equation}
where $0<T<1$ is a transmission coefficient which depend on $\Gamma$ and no dependence on the wavepacket charge $Q$ is present. The classical strong tunneling regime with excitations with the shape of the derivative of the incoming wavepacket (see figure \ref{fig:3_qp_strong}) is also absent, we just get $T\rightarrow 0$. \\
In figure \ref{fig:5_IQH_transm} we show the dependence on $\Gamma$. Note that the transmission coefficient is independent of the temperature $\beta$ as we found at DC consistently with an Ohmic behaviour. 
\\
The phenomenology we described here is quite interesting; even though the classical theory is completely wrong at $m=1$ the TWA successfully includes the quantum fluctuations of the quantum theory and gives qualitatively correct results. Notice in general that the range where we are now varying $\Gamma$ is at higher values with respect to the classical cases. A qualitative understanding of this is that the fluctuations somehow renormalize the strength of the coupling. For example higher temperatures means stronger fluctuations and so higher transmission $T$ in figure \ref{fig:5_IQH_transm}. However here is difficult to be quantitative since the coupling $\Gamma$ has a completely different effect in the classical and quantum cases. We will see later in the FQH that this reasoning can be expanded. 
\section{FQH case}
We can now discuss the FQH scenario where $m> 1$. In this regime as $m\rightarrow \infty$ the fluctuations of $q$ shrinks and if those are smaller we can expect to end up in the classical limit. The scale to which the fluctuations should be smaller is the periodicity of the cosine potential, the unit charge. Of course $m$ is not infinite and hence we expect some changes in the dynamics. We start to analyze the weak tunneling case.
\subsection{Weak tunneling: $\Gamma / v  < l^{-1} , \sigma_{wp}^{-1}$ }
\begin{figure}[!h]
\begin{subfigure}{.5\textwidth}
    \centering
  \includegraphics[width=\textwidth]{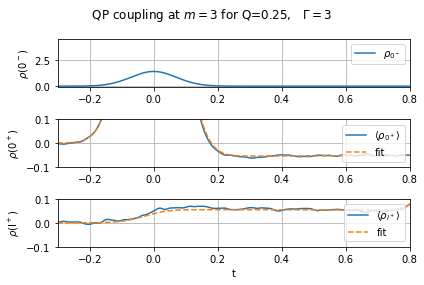}
    \caption{Q=0.25  }
    \label{fig:5_FQH_q025}
\end{subfigure}
\begin{subfigure}{.5\textwidth}
  \centering
  \includegraphics[width=\textwidth]{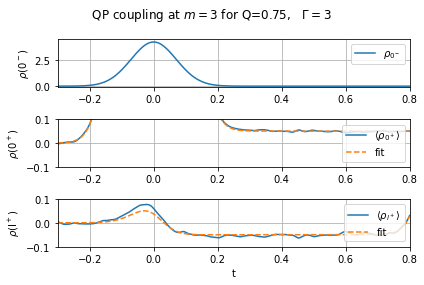}
  \caption{Q=0.75}
  \label{fig:5_FQH_q075}
\end{subfigure}
    \caption{Upper panels show the incident wavepacket, the center panels shows the transmitted current after the constriction at $x=0^+$ while the lower panel shows the backscattered current at $x=l^+$. The roundtrip time is $t_l=1$ so that only $t<t_l$ are shown. The fit follows equation \ref{eq:5_FQH_tail_fit} with $\Tilde{\Gamma}=0.05 \pm 0.01$ for $Q=0.25$ and  $\Tilde{\Gamma}=0.052\pm0.006$ for $Q=0.75$. $m=3$}
    \label{fig:5_FQH_tail}
\end{figure}

\begin{figure}
    \centering
    \includegraphics[width=10cm]{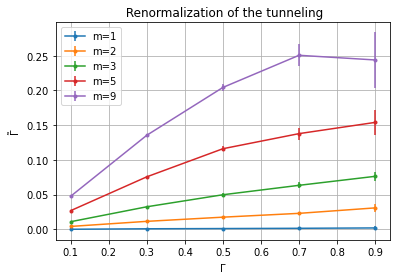}
    \caption{Renormalization of the coupling $\Tilde{\Gamma}$ as a function of $\Gamma$ in the weak coupling regime at zero temperature for different $m$. Calculated by fitting equation \ref{eq:5_FQH_tail_fit} with wavepackets of $\sigma_{wp}=0.1$ and $Q=0.75$. }
    \label{fig:5_FQH_tail_ren}
\end{figure}

\begin{figure}
\begin{subfigure}{.5\textwidth}
    \centering
  \includegraphics[width=\textwidth]{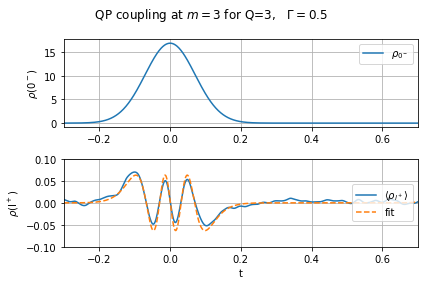}
    \caption{Q=3}
    \label{fig:5_FQH_q3}
\end{subfigure}
\begin{subfigure}{.5\textwidth}
  \centering
  \includegraphics[width=\textwidth]{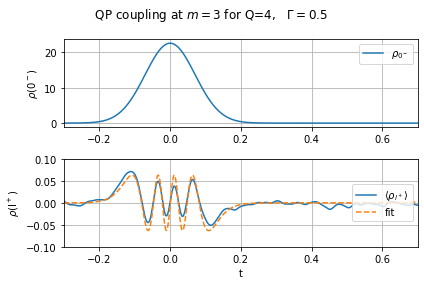}
  \caption{Q=4}
  \label{fig:5_FQH_q4}
\end{subfigure}
    \caption{Upper panels show the incident wavepacket while the lower panels shows the backscattered current at $x=l^+$. The roundtrip time is $t_l=1$ so that only $t<t_l$ are shown. The fit is the classical trajectory with $\Tilde{\Gamma}$ taken as half of the peak to valley width of the exitations in the lower panels,  $\Tilde{\Gamma}=0.06$ for $Q=3$ and  $\Tilde{\Gamma}=0.06$ for $Q=4$.  }
    \label{fig:5_FQH_crist}
\end{figure}

\begin{figure}[h!]
    \centering
    \includegraphics[width=10cm]{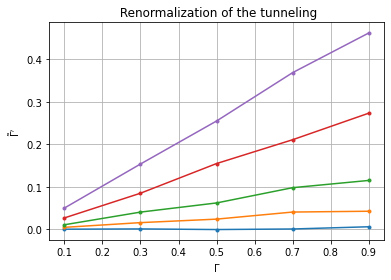}
    \caption{Renormalization of the coupling $\Tilde{\Gamma}'$ as a function of $\Gamma$ in the weak coupling regime at zero temperature for different $m$. Calculated by the peak to valley amplitude with the crystallization effect shown in figure \ref{fig:5_FQH_crist} with wavepackets of $\sigma_{wp}=0.1$ and $Q=3$.}
    \label{fig:5_FQH_crist_ren}
\end{figure}

\begin{figure}[h!]
    \centering
    \includegraphics[width=10cm]{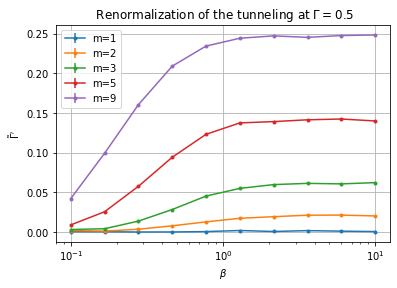}
    \caption{Renormalization of the coupling $\Tilde{\Gamma}'$ as a function of $\beta$ in the weak coupling regime for different $m$ and fixed $\Gamma$. Calculated by fitting equation \ref{eq:5_FQH_tail_fit} with wavepackets of $\sigma_{wp}=0.1$ and $Q=0.75$. }
    \label{fig:5_FQH_tail_renT}
\end{figure}

Differently from the $m=1$ here we have a difference between the weak and strong tunneling regimes, in particular for $m\gtrsim 3$. We start by analysing the scattering for the two charges $Q=0.25$ and $Q=0.75$ at a fixed value $m=3$ and zero temperature, shown in figure \ref{fig:5_FQH_tail}.
\\
The dynamics closely remember that of the classical theory and still show the charge commensurability effect we described in chapter \ref{chap:3} ! However the classical trajectory which best fit the quantum evolution has a different $\Gamma$ which we will call $\Tilde{\Gamma}$. In order to fit it we can take the current at $x=l^+$ and fit:
\begin{equation}\label{eq:5_FQH_tail_fit}
    \langle\rho(l^+,t) \rangle= \Tilde{\Gamma} \sin(2\pi \langle q(t)\rangle ) 
\end{equation}
which is the expression for the classical trajectories. As we anticipated in the $m=1$ case we are witnessing a renormalization of the coupling due to fluctuations. In figure \ref{fig:5_FQH_tail_ren} we show how $\Tilde{\Gamma}$ changes in the weak tunneling regime as a function of $\Gamma$ for different values of $m$. Note that for higher values of $m$ the renormalization is less effective because of the reduced fluctuations while at $m=1$ the fluctuations totally kill the classical trajectory. 
\\
Another feature of the classical trajectories was the crystallization (see figure \ref{fig:3_cristallization}) which showed consequent peaks and valleys in number exactly equal to the integer closest to the incident charge $Q$. In the quantum evolution this behaviour has already been predicted in \cite{Cristallization} using perturbation theory but here we are able to recover this effect in the context of the TWA and explain it in term of classical trajectories and small fluctuations around it. In figure \ref{fig:5_FQH_crist}  we show the results for the scattering of two wavepackets with charge $Q=3$ and $Q=4$ showing respectively 3 and 4 peaks. Interestingly also here we have a similar effective renormalization of the coupling which we can extract from the peak to valley width that in the classical trajectories is $2\Gamma$.
The renormalization $\Tilde{\Gamma}'$ depend both on $\Gamma$ and on $m$ as we show in figure \ref{fig:5_FQH_crist_ren} . 
Note that the renormalization $\Tilde{\Gamma}'$ and $\Tilde{\Gamma}$ in figures \ref{fig:5_FQH_crist_ren} and \ref{fig:5_FQH_tail_ren} are very similar and then we conclude that this phenomena does not depend on the incident wavepacket but only on the parameters of the system alone. 
\\
Once we described the physics at zero temperature we can look at finite temperatures and see what happens. As we already saw while discussing the equilibrium properties in section \ref{sec:4_equilibrium} we expect some effects only for $(\hbar\beta)^{-1} \gtrsim t_l$. Then from a physical point of view we expect temperature to increase the fluctuations and hence bring us away from the classical trajectories. Indeed this is what happens, in figure \ref{fig:5_FQH_tail_renT} we show that the renormalization of the coupling depends also on the temperature and that it decrease at high enough temperature $\hbar \beta< l/v$. The visibility of the classical trajectories remains in any case higher for higher $m$.
\\
\subsection{Strong tunneling: $\Gamma / v  < l^{-1} , \sigma_{wp}^{-1}$ }
\begin{figure}[!h]
    \centering
    \includegraphics[width=10cm]{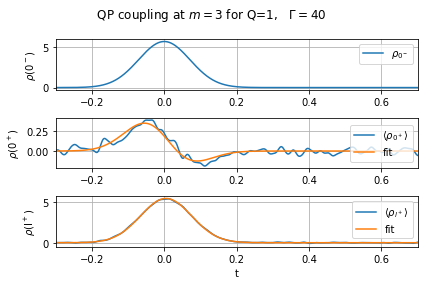}
    \caption{Upper panels show the incident wavepacket, the center panels shows the transmitted current after the constriction at $x=0^+$ while the lower panel shows the backscattered current at $x=l^+$. The roundtrip time is $t_l=1$ so that only $t<t_l$ are shown. The fit follows from the discussion in the text. Fitted parameters $T=0.05\pm0.01$ and $\Tilde{\Gamma}^{-1}=-0.028 \pm0.005$ }
    \label{fig:5_FQH_st}
\end{figure}

\begin{figure}[h!]
\begin{subfigure}{.5\textwidth}
    \centering
  \includegraphics[width=\textwidth]{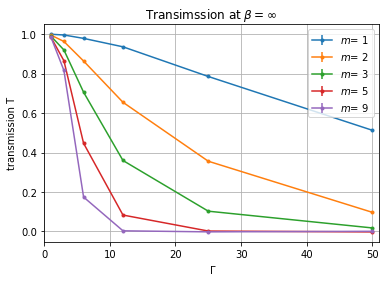}
    \caption{$T$}
    \label{fig:5_FQH_st_Tren}
\end{subfigure}
\begin{subfigure}{.5\textwidth}
  \centering
  \includegraphics[width=\textwidth]{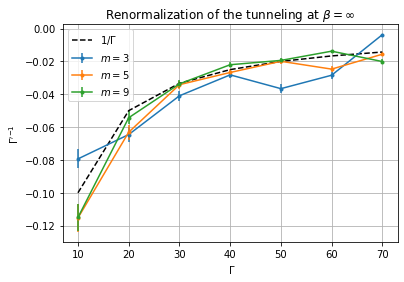}
  \caption{$\Tilde{\Gamma}$}
  \label{fig:5_FQH_st_Gammaren}
\end{subfigure}
    \caption{Renormalization of the classical coupling constant $\Tilde{\Gamma}$ (right) and transmission coefficient $T$ (left) at $\beta=\infty$. $Q=0.5$}
    \label{fig:5_FQH_st_ren}
\end{figure}

The strong tunneling regime for the classical trajectories generated a the peculiar phenomenology which was a total reflection of the excitations to the other part of the constriction and a zero charge excitation passing through the constriction with the shape of the derivative of the incoming wavepacket, reminiscent of a density-density coupling with a dual picture (see section \ref{sec:3_dd_scattering}). In figure \ref{fig:5_FQH_st} we show that indeed in the FQH case this picture survives. However in addition to it we also have a partial transmission of the wavepacket as in the IQH case (see equation \ref{eq:5_IQH_transm}). To fit the latter behaviour we look at the integral of the charge in the reflected channel at $x=l^+$ and compare it with the charge of the initial wavepacket so that:
\begin{equation}
    1-T = \frac{\int dt \, \langle \rho(l^+,t)\rangle }{Q} 
\end{equation}
Then to fit the "classical" effect which gives the derivative of the incident wavepacket we use the classical trajectory with a renormalized $\Tilde{\Gamma}$ plus the "quantum" transmission of the incident wavepacket:
\begin{equation}
    \rho(0^+,t) = \frac{1}{2\pi \Tilde{\Gamma}} \partial_t \rho(0^-,t) + T \rho(0^-,t)
\end{equation}
As we did for the weak tunneling case we can study the behaviour of $\Tilde{\Gamma}$ and $T$ varying the parameter $m$ and the coupling $\Gamma$. The transmission coefficient then goes down exponentially fast with $\Gamma$ for high $m$ as can be seen from \ref{fig:5_FQH_st_Tren} while it decrease slowly for $m=1$. In figure \ref{fig:5_FQH_st_Gammaren} we instead show $\Tilde{\Gamma}^{-1}$.
The effective density density coupling in the strong coupling regime seems to be changing sign when passing to the quantum case, further investigation of such phenomena are needed and for the moment we limit ourselves to noticing its existence.
\\
\\

In this last chapter we have seen how the TWA is able to recover the dynamics of free electrons of the IQH case at $m=1$ and at the same time identifies a regime of $m$ where the classical trajectories of chapter 3 are recovered. Our approach explains the crystallization of excitations with integer charges in simple terms and more in general can be used as a theoretical tool to explore the dynamics of time dependent excitations on FQH edges. The charge commensurability effect we described is a novel result which to our knowledge has never been described in the literature. Then also the density-density dual coupling arising in the strong tunneling regime has not been described before in the context of time dependent excitations. In the near future we plan to include the study of the noise properties in time dependent excitations settings as this is the main experimental tool available in electronic systems.

\chapter*{Conclusions}
\addcontentsline{toc}{chapter}{Conclusion}
The main result of this thesis is the development of a new framework to analyze the dynamics of scattering in $\chi$LL.
The TWA is a powerful tool that enabled us to recover both DC properties and dynamics of scattering in $\chi$LL using a mapping of the quantum problem onto a stochastic problem given by the classical theory plus fluctuations. It provides a completely new perspective, orthogonal to the approaches used in the past (perturbation theory, RG and integrability) which are best suited for DC properties. On the contrary, solving for the dynamics appears as a natural challenge in the TWA. The nature of this approximation is non-perturbative and it can predict some phenomenology also at strong quasi-particle tunneling. It is however still not clear to us which type of quantum fluctuations are not included in the TWA and further investigations of its limitations and its validity range will be object of future works. Nonetheless the picture emerging from our calculations is quite promising, quantum fluctuations corrections over the classical theory are well captured for the quasi-particle tunneling non-linearity in the whole regime of the FQH parameter $m$; from the $m=\infty$ limit where the TWA is exact up to $m=1$ where it clearly recovers the dynamics of free particles. \\
Thanks to the new physical intuition given by the TWA we were able to justify the surprising effect of crystallization (\cite{Cristallization}) in terms of classical trajectories\footnote{During the thesis work the chronological order of this was actually the opposite, at first we noticed the effect in the classical trajectories and then we discovered it in the literature. This was an important step as it made us realize that the classical trajectories must had been correct at least in some regime of the problem}. Then, thanks to the intuition of the classical theory, we also discovered a charge commensurability effect which to our knowledge has never been described in the literature. \\
This thesis sets up a novel method with which we will be able to inspect scattering of excitations in even more complex configurations. Among the others, we cite the possibility to treat anyonic interferometers like the one of \cite{Nakamura2020} with time dependent excitations whose scattering properties may depend on the fractional quantum statistics of quasi-particles in the FQH. The understanding of time dependent excitations scattering in FQH edges is in our opinion still at an embryonic stage and we are convinced that  our original work will give a new and important perspective on the problem. The importance of our achievements is made even stronger by the forthcoming realization of FQH physics in synthetic systems of atoms or photons. In these systems new probes and new experiments can be devised beyond what is usually possible in electronic systems with a more natural inclination to real time dynamics. \\

\newpage
\printbibliography
\newpage

\end{document}